\documentclass[utf8]{frontiersFPHY_mine} 

\setcitestyle{square} 
\usepackage{url,hyperref,lineno,microtype,subcaption}
\usepackage[onehalfspacing]{setspace}
\usepackage{amsmath}
\usepackage{color}
\usepackage{cancel}
\usepackage{ulem}

\def\keyFont{\fontsize{8}{11}\helveticabold }
\def\firstAuthorLast{Sinya Aoki {et~al.}} 
\def\Authors{Sinya Aoki\,$^{1,2,*}$ and Takumi Doi\,$^{2,3}$}
\def\Address{$^{1}$Center for Gravitational Physics, Yukawa Institute for Theoretical Physics,  Kyoto University, Kyoto 606-8502 , Japan\\
$^{2}$Quantum Hadron Physics Laboratory, RIKEN Nishina Center, Wako,  351-0198, Japan\\
$^{3}$Interdisciplinary Theoretical and Mathematical Sciences Program (iTHEMS), RIKEN, Wako,  351-0198, Japan
}
\def\corrAuthor{Sinya Aoki
}

\def\corrEmail{saoki@yukawa.kyoto-u.ac.jp}
\newcommand{\beqa}{\begin{eqnarray}}
\newcommand{\eeqa}{\end{eqnarray}}
\newcommand{\nn}{\nonumber}

\newcommand{\bx}{\mbox{\boldmath $x$}}

\def\dfrac#1#2{\displaystyle\frac{#1}{#2}}
\newcommand{\Pu}{p_{\uparrow}}

\newcommand{\Nu}{n_{\uparrow}}
\newcommand{\Nd}{n_{\downarrow}}

\begin{document}
\onecolumn
\firstpage{1}

\title[Lattice QCD and baryon-baryon interactions]
{Lattice QCD and baryon-baryon interactions: HAL QCD  method} 

\author[\firstAuthorLast ]{\Authors} 
\address{} 
\correspondance{} 

\extraAuth{}

\maketitle

\begin{abstract}
In this article, we review the HAL QCD method to investigate baryon-baryon interactions such as nuclear forces in lattice QCD. 
We first explain our strategy in detail to investigate baryon-baryon interactions by defining potentials in field theories such as QCD. 
We introduce the Nambu-Bethe-Salpeter (NBS) wave functions  in QCD
for two baryons below the inelastic threshold.  We then define the potential from NBS wave functions in terms of the derivative expansion, which is shown to reproduce the scattering phase shifts correctly below  the inelastic threshold.
Using this definition, we formulate a method to extract the potential in lattice QCD. 
Secondly, we discuss pros and cons of the HAL QCD method, by comparing it with the conventional method, where one directly extracts the scattering phase shifts from the finite volume energies through the L\"uscher's formula.
We give several theoretical and numerical evidences that the conventional method
combined with the naive plateau fitting for the finite volume energies in the literature
so far fails to work on baryon-baryon interactions due to contaminations of elastic excited states. 
On the other hand, we show that such a serious problem  can be avoided in the HAL QCD method
by defining the potential in an energy-independent way.
We also discuss systematics of the  HAL QCD method, in particular errors associated with a truncation of the derivative expansion.
Thirdly, we present several results obtained from the HAL QCD method, which include (central) nuclear force,  tensor force,  spin-orbital force, and three nucleon force.
We finally show the latest results calculated at
the nearly physical pion mass, $m_\pi \simeq 146$ MeV,
including hyperon forces which lead to form $\Omega\Omega$ and $N\Omega$ dibaryons. 

\tiny
\keyFont{ \section{Keywords:}
lattice QCD, nuclear forces, baryon-baryon interactions, dibaryons, equation of state, neutron stars
}  
\end{abstract}

\section{Introduction}
\label{sec:intro}

How do nuclear many-body systems emerge from the fundamental degrees of freedom,
quarks and gluons?
It has been a long-standing problem to establish a connection between
nuclear physics and the fundamental theory of strong interaction,
quantum chromodynamics (QCD).
In particular, nuclear forces serve as
one of the most basic constituents in nuclear physics,
which are yet to be understood from QCD.
While so-called realistic nuclear forces~\citep{Stoks:1994wp,Wiringa:1994wb,Machleidt:2000ge}
have been established
with a good precision,
they are constructed phenomenologically based on scattering data
experimentally obtained.
Recent development in effective field theory (EFT) provides
a more systematic approach for nuclear forces
from a viewpoint of chiral symmetry in QCD~\citep{Weinberg:1990rz,Weinberg:1991um,Epelbaum:2008ga,Machleidt:2011zz,Hammer:2019poc},
whose unknown low-energy constants, however, cannot be determined 
within its framework but are obtained only 
by the fit to the experimental data.
Under these circumstances,
it is most desirable to determine nuclear forces
as well as general baryon-baryon interactions
from first-principles calculations of QCD,
the lattice QCD method.
Once baryon forces are extracted from QCD,
we can solve finite nuclei, hypernuclei and nuclear/hyperonic matter
by employing various many-body techniques developed in nuclear physics.
The outcome is expected to make a significant impact on
our understanding of nuclear astrophysical phenomena
such as supernovae, binary neutron star merges and nucleosynthesis.

In this paper, we review the HAL QCD method to
determine baryon-baryon interactions
in lattice QCD.
In this method,
integral kernels, or so-called ``potentials'', are first extracted
from lattice QCD, and physical observables such as scattering phase shifts
and binding energies are calculated
by solving the Schr\"odinger equation with obtained potentials in the infinite volume.
We show that the notion of potential can be rigorously
introduced as a representation of the S-matrix in quantum field theories as QCD.
The essential point is that the potentials are defined through
the Nambu-Bethe-Salpeter (NBS) wave functions,
in which the information of phase shifts are encoded in their asymptotic behaviors.
We employ a non-local and energy-independent potential where the non-locality is defined
through the derivative expansion.
In particular, energy-independence of the potential is useful
since one can extract the potential from the ground state as well as elastic excited states
simultaneously. This enables us to avoid
the notorious signal-to-noise issue for multi-baryon systems in lattice QCD
(or the ground state saturation problem),
and to make a reliable determination of baryon-baryon interactions. 

In lattice QCD, there also exists a conventional method,
in which phase shifts are obtained from finite volume energies through the L\"uscher's formula.
For meson-meson systems,
  a number of works have been performed based on the L\"uscher's formula~\cite{Briceno:2017max},
  where finite volume energies are extracted utilizing the variational method~\citep{Luscher:1990ck}.
  The L\"uscher's formula has been generalized for various systems,
  such as boosted systems~\cite{Rummukainen:1995vs}, arbitrary spin/partial waves~\cite{Briceno:2013lba, Briceno:2014oea}
  and three-particle systems~\cite{Hansen:2014eka,Hansen:2019nir}.
While theoretical bases are well established for both conventional method and HAL QCD method,
numerical results for baryon-baryon systems at heavy pion masses have shown 
inconsistencies with each other.
In this paper, we make a detailed comparison between two methods,
scrutinizing possible sources of systematic errors.
In particular, we examine whether the systematic errors associated with
excited state contaminations are controlled or not
in the procedure of the conventional method in the literature (``the direct method''),
namely, simple plateau fitting for the ground state at early Euclidean times.
We also examine systematic errors in the HAL QCD method,
in particular,
the truncation error of the derivative expansion.
We show theoretical and numerical evidences that
the inconsistency between two methods originates from
excited state contaminations in the direct method.
We also demonstrate that the inconsistency can be actually resolved
if and only if finite energy spectra are properly obtained with an improved method rather than the naive plateau fitting in the conventional method.

After establishing the reliability of the HAL QCD method,
we present the numerical results of nuclear forces from the HAL QCD method
at various lattice QCD setups.
The results at heavy pion masses for central and tensor forces
are shown and their quark mass dependence
as well as physical implications are discussed.
The calculations of spin-orbit forces and three-nucleon forces
are also given.
Once nuclear forces are obtained, one can solve nuclear many-body systems
with the obtained potentials.
We study finite nuclei, nuclear equation of state and structure of neutron stars
based on lattice nuclear forces at heavy pion masses.
Finally, the latest results of nuclear forces near the physical pion mass are
presented,
as well as hyperon forces, which are shown to generate $\Omega\Omega$ and $N\Omega$ dibaryons.

This paper is organized as follows.
In Sec.~\ref{sec:lat},
we discuss methods to study baryon-baryon interactions from lattice QCD.
After briefly introducing the conventional method and its actual practice, called the ``direct method'',
we describe the detailed theoretical formulation
as well as its practical demonstration for the newly developed method, the HAL QCD method.
In Sec.~\ref{sec:comp}, we discuss pros and cons of these two methods,
and compare the numerical results at heavy pion masses.
We present evidences that the results from the direct method
suffer from uncontrolled systematic errors associated with the excited state contaminations.
In Sec.~\ref{sec:NN_pot},
we summarize results on nuclear potentials in the HAL QCD method.
After reviewing the results obtained at heavy pion masses
for central and tensor forces in the parity-even channel
as well as spin-orbit forces and three-nucleon forces,
we present
nuclear many-body calculations based on lattice nuclear forces
for double-magic nuclei, equation of state and the structure of neutron stars.
Latest results for nuclear forces near the physical pion mass are also given.
In Sec.~\ref{sec:dibaryon},
we present hyperon forces near the physical pion mass,
which lead to $\Omega\Omega$ and $N\Omega$ dibaryons.
Sec.~\ref{sec:conclusion} is devoted to the summary and concluding remarks.

\section{Two baryon systems in lattice QCD}
\label{sec:lat}
In lattice QCD, the 2-pt function for a hadron $H$, created by $O_H^\dagger$ and annihilated by
$O_H$, is expressed as 
\beqa
\langle 0\vert  O_H(\vec p, t)  O_H^\dagger(\vec p, 0) \vert 0 \rangle
&=& \sum_{n=0}^\infty Z_n(\vec p) e^{- E_n(\vec p) t} +\cdots,
\quad Z_n(\vec p) = \vert \langle 0 \vert O_H(\vec p, 0) \vert n, E_n(\vec{p}) \rangle  \vert^2,
\eeqa 
where
$\vert n, E_n(\vec p) \rangle$ is the $n$-th one-particle state with a mass $m_n$, a momentum $\vec p$ and an energy $E_n(\vec p) = \sqrt{m_n^2 +\vec p^2}$,
and ellipses represent contributions from multi particle states. We here assume $m_0 < m_{n>0}$, so that $m_0$ is the hadron mass for the ground state, which can be extracted from the asymptotic behavior of the  2-pt function in the large $t$ as
\beqa
\langle 0\vert  O_H(\vec p, t)  O_H^\dagger(\vec p, 0) \vert 0 \rangle &\simeq& Z_0 (\vec p) e^{- E_0(\vec p) t}+ O\left(e^{-E_{n>0}(\vec p)t} \right), \quad t\rightarrow \infty ,
\eeqa
where finite volume artifact is exponentially suppressed and can be eliminated by an infinite volume extrapolation.

So far, this method in lattice QCD (and the extension to lattice QCD+QED) has successfully reproduced  light hadron spectra~\citep{Durr:2008zz}
including the proton--neutron mass splitting~\citep{Borsanyi:2014jba}.
A simple application of the method, however, does not work well for an investigation of hadron interactions.
For example, the 2-pt function of two baryons in the center of mass system behaves in the large $t$ as
\beqa
\langle 0\vert O_{BB}(\vec 0, t)  O_{BB}(\vec 0, 0)^\dagger\vert 0 \rangle &\simeq&
Z_{BB} e^{- E_{BB}  t} +\cdots,
\label{eq:2pt_BB}
\eeqa
where we obtain  the lowest energy $E_{BB}$.
In the infinite volume limit, $E_{BB}$
behaves as
$E_{BB} = 2m_B$  or $E_{BB} = 2m_B - \Delta E$ depending on an absence or presence of bound state.
Here $m_B$ is the corresponding baryon mass and $\Delta E > 0$ is the binding energy of the lowest bound state.  Only the binding energy of the bound state  can be extracted by this simple method and thus more sophisticated methods are required.
Currently there are two methods to investigate hadron interactions in lattice QCD, the direct method (or finite volume method) and the HAL QCD method,
which are explained  in this section.

\subsection{Direct method}
\label{subsec:direct}
The method most widely used  to investigate hadron interactions in lattice QCD is to extract  scattering phase shifts from energy eigenvalues in 3-dimensional finite boxes through the L\"uscher's finite volume formula~\citep{Luscher:1990ux}.
For example, in the case of the $S$-wave scattering phase shift, $\delta_0(k)$, the formula reads
 \begin{equation}
  k \cot \delta_0 (k) = \frac{1}{\pi L} \sum_{\vec n\in \mathbf{Z}^3}\frac{1}{\vec n^2 -q^2},
  \qquad q=\frac{k L}{2\pi}, 
  \label{eq:kcot_delta}
\end{equation}   
where  $k$ is determined through  $E_{BB}(L) = 2\sqrt{k^2 + m_B^2}$ with $E_{BB}(L)$
being the energy of the two baryon measured in lattice QCD on a finite box with the spatial extension $L$ as in eq.~(\ref{eq:2pt_BB}).
We here neglect the partial wave mixing in the cubic group  and spin degrees of freedom, for simplicity.
Only the discrete sets of point $(k^2,k\cot\delta_0(k))$, which satisfies eq.~(\ref{eq:kcot_delta}),
are realized on a given volume $L^3$. Thus the scattering phase shift $\delta_0(k)$ at the corresponding $k$ can be extracted in lattice QCD, simply by  measuring the finite volume energy, $E_{BB}(L)$.
Note that the formula assumes that the hadron interaction is accommodated within the lattice box and is not distorted by the
  finite volume artifact, which condition should be examined numerically to be satisfied in actual calculations.

\begin{figure}[t]
\begin{center}
\includegraphics[width=8.5cm]{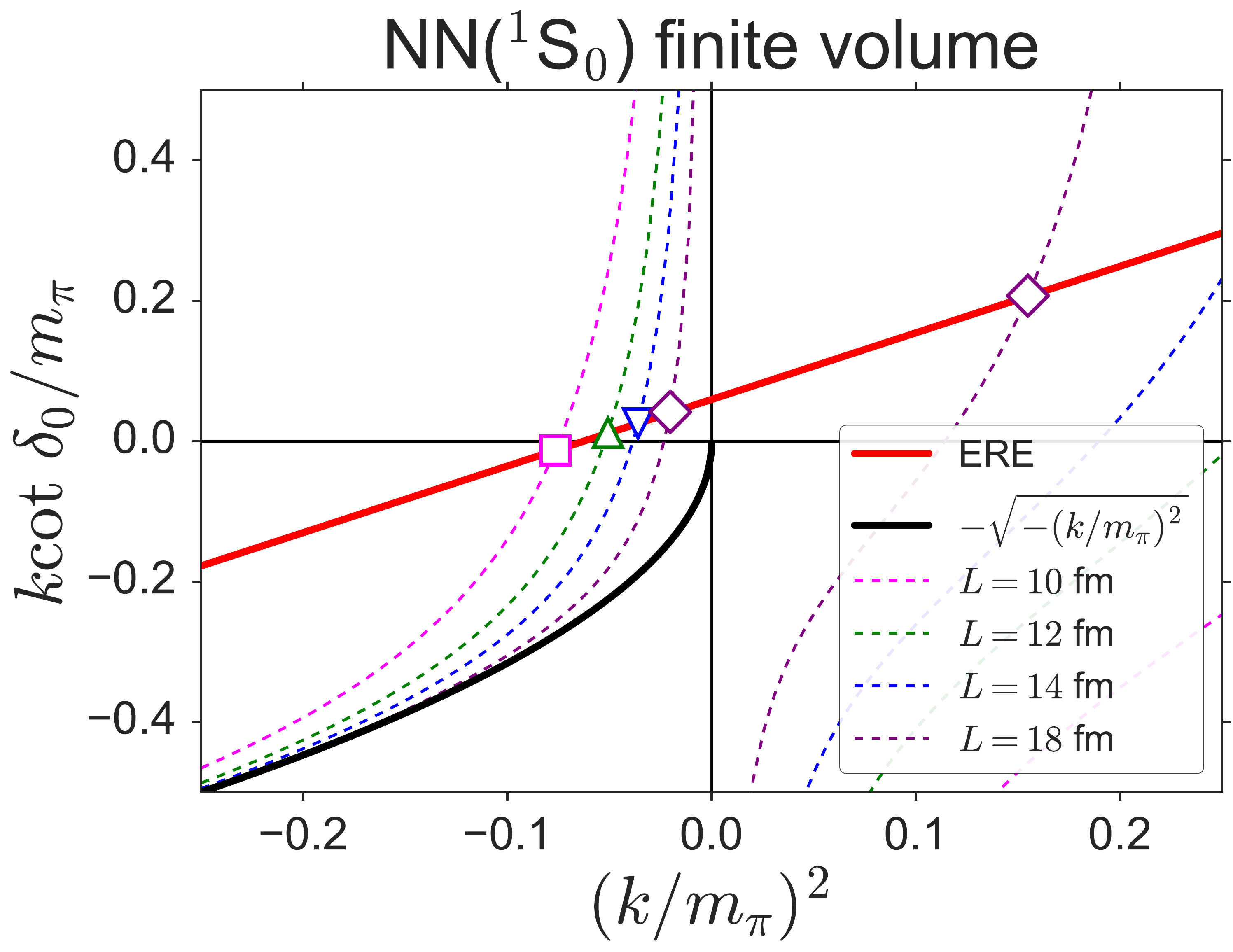}%
\hskip 0.5cm
\includegraphics[width=8.5cm]{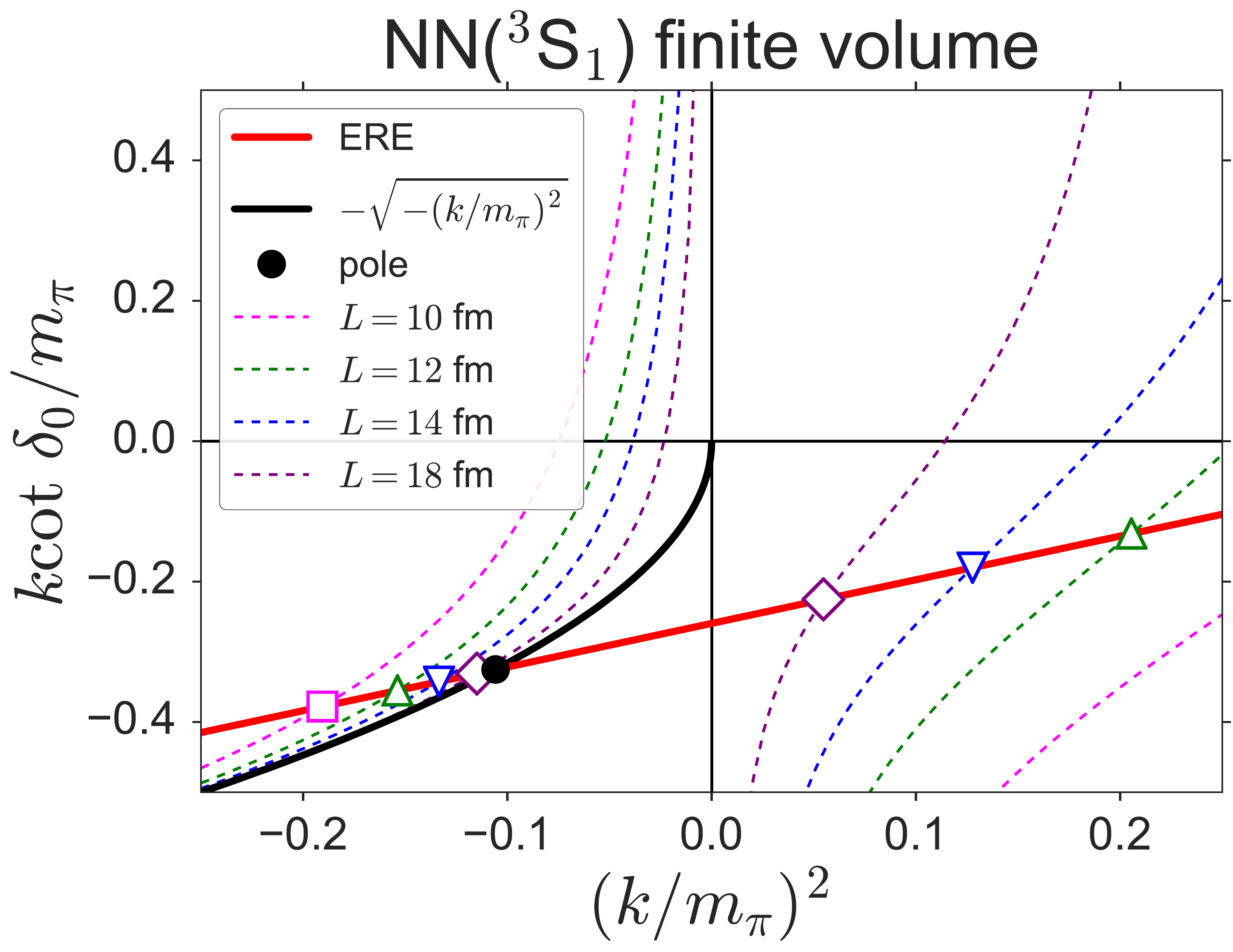}%
\end{center}
\caption{A determination of $k\cot\delta_0(k)/m_{\pi}$ from energies of the two nucleon state in the finite volume. Taken from \citep{Iritani:2017rlk}.
}
\label{fig:ERE_FV}
\end{figure}

In Fig.~\ref{fig:ERE_FV},
we illustrate how scattering phase shifts and the bound state energy can be extracted by this method  
in the case of  the $NN$ scatterings.
In the figure, the red solid line represents the effective range expansion (ERE) for $k\cot\delta_0(k)/m_\pi$ at the Next-to-Leading order (NLO) as
\beqa
 \frac{k}{m_\pi} \cot \delta_0 (k) &=& \frac{1}{a_0 m_\pi} +  \frac{r_0 m_\pi}{2} \frac{k^2}{m_\pi^2}
 \label{eq:ERE_nlo}
\eeqa
where the scattering length $a_0$ and the effective range $r_0$ are  taken to be
$a_0 m_\pi = 16.8$, $r_0 m_\pi = 1.9$ for $NN(^1S_0)$  (Left) or
 $a_0 m_\pi = -3.8$, $r_0 m_\pi = 1.3$ for $NN(^3S_1)$ (Right) with $m_\pi = 140$ MeV,
 while colored dashed lines represent the  L\"uscher's finite volume formula, eq.~(\ref{eq:kcot_delta})
 on $L=10, 12,14,18$ fm. Discrete points which satisfy both the L\"uscher's finite volume formula and the ERE are realized on each volume, as shown by the open squares, up/down triangles and diamonds.

A distribution of the allowed $k^2$ for $k^2 >0$ becomes denser as the volume increases, so as to be continuous in the infinite volume limit, while  a sequence of discrete points for $k^2 < 0$ leads to an accumulation point, which corresponds to the scattering state at $k^2=0$ in the left figure or the bound state pole, denoted by the  black solid circle in the right figure.
It is noted here that the bound state pole appears as the intersection between the ERE and the bound state condition,  $ -\sqrt{-(k/m_\pi)^2}$ (black solid line). To see this,  we first write
\beqa
k\cot \delta_0(k) &=& ik \cdot\frac{S(k)+1}{S(k)-1}, \qquad S(k) = e^{2i\delta_0(k)},
\eeqa
where $S(k)$ is the S-matrix for the $NN$ elastic scattering. The bound state energy $\kappa_b$ can be extracted from the pole of this S-matrix as
\beqa
S(k\sim i\kappa_b) \simeq \frac{-i\beta_b^2}{k-i\kappa_b},
\eeqa
where $\beta_{b}^2$ is real and positive for physical poles~\citep{Sitenko}.
Thus at $k^2 \simeq -\kappa_b^2$, we have
\beqa
\left. k\cot \delta_0(k)\right\vert_{k=i\kappa_b} &=& -\kappa_b =\left. -\sqrt{-k^2}\right\vert_{k=i\kappa_b}, 
\eeqa
which means that the {binding} momentum $k=i\kappa_b$ is given by an intersection between
$k\cot \delta_0(k)$ and $-\sqrt{-k^2}$.
Moreover, since
\beqa
\left.\frac{d}{d k^2}\left[k\cot\delta_0(k) - (-\sqrt{-k^2})\right]\right\vert_{k^2=-\kappa_b^2} &=& 
-\frac{1}{\beta_b^2} < 0,
\label{eq:phys_pole}
\eeqa
the slope of $k\cot\delta_0(k)$ must be smaller than that of $-\sqrt{-k^2}$ as a function of $k^2$ at the bound state pole, as in the case of Fig.~\ref{fig:ERE_FV} (Right).
The finite volume analysis thus provides not only an infinite volume extrapolation of the binding energy
but also a novel way to examine the normality of the result in the direct method~\citep{Iritani:2017rlk}. 

\subsection{HAL QCD method}
\label{subsec:HALQCD}
\subsubsection{Formulation}
\label{subsubsec:strategy}
The HAL QCD method, another method to investigate hadron interactions in lattice QCD, employs the equal time Nambu-Bethe-Salpeter (NBS) wave function, defined by
\beqa
\phi_{\bf k}({\bf r}) e^{-W_{\bf k} t}&\equiv& \langle 0 \vert N({\bf x} +{\bf r}, t) N({\bf x},t) \vert NN, W_k \rangle,
\eeqa 
where $\vert NN, W_k \rangle$ is the $NN$ eigenstate in QCD with the center of mass energy $W_k =2\sqrt{{\bf k}^2 + m_N^2}$
and the nucleon mass $m_N$,
and $N({\bf x},t)$ is a nucleon (annihilation) operator, made of quarks.
Other quantum numbers such as spin/isospin of two nucleons are suppressed for simplicity.
We mainly use 
\beqa
N_\alpha(x) &=& \varepsilon^{abc} \left( u^{a\, T} (x) C \gamma_5 d^b(x) \right) q_\alpha^c(x), \qquad x\equiv({\bf x},t),
\eeqa
where $C=\gamma_2\gamma_4$ is the charge conjugation matrix, $q=u (d) $ for proton (neutron).
Other choices such as smeared quarks are possible here, and such arbitrariness is considered to be a choice of the scheme
for the definition of the NBS wave function or the potential.  (See \citep{Kawai:2017goq} for such an example.)
Throughout this paper, we consider the $NN$ elastic scattering, so that $W_k < W_{\rm th} \equiv 2m_N + m_\pi$, where $m_\pi$ is the pion mass.
Note that this condition is also necessary for the finite volume method in the previous subsection.

Since interactions among hadrons are all short-ranged in QCD,
there exists some length scale $R$, beyond which ({\it i.e.}  $r \equiv \vert {\bf r}\vert > R$ )
the NBS wave function satisfies the Helmholtz equation as
\beqa
(k^2+\nabla^2) \phi_{\bf k} ({\bf r}) &\simeq & 0, \qquad k=\vert {\bf k}\vert .
\label{eq:HHE}
\eeqa
Furthermore, it behaves for large $r > R$ as
\beqa
 \phi_{\bf k} ({\bf r}) &\simeq & \sum_{l,m} Z_{l,m} \frac{\sin(kr - l \pi/2 +\delta_l(k))}{kr} Y_{lm}(\Omega_{\bf r}),
\eeqa
where $Y_{lm}$ is the spherical harmonic function for the solid angle $\Omega_{\bf r}$ of ${\bf r}$,
and we ignore spins of nucleon for simplicity\footnote{The formula becomes more complicated if the nucleon spins are considered~\citep{Ishizuka:2009bx,Aoki:2009ji}.}. 
Here it is important to note that the NBS wave function contains information of the phase $\delta_l(k)$ of the S-matrix for the orbital angular momentum $l$, which is a consequence of the unitarity of the S-matrix in QCD~\citep{Lin:2001ek,Aoki:2005uf}. 
 
In the HAL QCD method,  the non-local but energy-independent potential is defined from the NBS wave function through the following equation,
\beqa
(E_k - H_0)  \phi_{\bf k} ({\bf r}) &=& \int d^3 \, r^\prime U({\bf r}, {\bf r}^\prime )  \phi_{\bf k} ({\bf r}^\prime), \quad E_k = \frac{k^2}{2 m}, \quad H_0 =-\frac{\nabla^2}{2m}, \quad m=\frac{m_N}{2}, 
\label{eq:potential}
\eeqa
for $W_k < W_{\rm th}$, and  eq.~(\ref{eq:HHE}) implies $U({\bf r},{\bf r}^\prime) = 0$ for $r > R$.
While an existence of $U({\bf r},{\bf r}^\prime)$ has been shown in~\citep{Ishii:2006ec,Aoki:2009ji,Aoki:2012tk}, the non-local potential which satisfies eq.~(\ref{eq:potential}) is not unique.
Thus we have to define the potential uniquely, by specifying how to extract it.
For this purpose,  
we introduce the derivative expansion,
$U({\bf r},{\bf r}^\prime) = V({\bf r}, \nabla)\delta^{(3)}({\bf r} -{\bf r}^\prime)$, 
 whose lowest few orders for the $NN$ with a given isospin channel are written as
\beqa
 V({\bf r}, \nabla)  &=& \underbrace{V_0( r) + V_\sigma (r) ({\bf \sigma}_1 \cdot  {\bf \sigma}_2 )
+ V_T(r) S_{12}}_{\rm LO} + \underbrace{V_{\rm LS}(r) {\bf L}\cdot {\bf S}}_{\rm NLO} +O(\nabla^2) ,
\eeqa
where $V_0( r)$ is the central potential, $ V_\sigma (r)$ is the spin dependent potential
with ${\bf \sigma}_i$ being the Pauli matrix acting on the spinor index of the $i$-th nucleon,
$V_T(r)$ is the tensor potential with the tensor operator $S_{12} = 3 (\hat{\bf r}\cdot {\bf \sigma}_1) 
 (\hat{\bf r}\cdot {\bf \sigma}_2) -({\bf \sigma}_1 \cdot  {\bf \sigma}_2 )$ ( $\hat{\bf r} \equiv {\bf r}/r$), 
and $V_{\rm LS}(r)$ is the spin-orbit (LS) potential with the angular momentum ${\bf L} ={\bf r}\times {\bf p}$ and the total spin ${\bf S} =({\bf \sigma}_1+{\bf\sigma}_2)/2$. 
It is noted that an expansion of the non-local potential is not unique. For example, we may improve the convergence of the expansion by modifying the $\nabla$ operator~\citep{Sugiura:2017vwo}.
 
Once we obtain the approximated potential at lowest few orders, we can calculate the scattering phase shifts or the binding energies of possible bound states by solving the Schr\"odinger equation with this potential in the infinite volume.
As is the case for the finite volume method, it is necessary that the potential is not distorted by the finite volume artifact,
  but this can be checked easily since the potential itself is explicitly obtained. 
We can also check how good the approximated potential is, by
increasing the order of the expansion.
Needless to say,
the approximated potential depends on momenta of input wave functions.
As pointed out in~\citep{Aoki:2017yru},
these dependences of the approximated potentials
have been misidentified with those of  the  non-local potential
in the literature~\citep{Yamazaki:2017gjl}.
In the next subsection, we will explicitly demonstrate how this procedure works.

\subsubsection{Demonstration}
In order to see how the scattering phase shifts can be  obtained by the HAL QCD method, we consider the quantum mechanics for a spinless system with a separable potential, defined by
\beqa
U({\bf r}, {\bf r}^\prime) &=& \omega v({\bf r} ) v({\bf r}^\prime ), \quad v({\bf r}) \equiv e^{-\mu r} .
\label{eq:sep_pot}
\eeqa
The $S$-wave solution of the Schr\"odinger equation with this potential is given exactly by
\beqa
\phi_k^0(r) &=&
  \dfrac{e^{i\delta_0(k)}}{k r} \left[\sin\{ k r+\delta_0(k)\} -\sin\delta_0(k) e^{-\mu r}
\left( 1+\dfrac{r(\mu^2+k^2)}{2\mu}\right)\right], 
\eeqa 
where
\beqa
k\cot\delta_0(k) &=& -\frac{1}{4\mu^2}\left[2\mu(\mu^2-k^2) -\frac{3\mu^2+k^2}{4\mu^3}(\mu^2+k^2)^2 +\frac{(\mu^2+k^2)^4}{8\pi m\omega}\right]  ,
\eeqa
which is the 4-th order polynomials in $k^2$.
In order to make the scattering phase shift a more complicated function of $k^2$, 
we artificially modify the wave function
from $\phi^0_k(r)$ to $\phi_k(r)$ which is defined by
\beqa
\phi_k(r)
 &=& 
 \left\{
 \begin{array}{ll}
 \phi^0_k(r) & (r \le R)  \\
 C(k)  \dfrac{e^{i\delta_R(k)}}{k r} \sin\{ k r+\delta_R(k)\}  & (r > R), 
 \end{array}
 \right.
\eeqa 
where $R$ is an infrared cutoff,
and it is understood that the potential is modified accordingly.
 The continuity of $\phi_k(r)$ and $\phi^\prime_k(r)$ at $r=R$
 gives
\beqa
k\cot\delta_R(k) &=& k\frac{Y \cot(kR) +X }{X \cot(kR)-Y} , \quad X=\phi_k^0(R), \ \left. Y= \frac{d 
}{dr} [r  \phi_k^0(r) ]\right\vert_{r=R} ,
\eeqa
as well as $C(k) = X/\sin(kR+\delta_R(k))$.
  Hereafter, we study
  how the scattering phase shifts are obtained
  in the HAL QCD method.

The derivative expansion for the $S$-wave scatterings leads to
\beqa
 V({\bf r}, \nabla)  &=& V_0(r) + V_1(r) \nabla^2 +O(\nabla^2) ,
\eeqa
and we consider to determine the potential in each order from $\phi_k(r)$.

The leading order (LO) potential  is given by
\beqa
V^{\rm LO}({\bf r}, \nabla) = V_0^{\rm LO} (r;k) =\frac{ (E_k - H_0)\phi_k(r)}{\phi_k(r)} ,
\eeqa  
while the next-to-leading order (NLO) potential is extracted as
\beqa
V^{\rm NLO}({\bf r}, \nabla) &=& V^{\rm NLO}_0 (r;k_1,k_2) +  V^{\rm NLO}_1(r;k_1,k_2) \nabla^2, 
\eeqa
where
\beqa
\left(
\begin{array}{c}
V_0^{\rm NLO}(r;k_1,k_2) \\
V_1^{\rm NLO}(r;k_1,k_2) \\
\end{array}
\right) &=& \frac{1}{D(r;k_1,k_2)}
\left(
\begin{array}{c}
2m\left[V_0^{\rm LO}(r;k_2) E_{k_1} -  V_0^{\rm LO}(r;k_1) E_{k_2}\right]\\
 V_0^{\rm LO}(r;k_2)  -  V_0^{\rm LO}(r;k_1) \\
\end{array}
\right), \nn \\
D(r;k_1,k_2) &=& 2m \left[V_0^{\rm LO}(r;k_2)-  V_0^{\rm LO}(r;k_1) -( E_{k_2}- E_{k_1} )\right].
\eeqa
  Note that the potential in each order in the derivative expansion $\{V_0(r), V_1(r), \cdots\}$
  are defined to be $k$-independent,
  while the potentials approximately obtained in each LO/NLO analysis,
  $\{V_0^{\rm LO}(r;k)\}$ and $\{V_0^{\rm NLO}(r;k_1,k_2), V_1^{\rm NLO}(r;k_1,k_2)\}$,
  have implicit $k$-dependence
  due to the truncation error in the derivative expansion~\citep{Aoki:2017yru}.

\begin{figure}[t]
\begin{center}
\includegraphics[width=8.3cm]{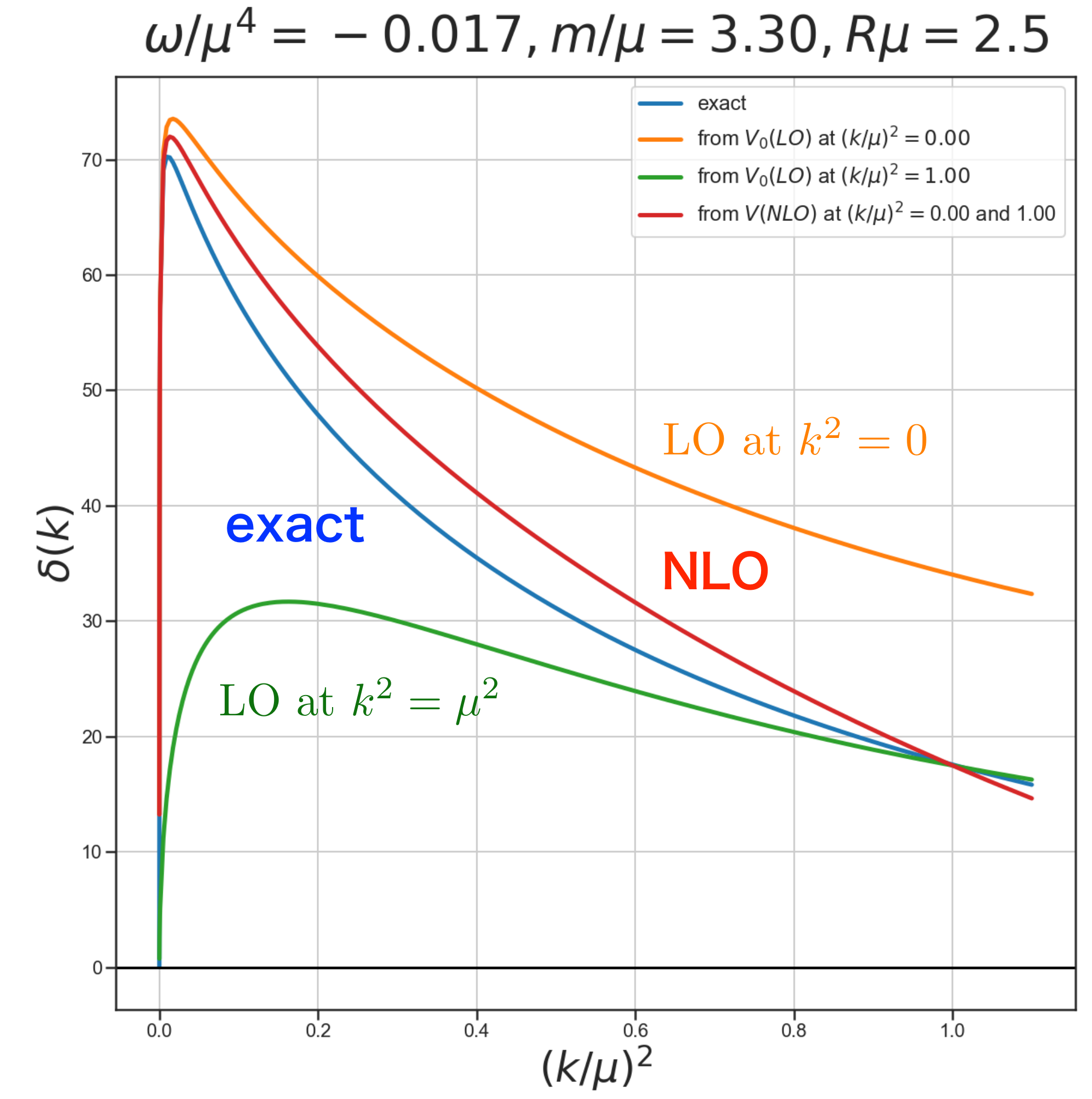}%
\hskip 0.5cm
\includegraphics[width=8.9cm]{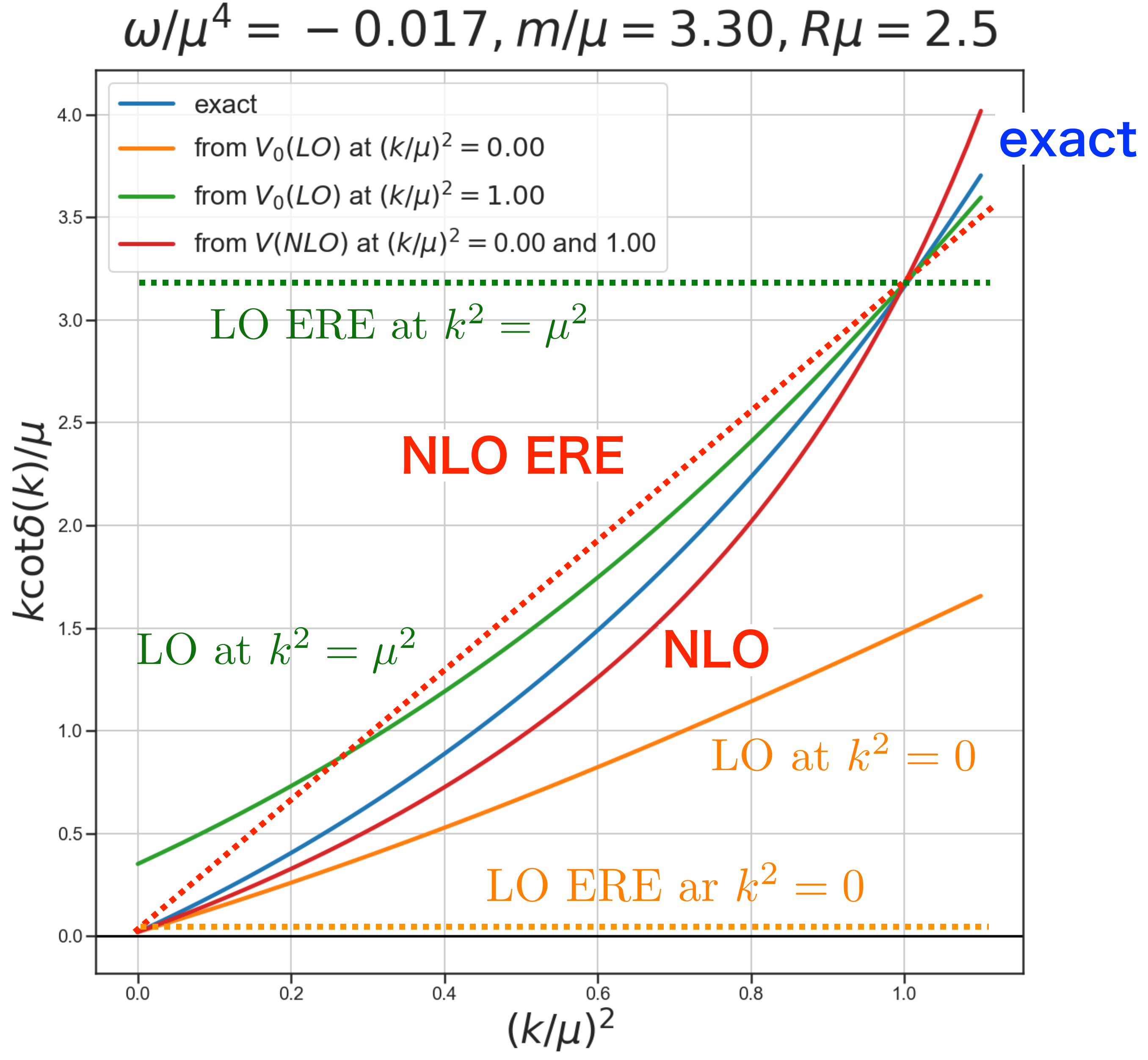}%
\end{center}
\caption{The scattering phase shifts $\delta(k)$ and $k\cot\delta(k)$ as a function of $k^2$. See the main text for more details.
}
\label{fig:separable}
\end{figure}

We calculate $S$-wave scattering phase shifts corresponding to these approximated potentials, and 
compare them with the exact phase shifts, $\delta_R(k)$.
Considering $\mu$ as  a typical inelastic threshold energy in this model,
we take $k= 0$ and/or $k=\mu$ for the following analysis.
Fig.~\ref{fig:separable} shows the $S$-wave scattering phase shift $\delta(k)$ (Left) and $k\cot\delta(k)$ (Right) as a function of $k^2$, where all (dimensionful) quantities are measured in units of $\mu$. In this example, we take $\omega = -0.017 \mu^4$, $m= 3.30\mu$ and $R=2.5/\mu$.
In the figures, the exact phase shift $\delta_R(k)$(Left) or $k\cot\delta_R(k)$ (Right) is given by the blue solid line, while the LO approximations at   $k=0$ or $k=\mu$ are represented by orange and green solid lines, respectively.
As seen from the figures, the LO approximation at $k=0$ (orange),  exact at $k^2=0$ by construction, gives a reasonable approximation at low energies ($k^2\simeq 0$) but deviates from the exact one at high energies near $k^2\simeq \mu^2$.
On the other hand,
the LO approximation at $k=\mu$ (green) becomes accurate at higher energies near $k^2\simeq\mu^2$ but inaccurate at low energies near $k^2\simeq 0$.
Combining two NBS wave functions, $\phi_{k_1=0}(r)$ and $\phi_{k_2=\mu}(r)$, one can determine the approximated potential at the NLO, $V^{\rm NLO}({\bf r},\nabla)$,
whose scattering phase shifts are represented by the red solid lines in the figures. 
The phase shifts at the NLO (red lines) gives reasonable approximations of the exact results (blue solid lines) in the whole range ($ 0\le k^2 \le \mu^2$), as
they are exact at $k^2=0$ and $k^2=\mu^2$ by construction. 
If we increase the order of the expansion more and more, the approximation becomes better and better.%
\footnote{
    A similar attempt to represent an arbitrary potential
    in terms of a separable potential is given in~\citep{Ernst:1973zzb,Ernst:1974zza}.
}

Using this model, let us compare the direct method and the HAL QCD method. At the LO, the direct method gives 
either $k\cot\delta(k)$ at $k^2= 0$ or $k^2=\mu^2$ without any information about the effective range, which only gives the LO ERE (an orange dashed line or a green dashed line in the right figure.) 
Thus the LO potentials approximate the exact $k\cot\delta(k)$ much better (the orange solid line or the green solid line).
In the direct method, the ERE at NLO is obtained by combining the data at $k^2=0$  and $k^2=\mu^2$ as
\beqa
k \cot \delta (k) &=& \frac{1}{a_0} + \frac{r_{\rm eff}}{2} k^2, 
\quad \frac{1}{a_0} =\lim_{k\rightarrow 0} k \cot \delta (k) , \quad
\frac{r_{\rm eff}}{2} = \frac{\cot\delta(\mu)}{\mu} -\dfrac{1}{\mu^2a_0},
\eeqa
which is given by a red dashed line in the right figure.
  By comparing the HAL QCD method with potentials at NLO (the red solid line)
  and the direct method with NLO ERE (the red dashed line),
  the former
leads to a better approximation of the exact result
than the latter,
since  higher order effects in ERE in terms of $k^2$ are included in the former. 
Note, however, that
sufficiently precise data in the direct method can also evaluate higher order ERE terms than NLO, in principle.

\subsubsection{Dependence of the LO $NN$ potential on energy and partial waves }

In this subsection, we consider effects of higher order terms in the derivative expansion for the $NN$ in QCD.
\begin{figure}[t]
\begin{center}
\includegraphics[width=8.5cm]{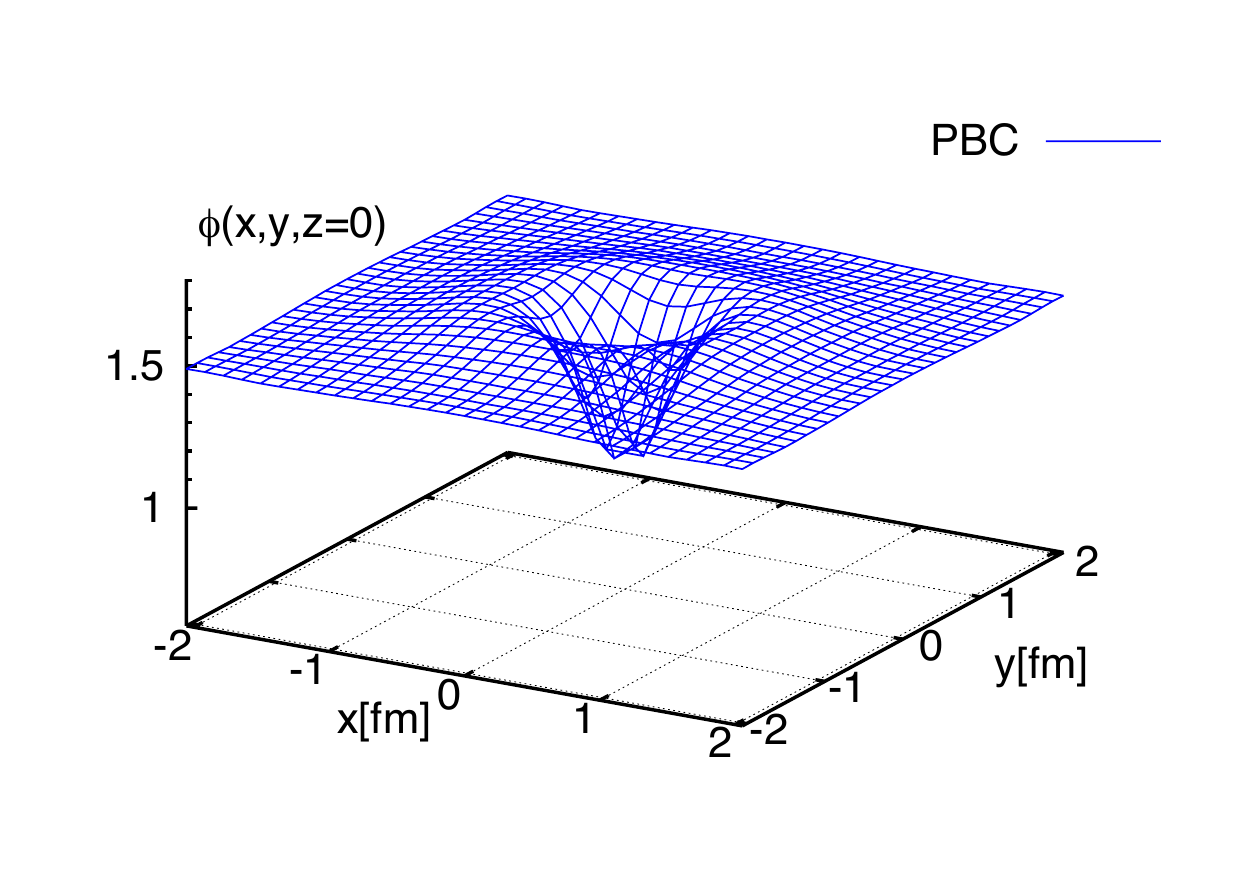}%
\includegraphics[width=8.5cm]{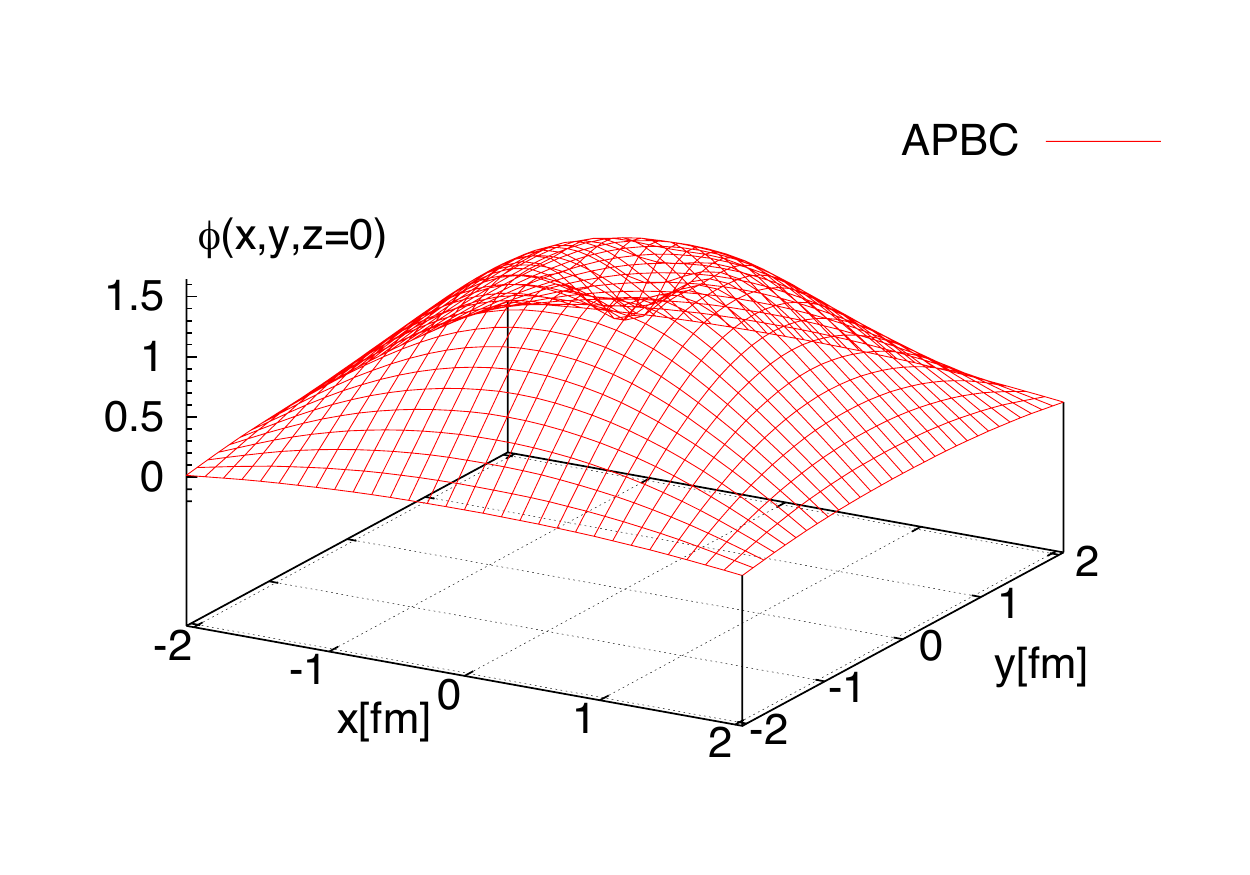}%
\end{center}
\caption{The NBS wave function for $NN(^1S_0)$  at $E_k\simeq 0$ MeV with the PBC (Left) and at $E_k\simeq 45$ MeV with the APBC (Right). Both are normalized to unity at $r=1$ fm. Taken from \citep{Murano:2011nz}.
}
\label{fig:PBC_APBC}
\end{figure}

Fig.~\ref{fig:PBC_APBC} shows three dimensional plots of the NBS wave functions
$\phi_{\bf k} (x,y,z=0)$ for $NN(^1S_0)$  with the periodic boundary condition (PBC) 
at $E_k\simeq 0$ MeV (Left) and with the anti-periodic boundary condition (APBC) at $E_k\simeq 45$ MeV (Right), in quenched lattice QCD at $a\simeq 0.137$ fm on $L\simeq 4.4$ fm with $m_\pi\simeq 530$ MeV~\citep{Murano:2011nz}.
As seen from the figure, two NBS wave functions look very different from each other.
In particular, the right one  vanishes on the boundary due to the APBC constraint.

\begin{figure}[t]
\begin{center}
\includegraphics[width=8.5cm]{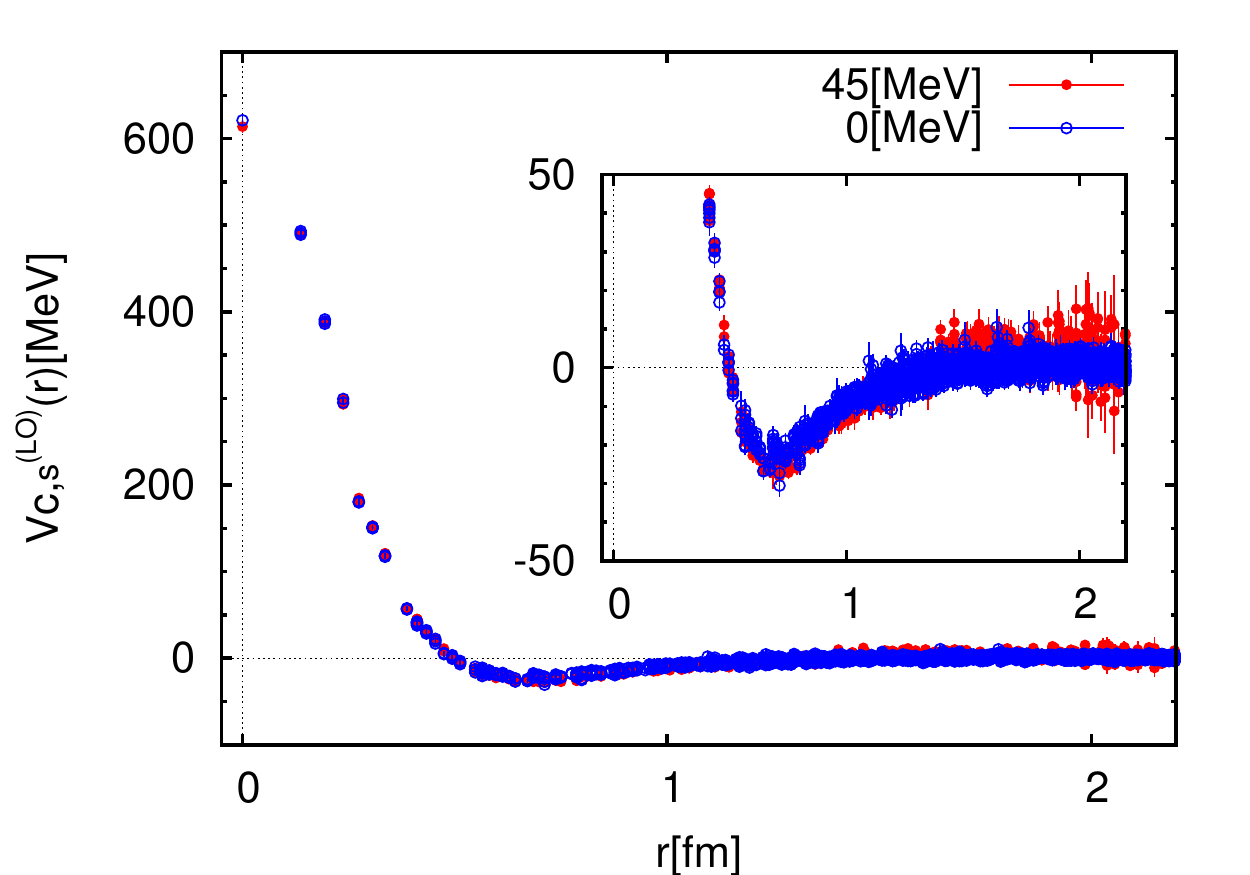}%
\includegraphics[width=8.5cm]{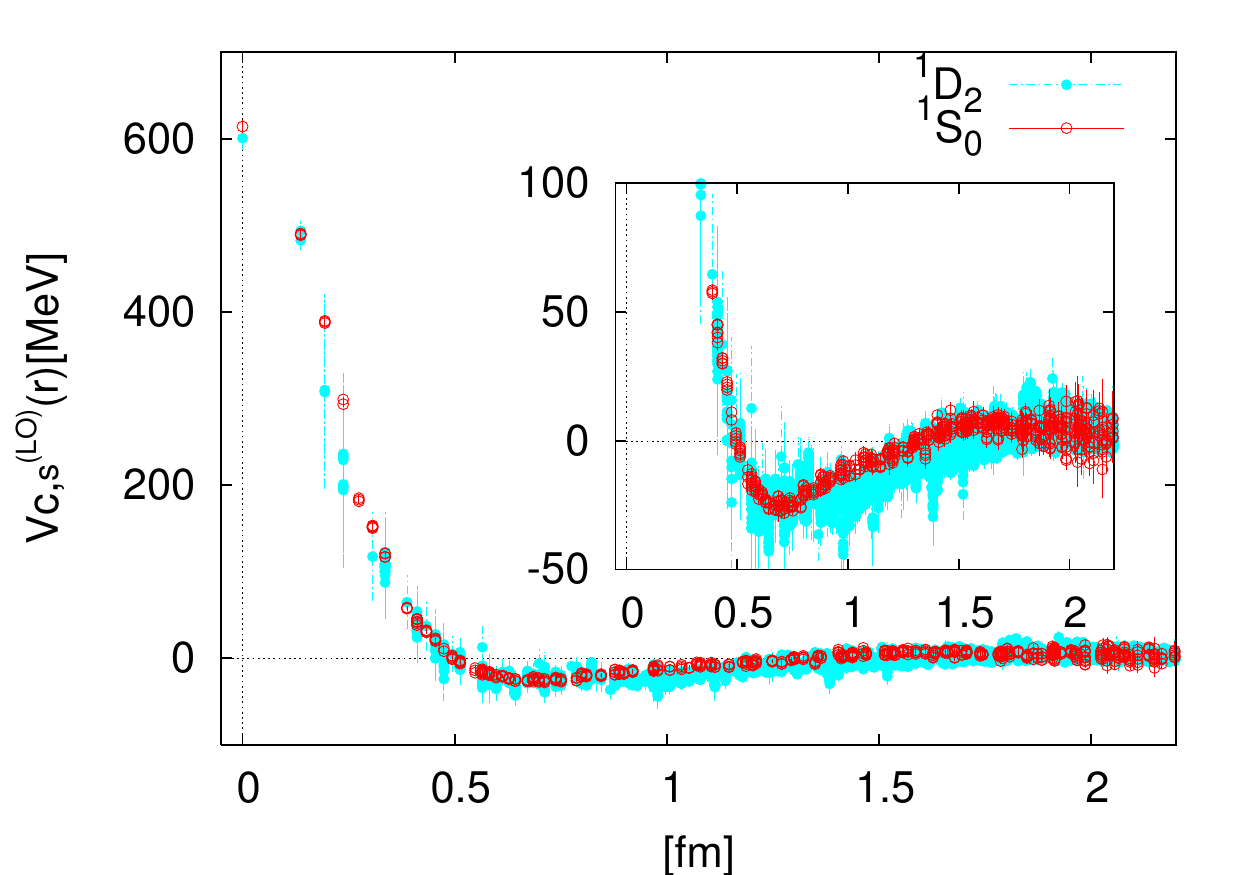}%
\end{center}
\caption{(Left) The LO potential for $NN(^1S_0)$ as a function of $r$ at $E_k\simeq 45$ MeV (red solid circles) and at $E_k\simeq 0$ MeV (blue open circles).
(Right) The LO potential as a function of $r$ at $E_k\simeq 45$ MeV for
$NN(^1S_0)$ (red open circles) and for $NN(^1D_2)$ (cyan solid circles).
Taken from \citep{Murano:2011nz}.
}
\label{fig:pot_PBC_APBC}
\end{figure}
Fig.~\ref{fig:pot_PBC_APBC} (Left) compares  the LO potentials for $NN(^1S_0)$ obtained from the corresponding NBS wave functions in Fig.~\ref{fig:PBC_APBC}.
While the NBS wave functions at different energies have different spatial structures, 
the potentials look very similar. This suggests that the higher order terms in the derivative expansion of the potential have negligible contributions at this energy interval, $ 0 \le E_k \le 45$ MeV.

Fig.~\ref{fig:pot_PBC_APBC} (Right) compares the LO potential for $NN(^1S_0)$ (red open circles) with the one for $NN(^1D_2)$ (cyan solid circles) at $E_k\simeq 45$ MeV.
Although statistical fluctuations are larger for the latter, they look similar, suggesting that ${\bf L}^2$ dependence of the potential is also small  in this setup. 
If more accurate data show a difference of potentials between  $NN(^1S_0)$ and $NN(^1D_2)$, one may determine the ${\bf L}^2$ dependent term of the potential in the spin-singlet channel.

\subsubsection{Time-dependent HAL QCD method}

In order to extract the NBS wave functions on the finite volume in lattice QCD, we consider the 4-pt function given by
\beqa
F^J ( {\bf r},t-t_0) &=& \langle 0\vert N({\bf x} +{\bf r}, t) N({\bf x},t) \bar J_{NN} (t_0)\vert 0\rangle
= \sum_{n} A_{n}^J \phi_{{\bf k}_n} ({\bf r}) e^{- W_{k_n} (t-t_0)} +\cdots,
\eeqa 
where $\bar J_{NN}(t_0)$ is an operator which creates two nucleon states
at time $t_0$, $A_{n}^J\equiv \langle NN, W_{k_n}  \vert \bar J_{NN}(0) \vert 0 \rangle$, and ellipses represent inelastic contributions, which become negligible at $ W_{\rm th} (t-t_0) \gg 1$.
Like the direct method, one can extract the NBS wave function for the ground state from the above 4-pt function as
\beqa
F^J ({\bf r},t ) &\simeq &   A_{0}^J \phi_{{\bf k}_0} ({\bf r}) e^{-W_{k_0} t} 
\eeqa
for $ (W_{k_1}-W_{k_0}) t \gg 1$,
where $W_{k_0}$ ($W_{k_1}$) is the lowest (second-lowest) energy on the finite volume.
The LO potential from the NBS wave function for the ground state is 
then extracted from $F^J({\bf r},t)$ at large $t$.
As will be discussed in the next section, however, it is numerically very difficult to determine $F^J({\bf r},t)$ for two nucleons
at such large $t$  due to the bad signal-to-noise (S/N) ratio.

Fortunately, an alternative extraction is available for the {HAL QCD} method~\citep{HALQCD:2012aa}.
Let us consider the ratio of 4-pt function to the 2-pt function squared as
\beqa
R^J({\bf r},t ) &\equiv & \frac{F^J({\bf r},t ) }{G_N(t)^2}, \quad
G_N(t) = \sum_{\bf x}\langle 0 \vert N({\bf x},t) N({\bf 0},0)\vert 0\rangle \simeq Z_N e^{-m_N t} +\cdots,
\eeqa
which behaves
\beqa
R^J ({\bf r},t ) &=&\sum_{n} \tilde A_{n}^J  \phi_{{\bf k}_n}({\bf r}) e^{- \Delta W_{k_n} t},
\quad
\tilde A_{n}^J \equiv \frac{ A_{n}^J}{Z_N^2}, 
\quad \Delta W_{k} \equiv W_k - 2 m_N, 
\eeqa
for $W_{\rm th} t \gg 1$, where inelastic contributions can be neglected.
Noticing that
\beqa
\Delta W_k &=& \frac{k^2}{m_N} -\frac{(\Delta W_k)^2}{4 m_N}, 
\quad
 \left( \frac{k^2}{m_N} - H_0\right) \phi _{\bf k} ({\bf r}) = V({\bf r},\nabla)   \phi_{\bf k}({\bf r}), 
\eeqa
we obtain
\beqa
\left\{ - H_0 - \frac{\partial}{\partial t} + \frac{1}{4 m_N} \frac{\partial^2}{\partial t^2} \right\} R^J({\bf r},t) 
&=& V({\bf r}, \nabla) R^J({\bf r},t)  .
\label{eq:t-dep_HAL}
\eeqa
We can approximately extract $V({\bf r}, \nabla)$  from $R^J({\bf r},t)$ for (different) $J$'s, as long as $t$ satisfies the condition that $W_{\rm th}\, t \gg 1$ {(elastic state saturation)}, which is much easier {than} to achieve  $(W_{k_1} -W_{k_0}) t \gg 1$ {(ground state saturation)}. We call this alternative extraction the time-dependent HAL QCD method. 

\section{A comparison of the two methods at heavier pion masses}
\label{sec:comp}
\begin{table}[hbt]
\centering
\begin{tabular}{|cr|c|c|c|c|c|}
\hline
Collaboration & Ref. & $N_f$ & $m_\pi$ & $-\Delta E(^1S_0)$  & $-\Delta E(^3S_1)$ 
& $-\Delta E(H)$   \\
\hline
\multicolumn{7}{|c|}{The direct method} \\
\hline
 YKU2011 &\citep{Yamazaki:2011nd}  & 0 & 800 & 4.4(1.2)  & 7.5(1.0)&---  \\
 YIKU2012&\citep{Yamazaki:2012hi}  & 2+1 & 510 & 7.4(1.4) & 11.5(1.3)& --- \\
  NPL2015 &\citep{Orginos:2015aya} & 2+1 & 450 & 12.5(${}^{+3.0}_{-5.0}$) & 14.4(${}^{+3.2}_{-2.6}$) & ---  \\
   NPL2012 & \citep{Beane:2011iw}    & 2+1    & 390 & 7.1(9.0) & 11(13)& 13.2(4.4) \\
 YIKU2015&\citep{Yamazaki:2015asa} & 2+1 & 300 & 8.5(${}^{+1.7}_{-0.9}$) &14.5(${}^{+2.5}_{-1.1}$) &---  \\
 NPL2013 &\citep{Beane:2012vq}  & 3  & 810 & 15.9(3.8)  & 19.5(4.8) & 74.6(4.7)\\ %
 NPL2017 &\citep{Wagman:2017tmp}  & 3  & 810 & 20.6(${}^{+3.3}_{-2.9}$)  & 27.9(${}^{+3.8}_{-2.7}$) & --- \\ %
 CalLat2017  & \citep{Berkowitz:2015eaa}   & 3 & 810 & 21.8(${}^{+3.3}_{-5.8}$) & 30.7(${}^{+2.5}_{-3.0}$) & --- \\ 
                     &                                             & 3   &       & 8.35(1.1)* & 3.3(${}^{+1.2}_{-0.9}$) & --- \\ 
Mainz2018 &\citep{Francis:2018qch} & 3$^{{\dagger}}$ & 960 & 0 & --- & 19(10) \\
                  &                                      & 2+1$^{{\dagger}}$ & 440 &   --- & --- & 18.8(5.5)* \\
                  \hline
\multicolumn{7}{|c|}{The {HAL QCD} method} \\
\hline
IAH2007 & \citep{Ishii:2006ec} & 0 & 530 & 0 & 0 & --- \\
AHI2009 & \citep{Aoki:2009ji} & 0 &380, 530, 730 & 0 & 0 & --- \\ 
  HAL2012 &\citep{Inoue:2011ai} & 3 & 1171 & 0 & 0 & 49.1(6.5) \\ 
   & & 3 & 1015 & 0 & 0 & 37.2(4.4) \\  
     & & 3 & 837 & 0 & 0 & 37.8(5.2) \\   
     & & 3 & 672 & 0 & 0 & 33.6(5.9) \\ 
     & & 3 & 469 & 0 & 0 & 26.0(6.5) \\  
 HAL2012a &\citep{HALQCD:2012aa} & 2+1 & 701 & 0 & --- & --- \\ 
 HAL2013  &\citep{Ishii:2013ira} & 2+1 & 411, 570, 701 & 0 & --- & --- \\  
 \hline
\end{tabular}
\caption{Summary of binding energies [MeV] for $NN(^1S_0)$, $NN(^3S_1)$ and $H$-dibaryon in lattice QCD.
NPL2013, NPL2017 and CalLat2017  employed the same set of gauge configurations.
CalLat2017 found two  states in each channel.
In Mainz2018, dynamical 2-flavor with quenched strange quark configurations are employed
and $N_f$ in the table (with $\dagger$ symbol) denotes the information in the valence quark sector.
All values of $\Delta E$ 
correspond to those in the infinite volume limit
except ones with $*$, which are values on the finite volumes.
The number $0$ in $\Delta E$  indicates
the system is unbound
in this channel.
}
\label{tab:summaryNN}
\end{table}
It is interesting to ask whether the attractions of the nuclear forces at low energies would become weaker or stronger if the pion mass were larger than the value in Nature.
In principle, such a question can be answered  by employing either the direct method or the {HAL QCD} method in lattice QCD.
There exists, however,  a qualitative discrepancy between the two methods on the answer to this question. 
As summarized in Table~\ref{tab:summaryNN},
the direct method tends to indicate that attractions between two nucleons become
stronger as the pion mass increases, so that both deuteron and di-neutron form bound states,
while the HAL QCD method suggests that the attractions become weaker  and the bound deuteron does not exist at heavier pion masses.
  Note that the results from the direct method in the flavor SU(3) limit ($N_f=3$ in the table),
  NPL2013/NPL2017, CalLat2017 and Mainz2018,
  exhibit discrepancies with each other~\citep{Iritani:2017rlk}.
In addition, while  both methods lead to the bound $H$-dibaryon
at heavier pion masses, in particular, in the flavor SU(3) limit, 
the predicted binding energies differ even within the direct method:
 NPL2013~\citep{Beane:2012vq} gives 75(5)MeV at $m_\pi = 810$ MeV, which is much larger than 19(10) MeV at $m_\pi=960$ MeV by Mainz2018~\citep{Francis:2018qch}.
 On the other hand,  HAL2012~\citep{Inoue:2011ai}
 gives 38(5) MeV at $m_\pi=837$ MeV from the {HAL QCD} method.
  These deviations seem to be too large to be explained by lattice artifacts.

In order to understand origins of these discrepancies, we have performed extensive investigations,
whose results  have been published in a series of papers~\citep{Iritani:2016jie,Iritani:2017rlk,Iritani:2018zbt,Iritani:2018vfn},
which will be explained in the following subsections.

\subsection{Operator dependence in the direct method}
\label{subsec:op_dep}

In the direct method, reliable extractions of the two nucleon ground state energies
are crucially important.
As long as $(W_{k_1} -W_{k_0}) t \gg 1$,
the two nucleon correlation function is dominated by the ground state as 
\beqa
G_{NN}(t) &=& \langle 0 \vert J_{NN}(t) \bar J^\prime_{NN}(0) \vert 0\rangle \simeq Z_{ {k_0}}^J \bar Z_{ {k_0}}^{J^\prime} e^{-W_{k_0} t}, \quad Z_{{k_0}}^{J(J^\prime)} \equiv \langle 0 \vert J_{NN} (J^\prime_{NN}) \vert NN, W_{k_0} \rangle,
\eeqa
so that the extracted ground state energy $W_{k_0}$ depends neither the source operator
$\bar J_{NN}^\prime$ nor the sink operator $J_{NN}$, while magnitudes of contaminations from excited states are affected by the choices of these operators.
Since $W_{k_1} -W_{k_0} \simeq (2\pi/L)^2/m_N$ on the finite box with the spacial extension $L$,  
$ t \gg 4$ fm is required,  for example,  for $L\simeq$ 4 fm  and $m_N\simeq 2$ GeV at heavier pion masses.
Due to the bad S/N ratio at such large $t$, however,  
authors in previous literature
extracted the ground state energies at much smaller $t$, $t\sim 1$ fm,
by tuning the source operators $\bar J_{NN}^\prime$
in order to achieve a plateau of the effective energy shift $\Delta E_{NN}^{\rm eff}(t)$
at such a small $t$,
where
\beqa
\Delta E_{NN}^{\rm eff}(t) &=& -\dfrac{1}{a}\log \dfrac{R_{NN}(t+a)}{R_{NN}(t)}, \quad
R_{NN}(t) \equiv \dfrac{G_{NN}(t)}{G_N(t)^2} ,
\eeqa

\begin{figure}[t]
\centering
\includegraphics[width=0.45\textwidth]{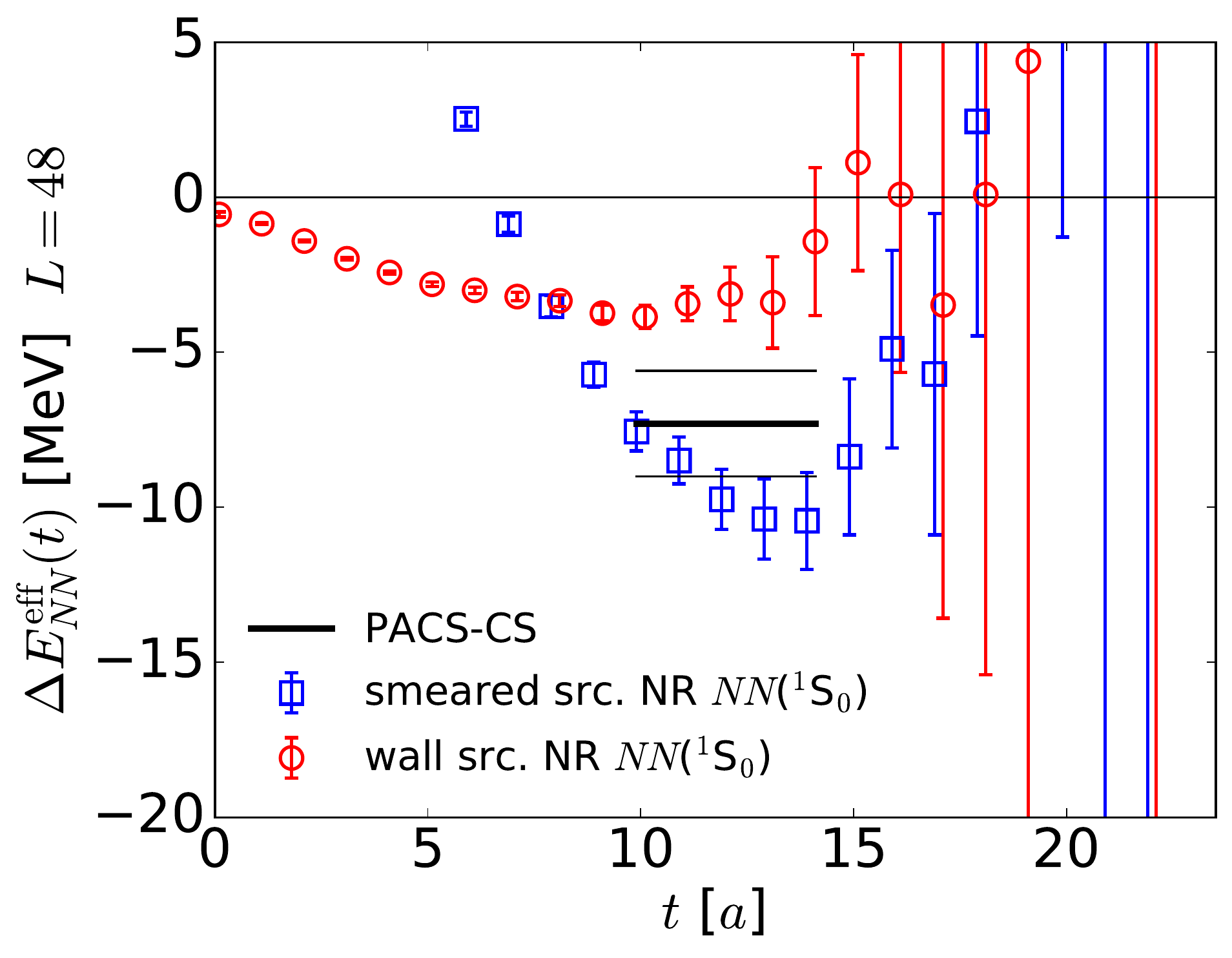}
\includegraphics[width=0.45\textwidth]{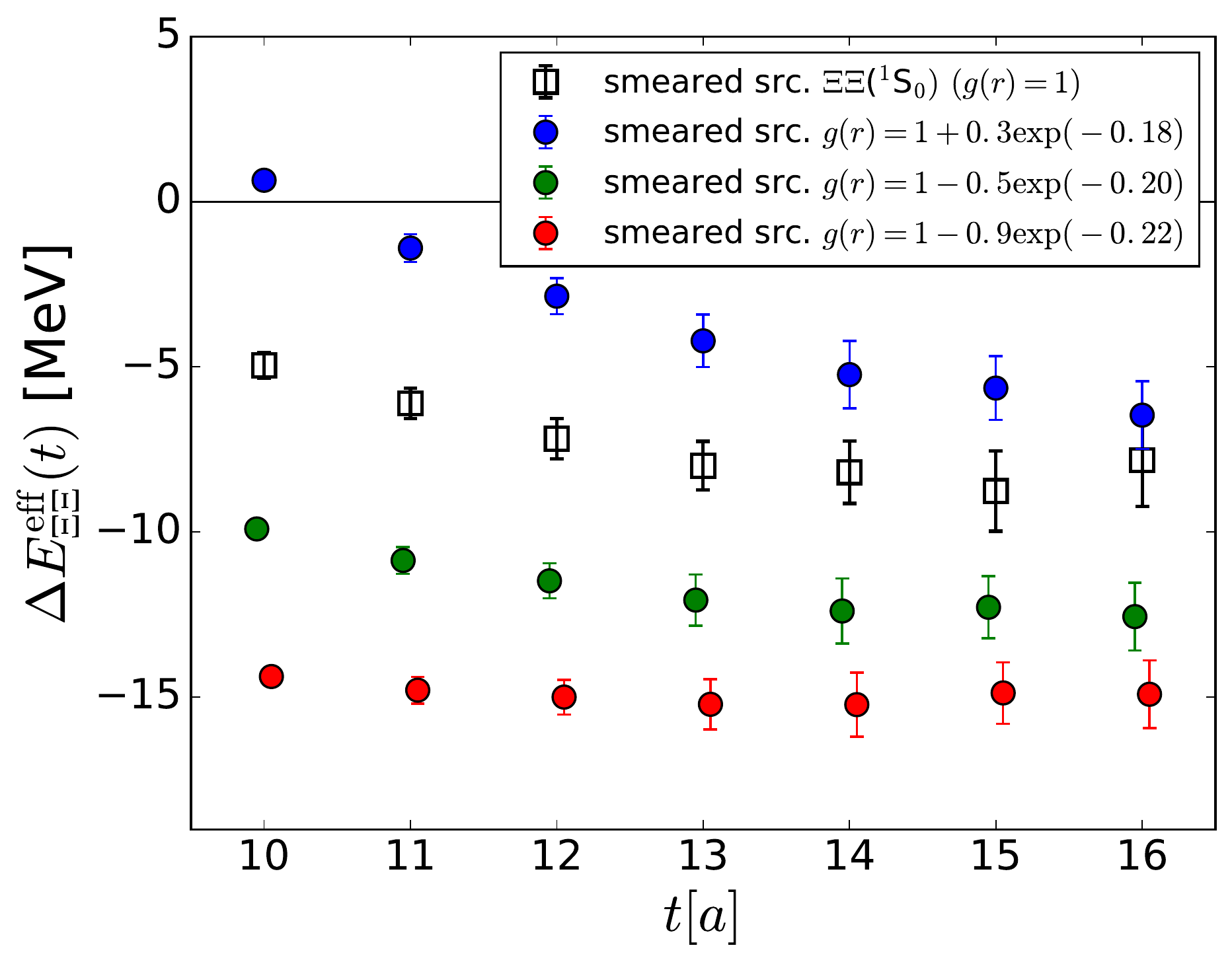}
 \caption{
(Left) The effective energy shift $\Delta E_{NN}^{\rm eff}(t)$ for $NN(^1S_0)$ from the wall source (red circles) and the smeared source (blue squares) on $L=48 a \simeq 4.3$ fm at
$m_\pi = 0.51$ GeV,  $m_N=1.32$ GeV and $m_\Xi = 1.46$ GeV~\citep{Iritani:2016jie}.
(Right) The effective energy shift $\Delta E^{\rm eff}_{\Xi\Xi}(t)$ {for $\Xi\Xi(^1S_0)$}
from the smeared source with different sink operators on {the} same gauge configurations~\citep{Iritani:2016jie}. 
}
 \label{fig:op_dep}
\end{figure}

Unfortunately, such a naive plateau fitting at earlier $t$  may not be reliable
due to contaminations from nearby excited states, which may easily produce (incorrect) plateau-like behaviors 
in effective energies.
It was indeed demonstrated that
plateau-like behaviors in effective energy shifts at small $t$ can depend
not only on the source operator  but also on the sink operator: 
Plateaux disagree between the wall source (red circle) and the smeared source (blue square)
in the left of Fig.~\ref{fig:op_dep},
while  plateaux depend on sink operators for the same smeared source in the right figure.

\begin{figure}[t]
\centering
\includegraphics[width=0.45\textwidth]{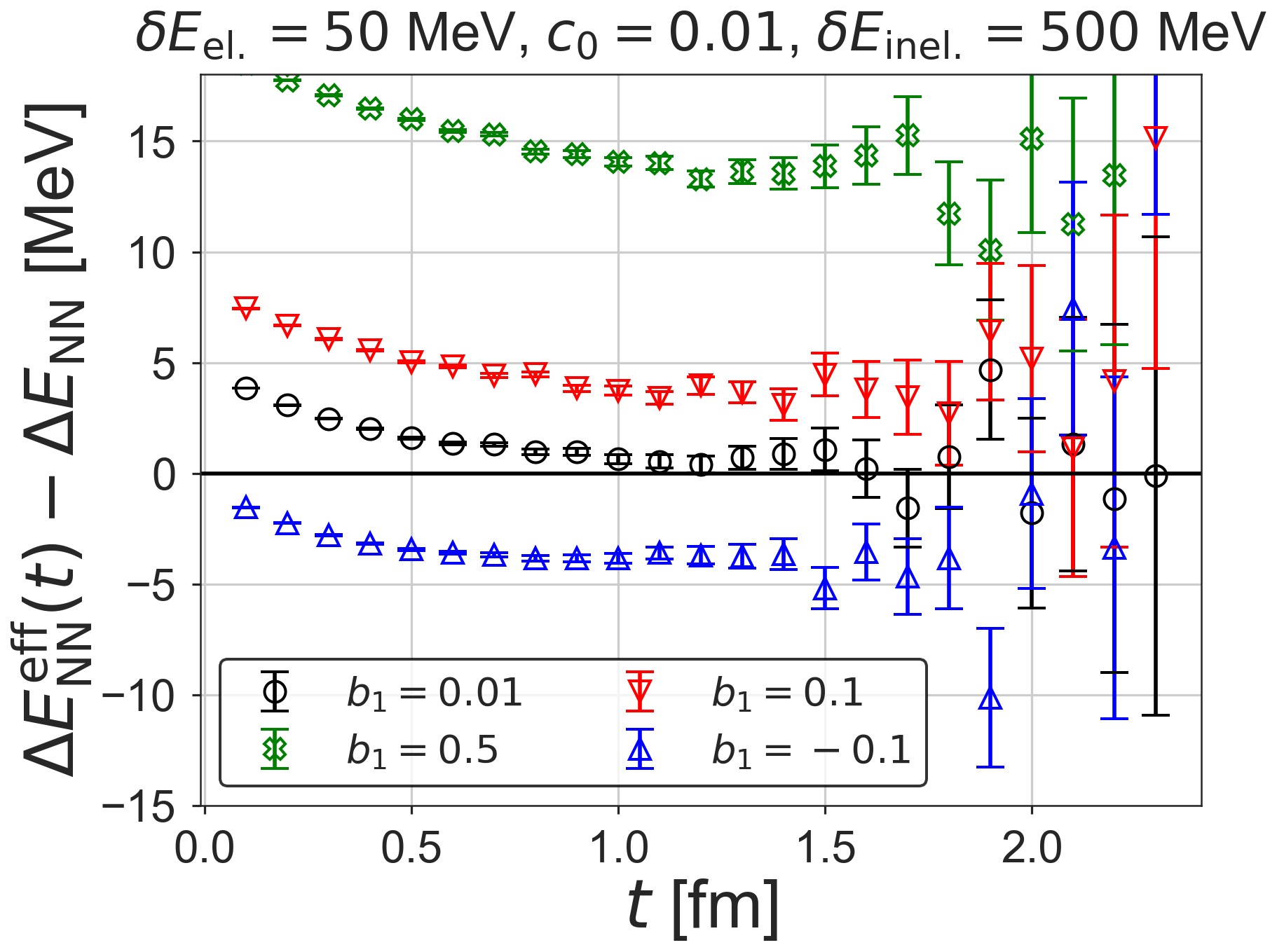}
\includegraphics[width=0.45\textwidth]{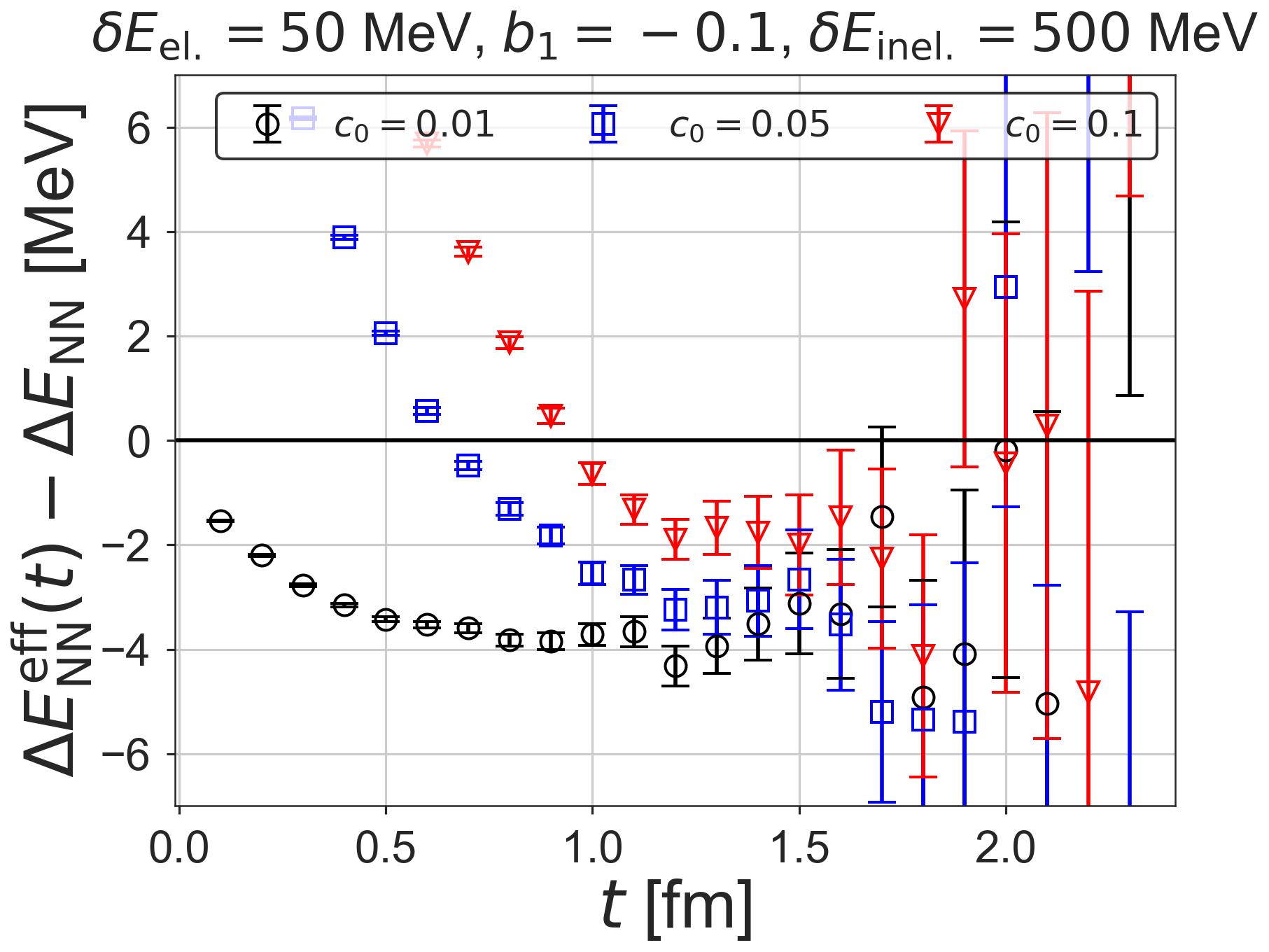}
 \caption{
$\Delta E_{NN}^{\rm eff}(t)-\Delta E_{NN}$ from the {mockup} data $R^{\rm mockup}_{NN}(t)$ with fluctuations and errors as a function of $t$.
(Left) $b_1=0.01, \pm 0.1,0.5$ and $c_0={0.01}$.
(Right) $c_0 =0.01, 0.05, 0.1$ and $b_1=-0.1$.
}
 \label{fig:mockup}
\end{figure}

In order to see how easily contaminations from elastic-excited states can produce
plateau-like behaviors at earlier $t$, 
let us consider the effective energy shift from
the mockup data for $R_{NN}(t)$, given by
\beqa
R_{NN}^{\rm mockup}(t) &=& e^{-\Delta E_{NN} t} \left(1 + b_1 e^{-\delta E_{\rm el.} t} + c_0 e^{-\delta E_{\rm inel.} t}\right), 
\eeqa
where
we take $\delta E_{\rm el.} = 50$ MeV for the typical lowest elastic excitation energy on $L\simeq 4$ fm at $m_N \simeq  1.5$ GeV, and $\delta E_{\rm inel.} \simeq m_\pi \simeq 500$ MeV for the lowest inelastic energy.   
Naively, it is expected that the correct plateau at $\Delta E_{NN}$ for the ground state
appears at $t {\gg} 1/\delta E_{\rm el.} \simeq 4$ fm, which however is too large to have good signals for two baryons such as $NN$.
By tuning the source operator, one may reduce coefficients $b_1$ and $c_0$.
Since the  $NN$ operator does not strongly couple to $NN\pi$ state, we expect small $c_0$
and take $c_0=0.01$. 
On the other hand, $NN$ operators easily couple to both  
ground and 1st elastic excited states as they become almost identical to each other in the infinite volume limit.
We therefore take  $b_1=0.01$ (the highly tuned operator) , $b_1 =\pm 0.1$ (the tuned ones) as well as $b_1  = 0.5$ (the untuned one). Fig.~\ref{fig:mockup} (Left) shows $\Delta E_{NN}^{\rm eff}(t)$ for these 4 examples with $c_0=0.01$, where random fluctuations and errors 
whose magnitude increase exponentially in $t$
are assigned to $R_{NN}^{\rm mockup}(t) $. 
All examples show plateau-like behaviors at $t\simeq 1$ fm, but these four plateaux disagree with each other. As $\vert b_1\vert$ increases, {the} deviation {between the values of these ``pseudo plateaux'' and} the true value becomes larger. 
Contaminations of the elastic excited states can easily produce the plateau-like behavior at earlier  $t$, and the $t$ dependence of data alone {cannot} tell us which plateau is correct, or in other words, {cannot} tell 
which tuning is good. 

Contaminations from inelastic states seem unimportant to produce the plateau-like behavior, 
as shown in Fig.~\ref{fig:mockup} (Right), where the effective energy shift for $c_0=0.01, 0.05, 0.1$
with $b_1=-0.1$ is plotted. All cases converge to almost the same {pseudo plateau,}
{while a pseudo plateau starts at later $t$ for larger $c_0$.}
It is noted that the multi-exponential fit
does not work in this case at $t\simeq 1.0$ fm,
which is much smaller than the necessary $t {\gg} 1/\delta E_{\rm el.}$.
The multi-exponential fit
at such small $t$ 
only separates the {pseudo}
plateau from the inelastic contributions but 
is difficult to distinguish the ground state and the 1st excited state for the elastic states.

\subsection{Normality check in the direct method}
\label{subsec:normality}

\begin{figure}[tbh]
\centering
  \includegraphics[width=0.48\textwidth]{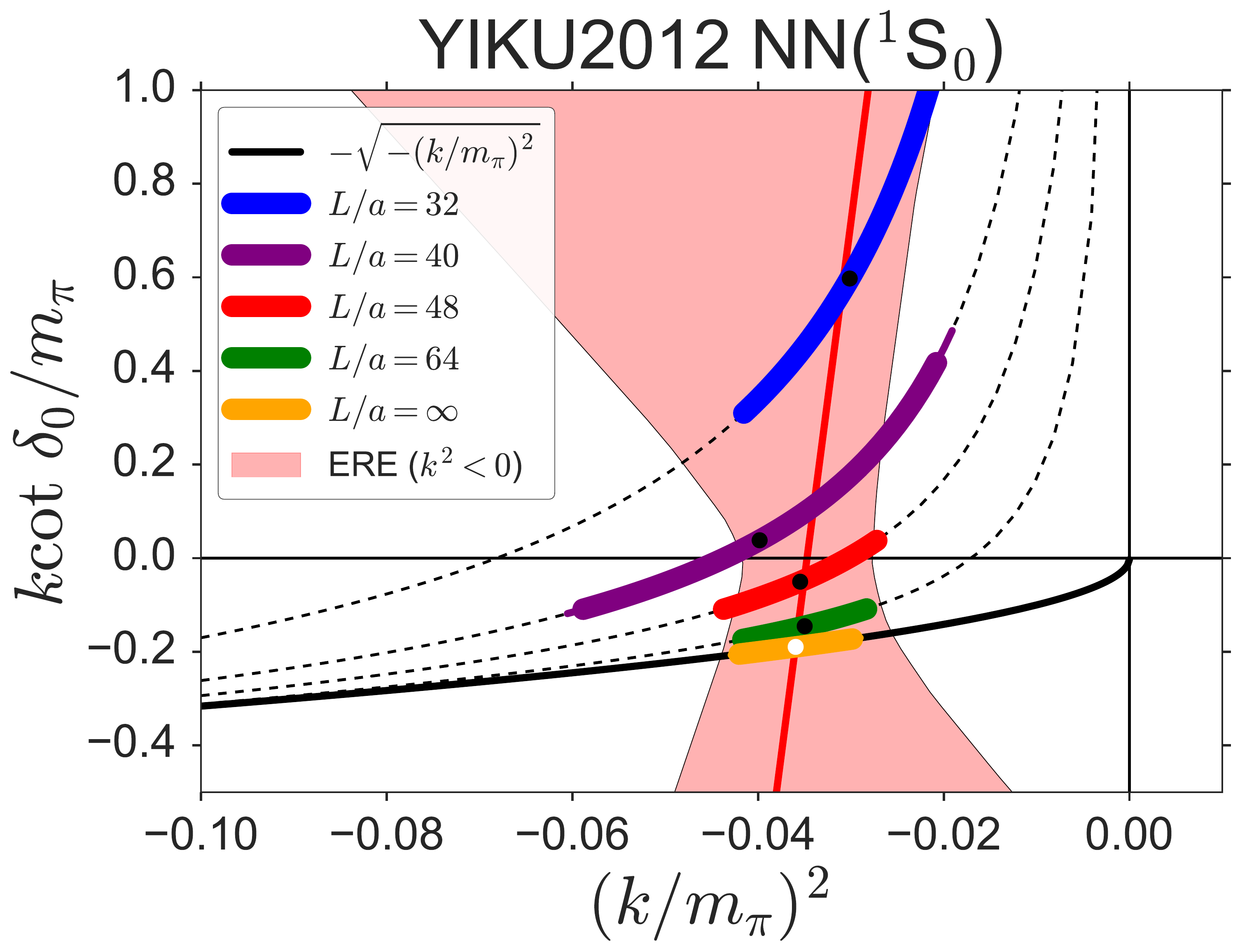}
  \includegraphics[width=0.48\textwidth]{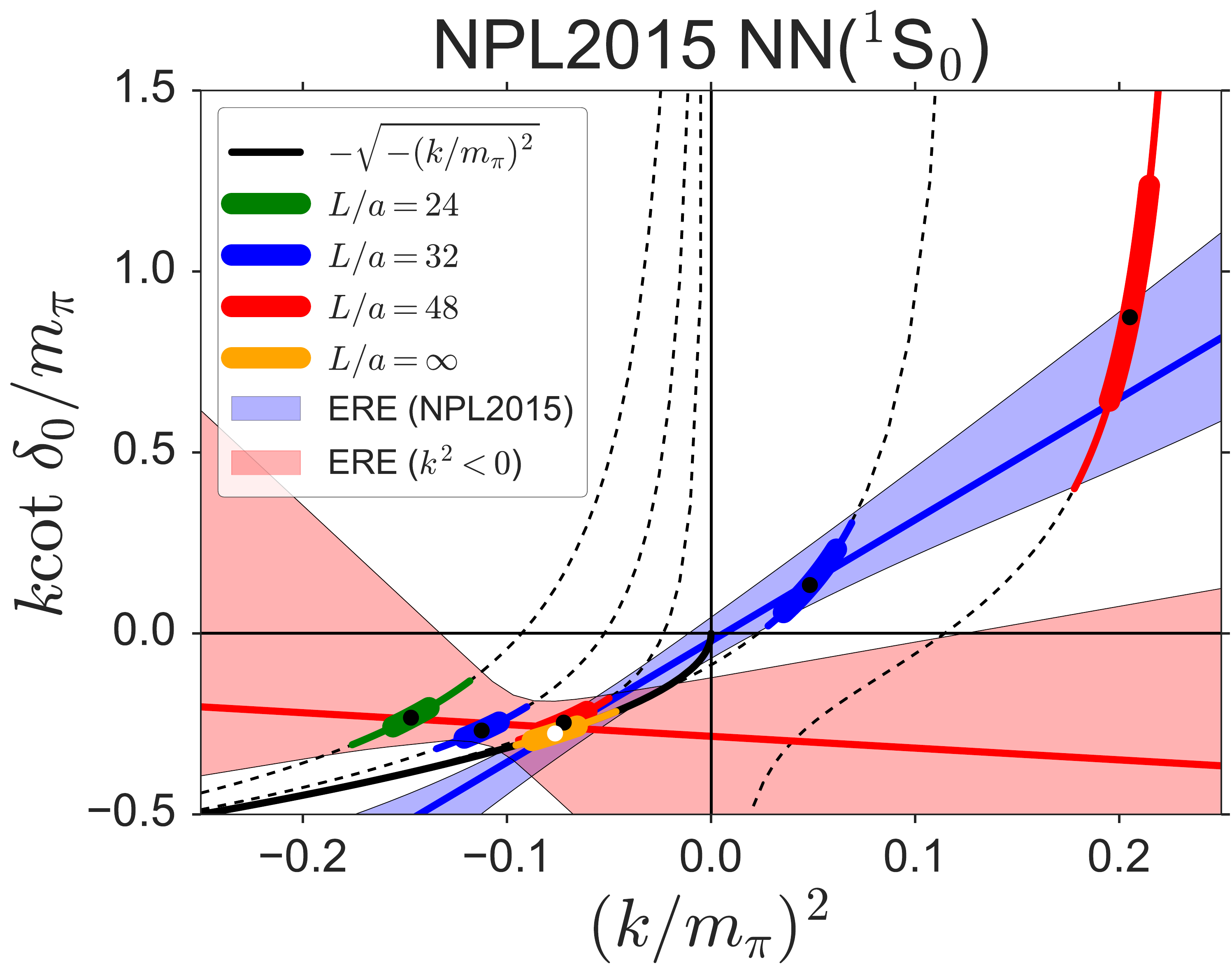}
 \caption{
   (Left)  $k\cot\delta_0(k)/m_\pi$ in YIKU2012~\citep{Yamazaki:2012hi}
   for $NN (^1S_0)$ as a function of $(k/m_\pi)^2$. 
   The solid red line and light red band
   represent the ERE fit 
   and the corresponding error (statistical and systematic added in quadrature),
     respectively.
The dashed lines are the finite volume formula for the corresponding volume.
(Right)   $k\cot\delta_0(k)/m_\pi$ in NPL2015~\citep{Orginos:2015aya} for $NN (^1S_0)$ as a function of $(k/m_\pi)^2$.
Two ERE fits are performed depending on the lattice data to be used for the fit.
The red line with the band represents the fit made by the authors in~\citep{Iritani:2017rlk},
while the blue line with the band is plotted by  the authors in~\citep{Iritani:2017rlk} using the fit result of NPL2015.
Both figures are taken from \citep{Iritani:2017rlk}.
}
 \label{fig:ERE_A}
\end{figure} 
While the check through operator dependence is useful,  it requires extra calculations.
We find that
the finite volume formula in eq.~(\ref{eq:kcot_delta}) provides a simpler test, which 
tells us whether the ground state energies extracted by the plateau fitting give a  reasonable ERE or not
without extra calculations. We call this test a normality check~\citep{Iritani:2017rlk}.
Fig.~\ref{fig:ERE_A} (Left) shows $k\cot \delta_0(k)/m_\pi$ in YIKU2012~\citep{Yamazaki:2012hi} as a function of $k^2/m_\pi^2$ for $NN(^1S_0)$, 
where the solid red line represents the NLO  ERE fit in eq.~(\ref{eq:ERE_nlo}), and the light red bands shows statistical and systematic errors added in quadrature~\citep{Iritani:2017rlk}. 
Contrary to a naive expectation from non-singular ERE behaviors, data align almost vertically, since $\Delta E_{NN}$ is almost independent of the volume. In other words, according to the finite volume formula, the claimed ``binding energy" (open circle) is too shallow to have such volume independent   $\Delta E$.
Not only the central value of the NLO ERE fit gives singular parameters as $( (a_0 m_\pi)^{-1}, r_0 m_\pi ) = (5.27, 303.6)$ but also  it violates the physical pole condition, eq.~(\ref{eq:phys_pole}), at the crossing point (open circle).
The singular and unphysical behaviors, in addition to the operator dependence of these data,
strongly indicate that the naive plateau fitting employed in the direct method is unreliable.   
  Another example is shown in Fig.~\ref{fig:ERE_A} (Right) for $NN(^1S_0)$ from NPL2015~\citep{Orginos:2015aya}.
  In this case,
  two different NLO ERE fits (red line/band and blue line/band) are performed
  depending on the lattice data to be used for the fit.
  It turns out that two ERE are inconsistent with each other,
  indicating that their lattice data themselves are ``self-inconsistent''.
  In addition, one of ERE (blue line/band) is found to violate the physical pole condition,
  eq.~(\ref{eq:phys_pole}), at the crossing point (open circle).
Similar symptoms are observed for all other data in the direct method
  claiming the existence of $NN$ bound states at heavy quark masses~\citep{Iritani:2017rlk}.
\footnote{After these problems were pointed out in ~\citep{Iritani:2017rlk},
revised data of NPL2013 have been presented in ~\citep{Wagman:2017tmp}, whose EREs 
are still marginal to satisfy/violate the physical pole condition.
}

\subsection{The source dependence and the derivative expansion in the HAL QCD method}
\label{subsec:HAL_sys}

\begin{figure}[t]
\centering
\includegraphics[width=0.48\textwidth]{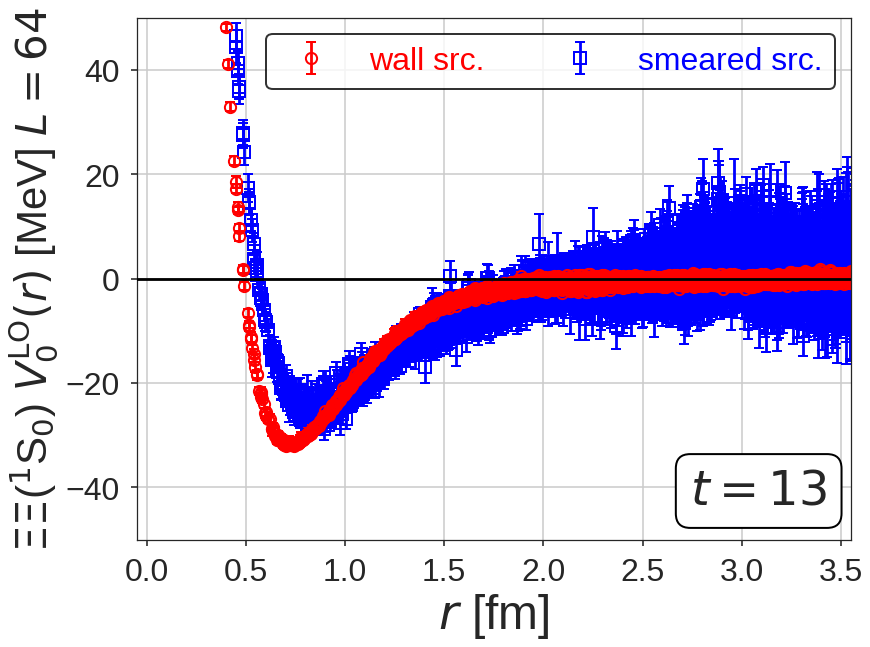}
 \includegraphics[width=0.48\textwidth]{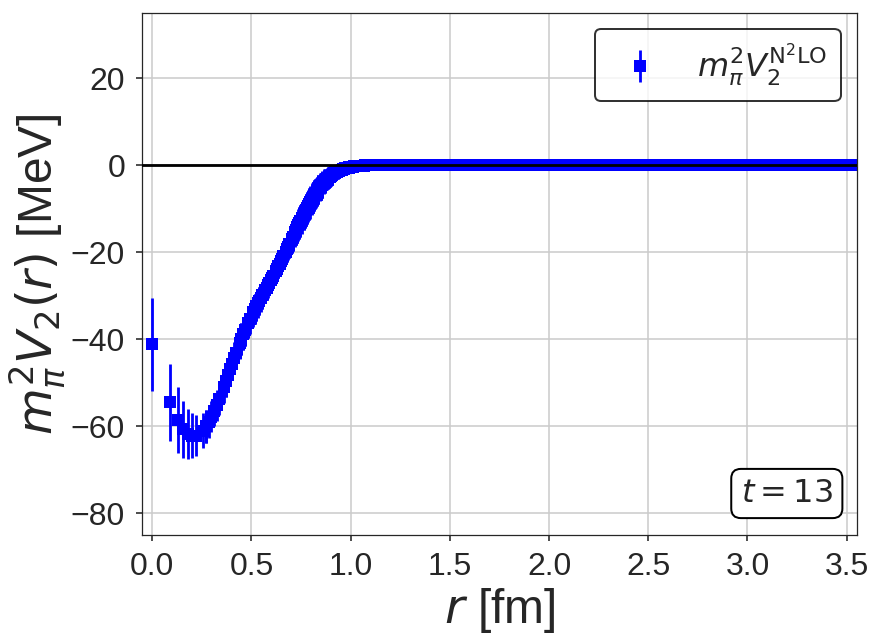}
 \caption{ 
(Left)   The LO  potential, {$V^{\rm LO}_0(r)$,} {for $\Xi\Xi (^1S_0)$} from the wall source (red open circles) and the smeared source (blue open square).  
(Right)   The second order term, $ V^{\rm N^2LO}_2(r)$ (blue solid squares), in the N$^2$LO potential  $V^{\rm N^2LO}({\bf r},\nabla) = V^{\rm N^2LO}_0(r) + V_2^{\rm N^2LO}(r) \nabla^2$
for $\Xi\Xi (^1S_0)$.
Both are taken from \citep{Iritani:2018zbt}.
}
 \label{fig:pot_NNLO}
\end{figure}

The source operator dependence of the HAL QCD potential has been investigated in \citep{Iritani:2018zbt}.
Fig.~\ref{fig:pot_NNLO} (Left) compares the LO potentials, $V^{\rm LO}_0(r)$, for $\Xi\Xi (^1S_0)$ between
the wall source (red open circles) and the smeared source (blue open squares).
We observe a small difference at short distances, from which 
one can determine the N$^2$LO potential,
$
V^{\rm N^2LO}({\bf r},\nabla) = V^{\rm N^2LO}_0(r) + V_2^{\rm N^2LO}(r) \nabla^2.
$
{Note that the NLO term, $V^{\rm N^2LO}_1(r) \nabla = V^{\rm N^2LO}_{\rm LS}(r) {\bf L}\cdot {\bf S}$
is absent in the $^1S_0$ channel.}
Fig.~\ref{fig:pot_NNLO} (Right) shows $V_2^{\rm N^2LO}(r)$, which 
is nonzero only at $r < 1.0$ fm, where two LO potentials differ.
We then extract the scattering phase shifts, using this N$^2$LO potential. 

\begin{figure}[t]
\centering
 \includegraphics[width=0.48\textwidth]{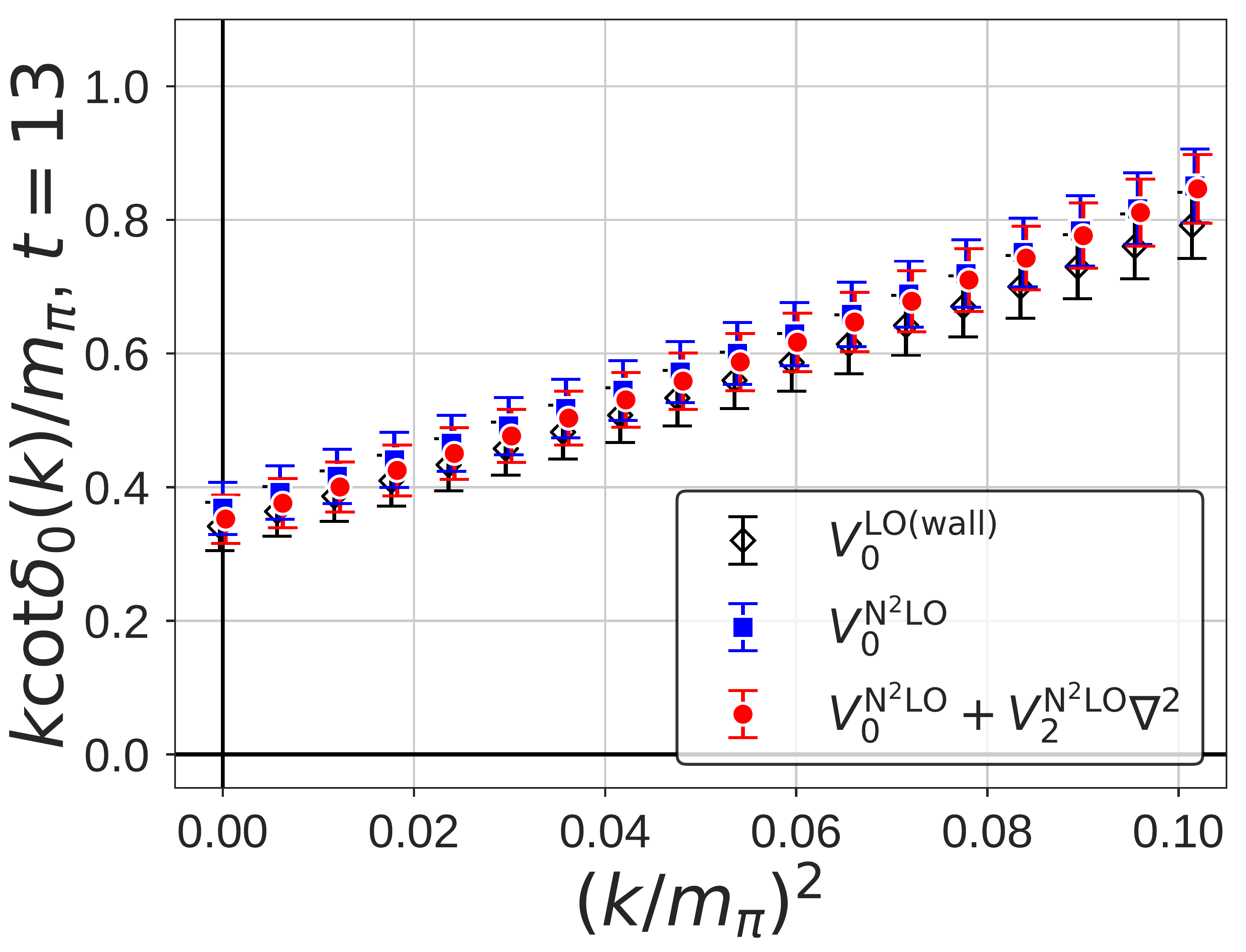}
   \includegraphics[width=0.48\textwidth]{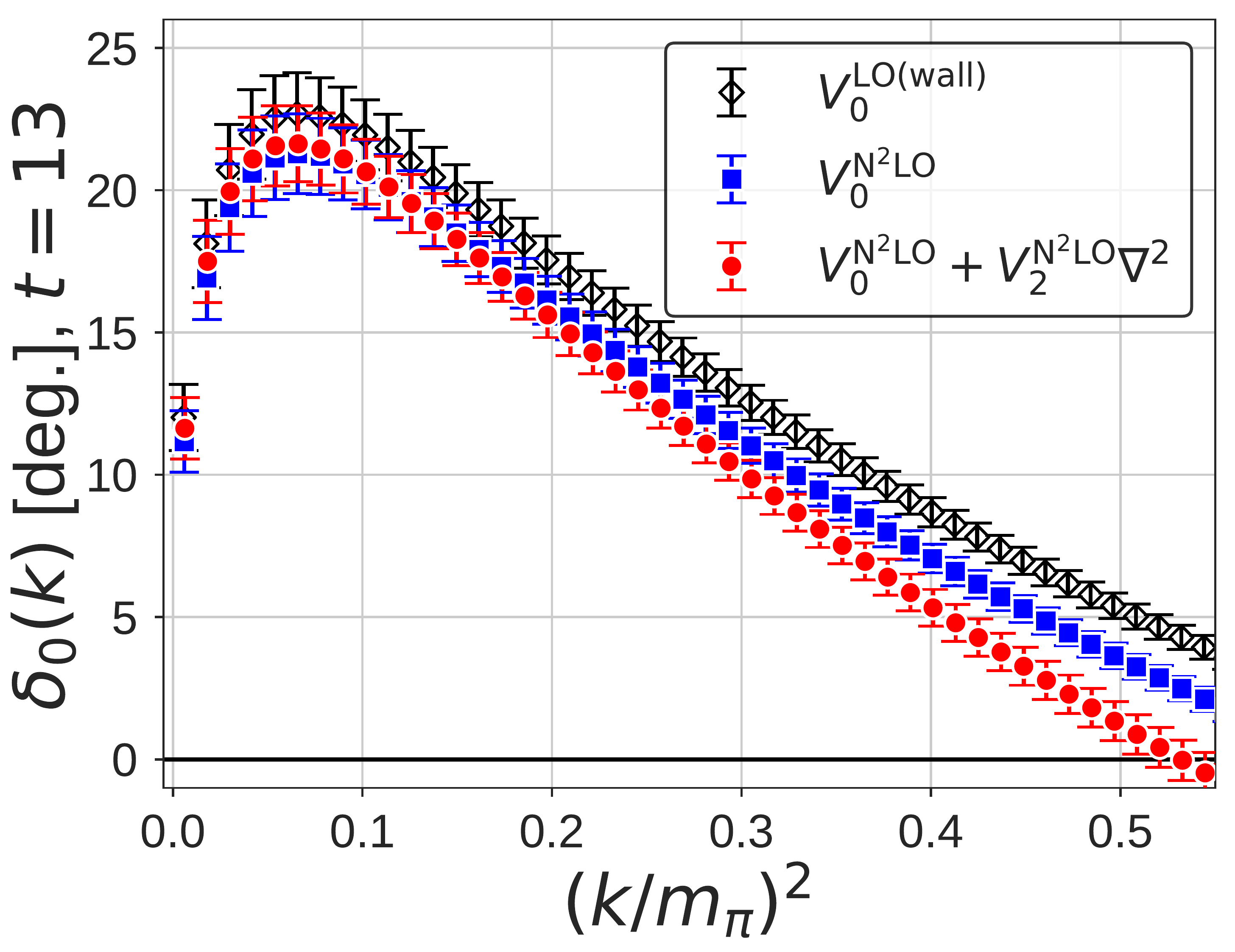}
 \caption{
 (Left)  $k\cot\delta_0(k)/m_\pi$ as a function of $(k/m_\pi)^2$ at low energies,
 where $\delta_0(k)$ is the scattering phase shift for $\Xi\Xi (^1S_0)$, 
 calculated from $V^{\rm N^2LO}({\bf r},\nabla)$ (red solid circles), $V^{\rm N^2LO}_0(r)$ (blue solid squares) and $V^{\rm LO(wall)}_0(r)$ (black open diamond).
(Right) The corresponding $\delta_0(k)$.
Both are taken from \citep{Iritani:2018zbt}.
}
 \label{fig:delta_NNLO}
\end{figure}

The N$^2$LO corrections {turn} out to be negligible at low energies, 
as  shown in Fig.~\ref{fig:delta_NNLO} (Left),
where $k\cot\delta_0(k)$ is almost identical between $V^{\rm N^2LO}({\bf r},\nabla)$ (red solid circles) and $V^{\rm N^2LO}_0(r)$ (blue solid squares). 
Furthermore, even the LO analysis for the wall source, $V^{\rm LO(wall)}_0(r)$ (black open diamond), is sufficiently good at low energies.
As energy increases, 
the N$^2$LO corrections become visible as seen in Fig.~\ref{fig:delta_NNLO} (Right),
where $(k/m_\pi)^2 = 0.5$ corresponds to $\Delta E \simeq 90$ MeV for
the energy shift from the threshold. It is noted that  $V^{\rm N^2LO}_0(r)$ (blue solid squares) gives a little closer results 
to N$^2$LO results  (red solid circles) 
than  $V^{\rm LO(wall)}_0(r)$ (black open diamond) does.

\subsection{Understanding {pseudo} plateaux}
\label{subsec:anatomy}
In this subsection, we  explain why the wall source and the smeared source give inconsistent plateau behaviors, in the case of $\Xi\Xi$ correlation functions as an example.

{To this end, we consider the Hamiltonian $H = H_0 + V_0^{\rm LO(wall)}$,
  where we employ $V^{\rm LO (wall)}_0(r)$, the LO potential from the wall source,
  since it works rather well at low energies as shown in the previous subsection.}
We first decompose $R_{\Xi\Xi}^{J}({\bf r},t)$ for ${J}=$ wall/smear in terms of  
finite volume eigenfunctions of {$H$} as
\beqa
R_{\Xi\Xi}^{{J}}({\bf r},t) &=& \sum_n a_n^{{J}}(t) \Psi_n({\bf r}) e^{-\Delta E_n t}, 
\quad  a_n^{{J}}(t) =\sum_{\bf r} \Psi_n^\dagger({\bf r}) R^{{J}}_{\Xi\Xi} ({\bf r},t) e^{\Delta E_n t} . 
\eeqa
where $\Psi_n({\bf r})$ and $\Delta E_n$ are normalized-eigenfunction and eigenenergy in the finite volume, respectively, and  {$a_n^{{J}}(t)$ is the overlapping coefficient
  extracted at $t$.}

Then the correlation function for the source ${J}$ in the direct method is given by
\beqa
R^{{J}}_{\Xi\Xi}(t) &=& \sum_{\bf r} R_{\Xi\Xi}^{{J}}({\bf r},t) = \sum_n b_n^{{J}}(t) e^{-\Delta E_n t}
\quad b_n^{J}(t) = a_n^{ J}(t)  \sum_{\bf r} \Psi_n({\bf r}) .
\eeqa 

Finally, approximating a sum over $n$ by the lowest few orders, we reconstruct the behavior of the effective energy shift as a function of $t$ as
\beqa
\overline{\Delta E}_{\rm eff}^{{J}} (t,t_0) &=& \frac{1}{a}
\log\left( \frac{R^{{J}}(t, t_0)}{R^{{J}}(t+a, t_0)}\right), \quad
R^{{J}}(t,t_0) = \sum_{n=0}^{n_{\rm max}} b_n^{{J}}(t_0) e^{-\Delta E_n  t}, 
\eeqa 
where we fix the overlapping coefficient $b_n^{{J}}(t_0)$ at $t=t_0$, 
and $n_{\rm max} $ is a number of excited states used in the approximation.

\begin{figure}[t]
\centering
 \includegraphics[width=0.48\textwidth]{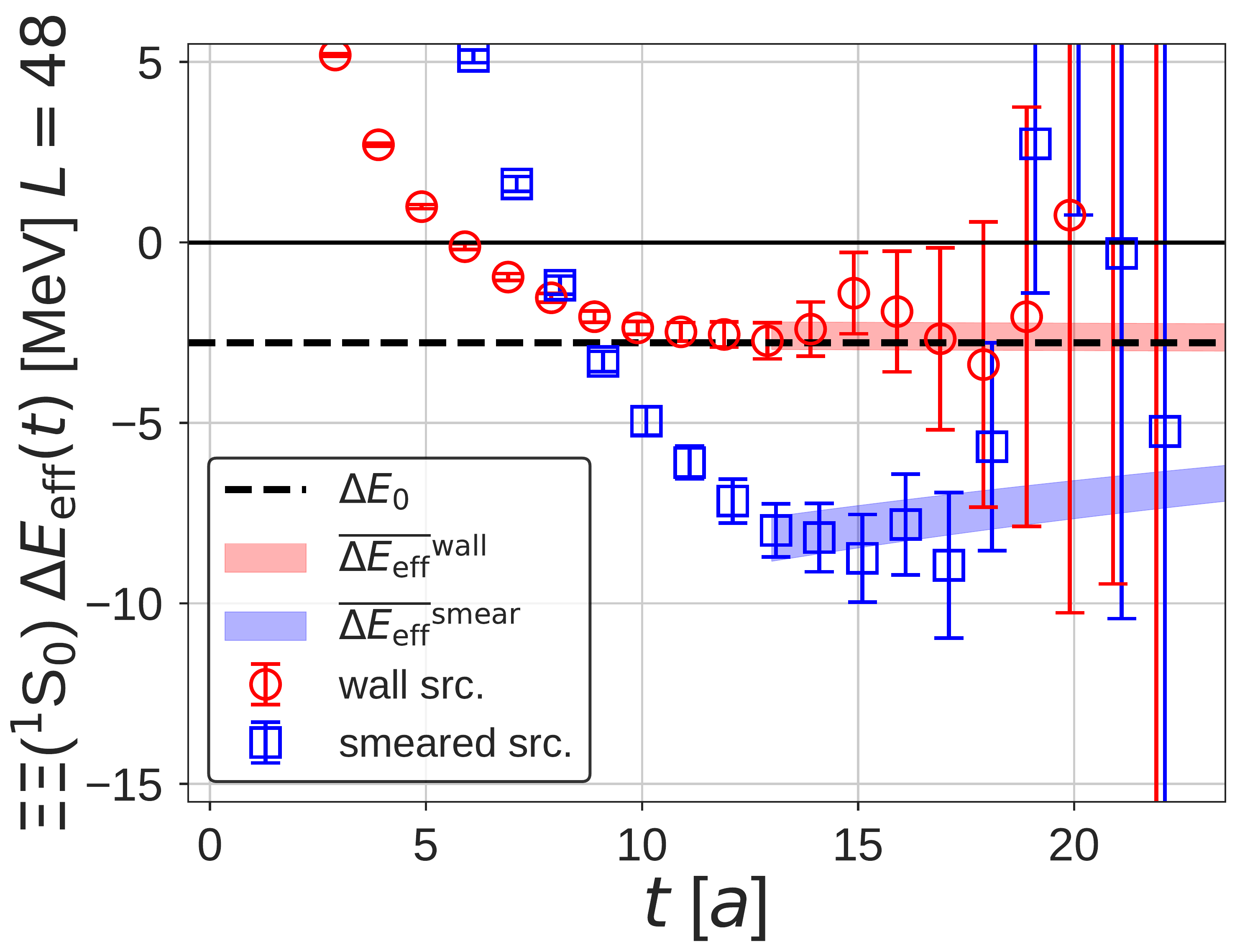}
   \includegraphics[width=0.48\textwidth]{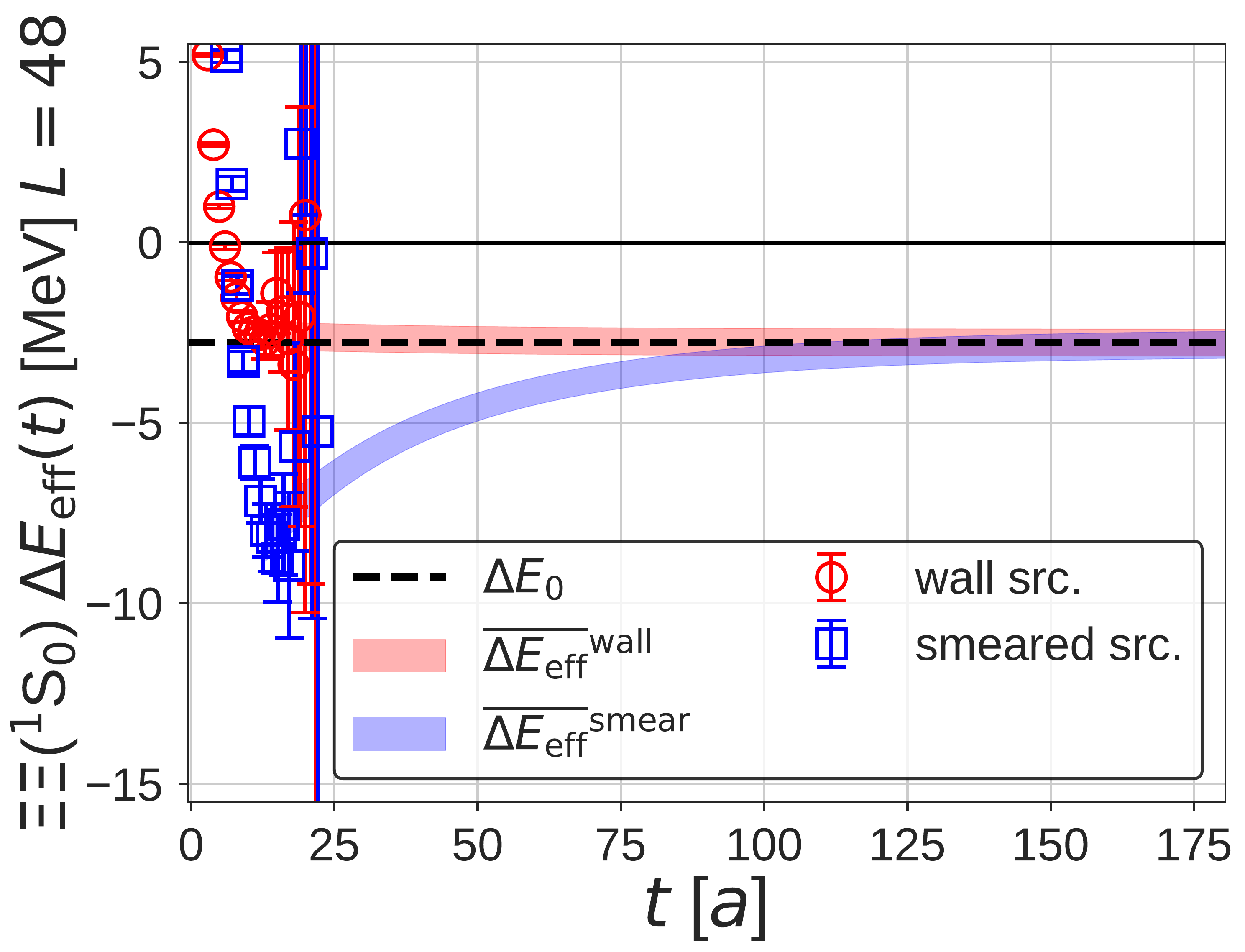}
 \caption{
 The reconstructed effective energy shift $\overline{\Delta E}_{\rm eff}^{{J}} (t,t_0=13a)$ for the wall source (red bands) and the smeared source (blue bands) on $L=48 a$, while
 the effective energy shifts directly from $R_{\Xi\Xi}^{J}(t)$ are shown
 for ${J}=$ wall (red open circles) and ${J}=$ smear (blue open squares).
 The black dashed lines are the energy shifts for the ground state of $H$ in the finite box. (Left) $ 0 \le t/a \le 24$. (Right) $0\le t/a \le 175$.
 Taken from \citep{Iritani:2018vfn}.
}
 \label{fig:Eeff_reconst}
\end{figure} 
In Fig.~\ref{fig:Eeff_reconst}, 
we show reconstructed effective energy shift $\overline{\Delta E}_{\rm eff}^{{J}} (t,t_0=13a)$ on $L=48a$ {with $n_{\rm max} =4$,}
together with the effective energy shifts from $R_{\Xi\Xi}^{{J}} (t)$,
for the wall source (red bands and red open circles) and the smeared source (blue bands and blue open squares).
The black dashed line represents the energy shift for the ground state of $H=H_0 + V_0^{\rm LO(wall)}$ on $L=48a$. 

We find that the plateau-like structures in the direct method around $t/a =15$ are well reproduced by
$\overline{\Delta E}_{\rm eff}^{{J}} (t,t_0=13a)$ for both sources in Fig.~\ref{fig:Eeff_reconst} (Left).
This indicates that the plateau-like structures in the direct method at this time interval 
are explained by the contributions from  several low-lying states.

These plateau-like structures of course do not necessarily correspond to the true energy shift of the ground state.
The fate of these structures is shown in Fig.~\ref{fig:Eeff_reconst} (Right), where
we plot $\overline{\Delta E}_{\rm eff}^{{J}} (t,t_0=13a)$ at asymptotically large $t$.
While the plateau-like structure for the wall source is almost unchanged, 
$\overline{\Delta E}_{\rm eff}^{{J}} (t,t_0=13a)$ for the smeared source gradually increases and reaches to the true value at $t/a \sim 100$.

The above results clearly reveal that the plateau-like structures  at $t/a\sim 15$ for the smeared source are
pseudo-plateaux caused by the contaminations of the excited states.
Large contaminations from excited states in the case of the smeared source are not caused by the smearing, but are indeed implied by putting two baryon operators on the same space-time point
as
\beqa
\frac{1}{L^3} \sum_{\bf x} B({\bf x}, t) B({\bf x}, t) = \sum_{\bf p} \tilde B({\bf p},t) \tilde B(-{\bf p},t) ,
\quad \tilde B({\bf p},t) \equiv \sum_{\bf x} B({\bf x},t) e^{- i{\bf p}\cdot{\bf x}},
\eeqa
where the above source operator couples to all momentum modes with almost equal weight.
Since almost all previous studies on $NN$ interactions in the direct method  
employed this type of the source operator, their conclusions on the existences of both deuteron  and di-neutron are not valid due to large contaminations.%
\footnote{
 Note that Mainz2018 employed a source operator as $\tilde B({\bf p=0},t) \tilde B(-{\bf p=0},t)$
    and they reported that ``In the 27-plet (dineutron) sector,  the finite volume analysis suggests that the existence of a bound state is unlikely.'' .
}

\subsection{Consistency between the two methods}
\label{subsec:consistency}

\begin{figure}[t]
\centering
 \includegraphics[width=0.48\textwidth]{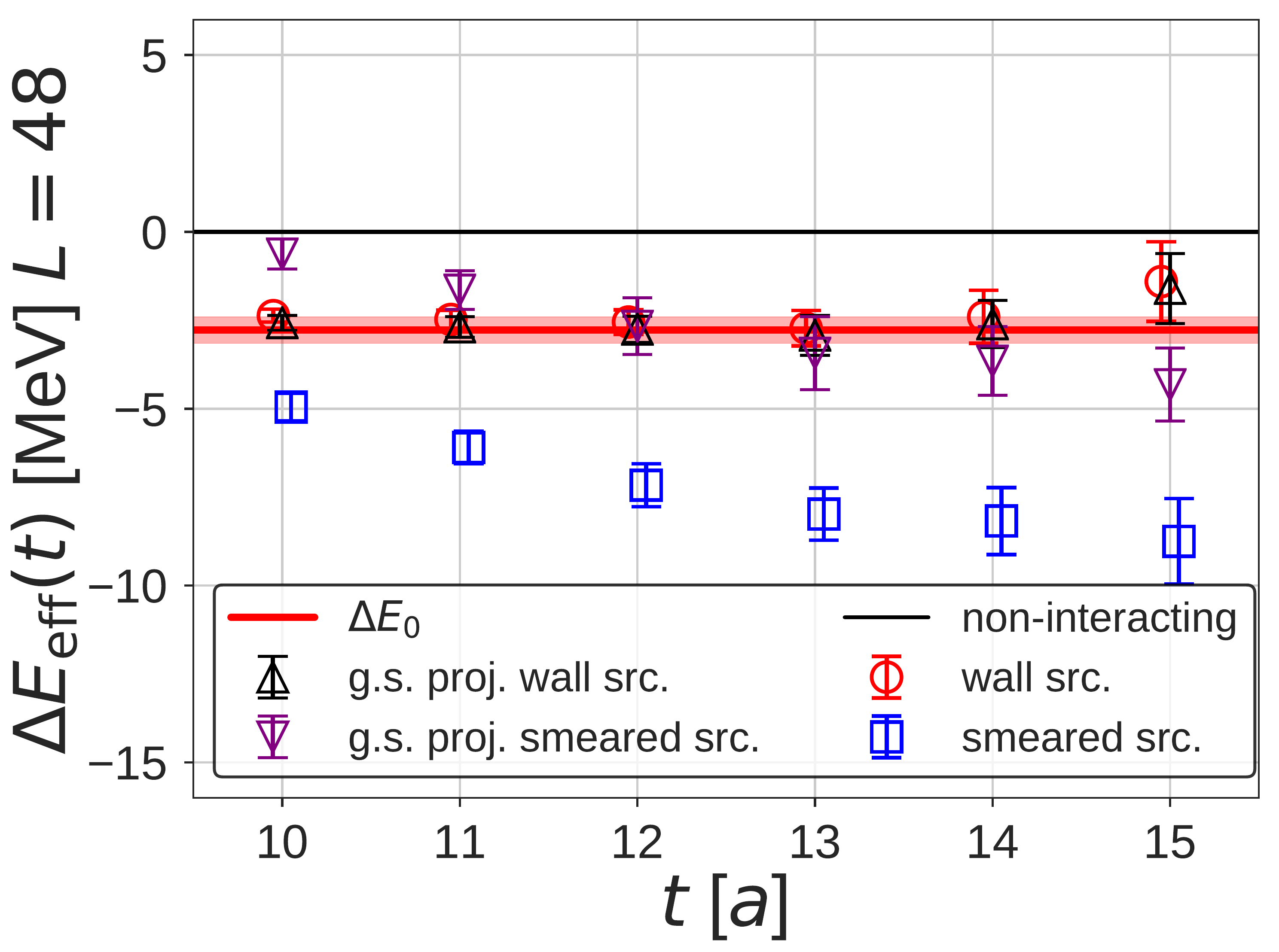}
   \includegraphics[width=0.48\textwidth]{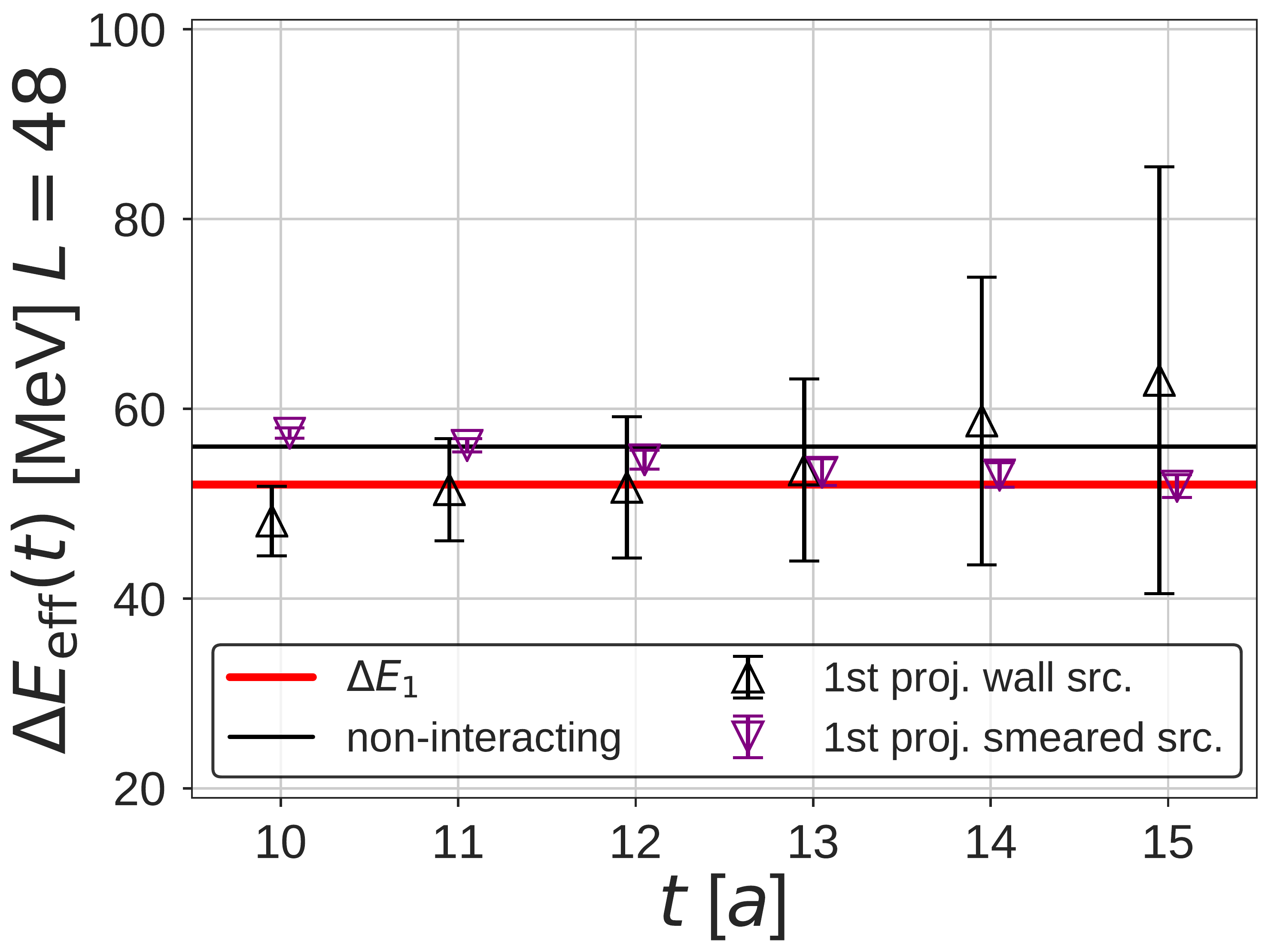}
 \caption{
 The  effective energy shift $\Delta E_{\rm eff}^{{J},n}(t)$ from  $R^{{J},n}_{\Xi\Xi}(t)$, the correlation function projected to the $n$-th eigenstate at the sink on $L=48 a$, for ${J}=$ wall (black open up-triangles) and ${J}=$ smear
(purple open down-triangle). 
Red bands represent the energy shifts from the eigenvalues of $H$ in the finite box, while black lines denote those of a free Hamiltonian $H_0$.  
(Left) The projection to the ground state ($n=0$), together with the effective energy shift in the direct method
without projection for the wall source (red open circles) and the smeared source (blue open squares).  
(Right) The projection to the 1st excited state ($n=1$).
Taken from \citep{Iritani:2018vfn}.
}
 \label{fig:Eeff_proj}
\end{figure} 
Once eigenmodes of $H$ in the finite box are obtained, we can construct an improved sink operator for a particular eigenstate, whose correlation function with the ${J}$ source  is given by
\beqa
R^{{J},n}_{BB}(t) &=& \sum_{\bf r} \Psi^\dagger_n({\bf r})  R^{{J}}_{BB}({\bf r}, t) .
\eeqa
Fig.~\ref{fig:Eeff_proj} shows the effective energy shift $\Delta E_{\rm eff}^{{J},n}(t)$ calculated from
$R^{{J},n}_{\Xi\Xi}(t)$ on $L=48a$ with ${J}=$ wall (black open up-triangles) and ${J}=$ smear (purple open down-triangle), for the ground state (Left) and the 1st excited state (Right),
together with $\Delta E_0$ or $\Delta E_1$, eigenvalues of $H$ in the finite box (red bands)
as well as those of $H_0$ (black lines). 
For the ground state in Fig.~\ref{fig:Eeff_proj} (Left), the effective energy shift in the direct method without projection are also plotted for the wall source (red open circles) and the smeared source (blue open squares).

After the sink projection, the effective energy shifts agree well between wall and smeared sources 
around $t/a \sim 13$, not only for the ground state but also for the 1st excited state.
while the effective energy shifts for the ground state in the direct method without projection disagree between two sources.
In particular, an agreement between two sources with sink projection for the 1st excited state
is rather remarkable, since variational methods, usually mandatory for excited states in  lattice QCD, are not used here.
Furthermore, the plateaux of the effective energy shifts after the sink projection also agree with  
$\Delta E_{0,1}$ of $H$ (red bands).
Note that the effective energy shift for the 1st excited state,  $\Delta E_{\rm eff}^{{\rm wall},1}(t)$,
has larger errors since the contribution of the 1st excited state in $R^{\rm wall}_{\Xi\Xi}(t)$
is much smaller.

Although the sink operator projection utilizes the information of the {HAL QCD} potential to construct eigenfunctions, agreements in the effective energy shifts for the ground state as well as the 1st excited state
provide a non-trivial consistency check between the HAL QCD  method and
{the L\"uscher's finite volume formula}
(with proper projections {to extract the finite volume spectra}).
We thus conclude from Fig.~\ref{fig:Eeff_proj} not only that
the HAL QCD potential correctly describes the energy shifts of two baryons in the finite box for both ground and excited states 
but also that
these energy shifts can be extracted even for baryon-baryon systems
if and only if the sink/source operators are highly improved.     
  We emphasize that improvement of operators has to be performed
  not by the tuning of the plateau-like structures
  but by a sophisticated method such as the variational method~\citep{Luscher:1990ck}
\footnote{
    In lattice QCD studies for the meson-meson scatterings~\cite{Briceno:2017max},
    serious systematics from the excited state contaminations
    in the simple plateau fitting have been widely recognized
    and the variational method has been utilized to obtain the finite volume spectra rather reliably,
    which can be combined with the L\"uscher's finite volume formula to extract phase shifts.
}
  (or a method presented here).
  See \citep{Francis:2018qch} for a recent study toward such a direction.

\section{Nuclear potential}
\label{sec:NN_pot}
In this section, we summarize results on nuclear potentials in the HAL QCD method.

\subsection{Parity-even channel with LO analysis at heavy pion masses}
\label{subsec:parity_even}
We first show the results of nuclear forces in the parity-even channel
($^1S_0$ and $^3S_1$-$^3D_1$ channels)
at heavy quark masses obtained by the LO analysis for the derivative expansion of the potential.
Since the statistical fluctuations are smaller at heavier quark masses
in lattice QCD,  this study is a good starting point to grasp the nature of
lattice QCD nuclear forces. In addition, quark mass dependence of nuclear forces is
of fundamental importance from a point of view of, e.g., anthropic principle,
which cannot be studied by experiments.

In the case of $^1S_0$ channel,
we obtain the LO central force following Eq.~(\ref{eq:t-dep_HAL}).
In the case of $^3S_1$-$^3D_1$ channel, the LO potentials consist of
the central and tensor forces, which can be obtained from
the coupled channel analysis between the $S$- and $D$-wave components as
\beqa
\left\{ -H_0 -\frac{\partial}{\partial t} +\frac{1}{4m_N} \frac{\partial^2}{\partial t^2}\right\}
R^J({\bf r},t) &=&\left[ V_C(r) + V_T(r) S_{12} +\cdots \right] R^J({\bf r},t),
\eeqa
where ellipses represent higher order terms in the derivative expansion.
Using the projection to the $A_1^+$ representation of the cubic group ($S$-wave projection), $\mathcal{P}^{A_1^+}$, and
the orthogonal one ($D$ wave projection), $(1-\mathcal{P}^{A_1^+})$, 
the above equation reduces to two independent equations, from which $V_C(r)$ and $V_T(r)$ can be obtained~\citep{Aoki:2009ji}.
Since the $A_1^+$ representation couples to
the angular momentum $l=0, 4, 6, \cdots$,
these projections are expected to serve as the relevant partial wave decomposition at low energies.
We find that  the NBS correlation functions after 
$\mathcal{P}^{A_1^+}$ and $(1-\mathcal{P}^{A_1^+})$
are dominated by $S$-wave and $D$-wave components, respectively,
indicating that the contaminations from $l\geq 4$ components are indeed small.
For a more advanced partial wave decomposition, see~\citep{Miyamoto:2019jjc}.

We perform the calculations in
quenched~\citep{Ishii:2006ec,Aoki:2009ji},
dynamical 2-flavor~\citep{Murano:2013xxa},
dynamical 3-flavor~\citep{Inoue:2010hs,Inoue:2010es,Inoue:2011ai}
and
dynamical (2+1)-flavor~\citep{Ishii:2009zr, HALQCD:2012aa, Ishii:2013ira, Iritani:2018zbt}
lattice QCD
with various quark masses.
We here present the results
obtained in 3-flavor  lattice QCD
at $(M_{\rm ps}, M_{\rm oct}) = (1171, 2274), (1015, 2031), (837, 1749), (672, 1484), (469, 1161)$
MeV~\citep{Inoue:2010hs,Inoue:2010es,Inoue:2011ai}.%
\footnote{$M_{\rm ps} = m_\pi = m_K$ and $M_{\rm oct} = m_N = m_\Lambda = m_\Sigma = m_\Xi$ in 3-flavor QCD.}
In the case of $(M_{\rm ps}, M_{\rm oct}) = (837, 1749)$,
the value of quark masses $m_u=m_d=m_s$ nearly correspond to the physical strange quark mass.
We generate gauge configurations 
with the RG-improved Iwasaki gauge action and
non-perturbatively $\mathcal{O}(a)$-improved Wilson quark action
on a $L^3\times T = 32^3\times 32$ lattice.
The lattice spacing is $a = 0.121(2)$ fm
and hence lattice size $L$ is 3.87 fm.
In the calculation of the NBS correlation function,
parity-even states are created by a two-baryon operator with a wall quark source,
while a point operator is employed for each baryon at the sink.

\begin{figure}[t]
  \centering
  \includegraphics[angle=0,width=0.32\textwidth]{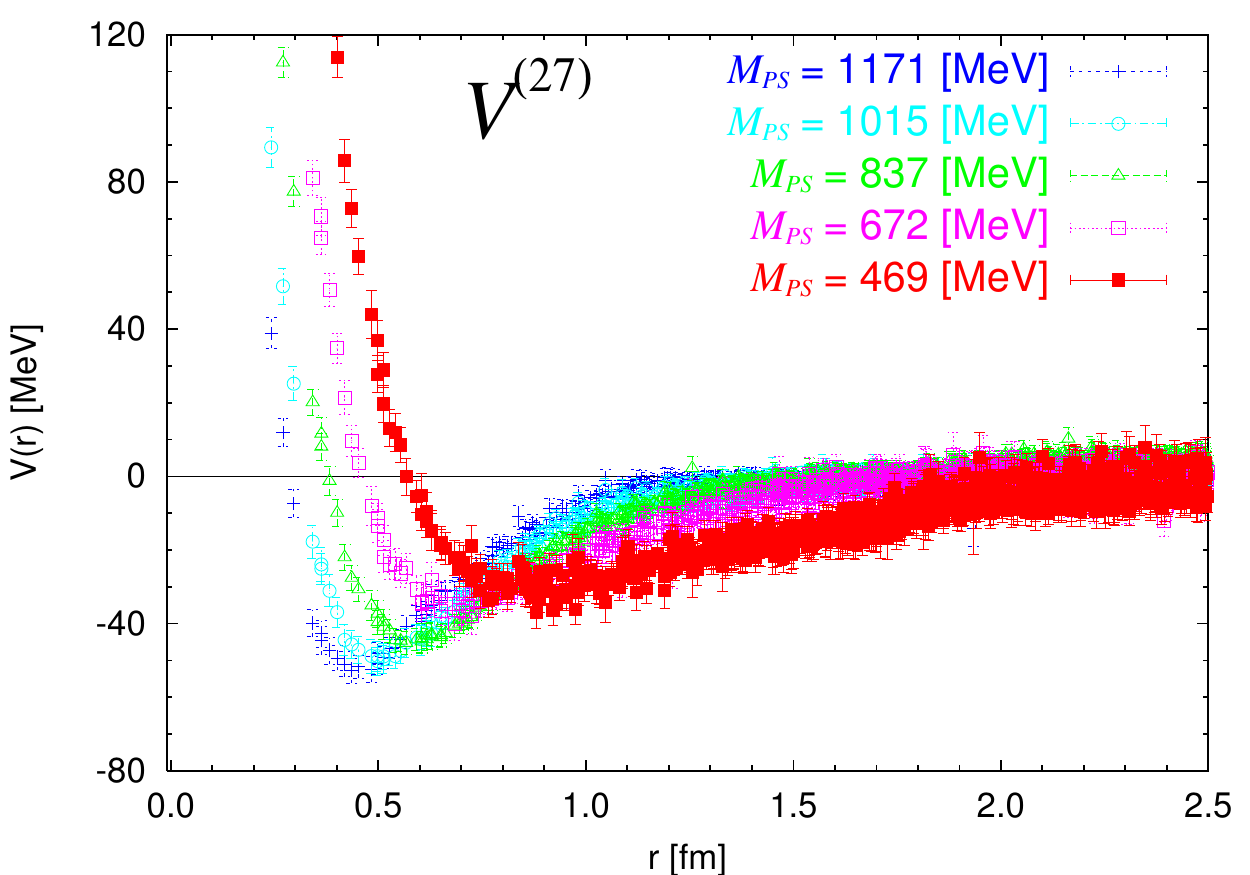}
  \includegraphics[angle=0,width=0.32\textwidth]{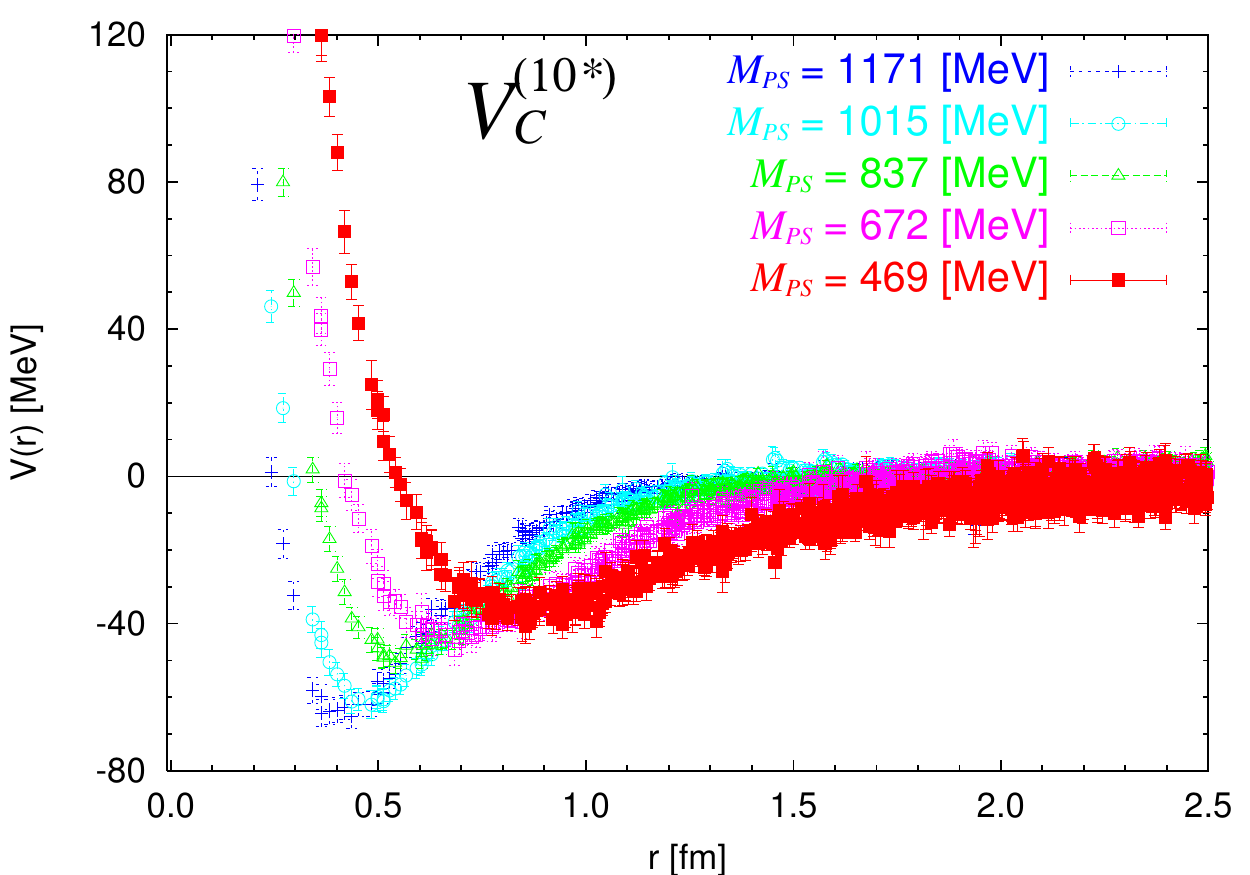}
  \includegraphics[angle=0,width=0.32\textwidth]{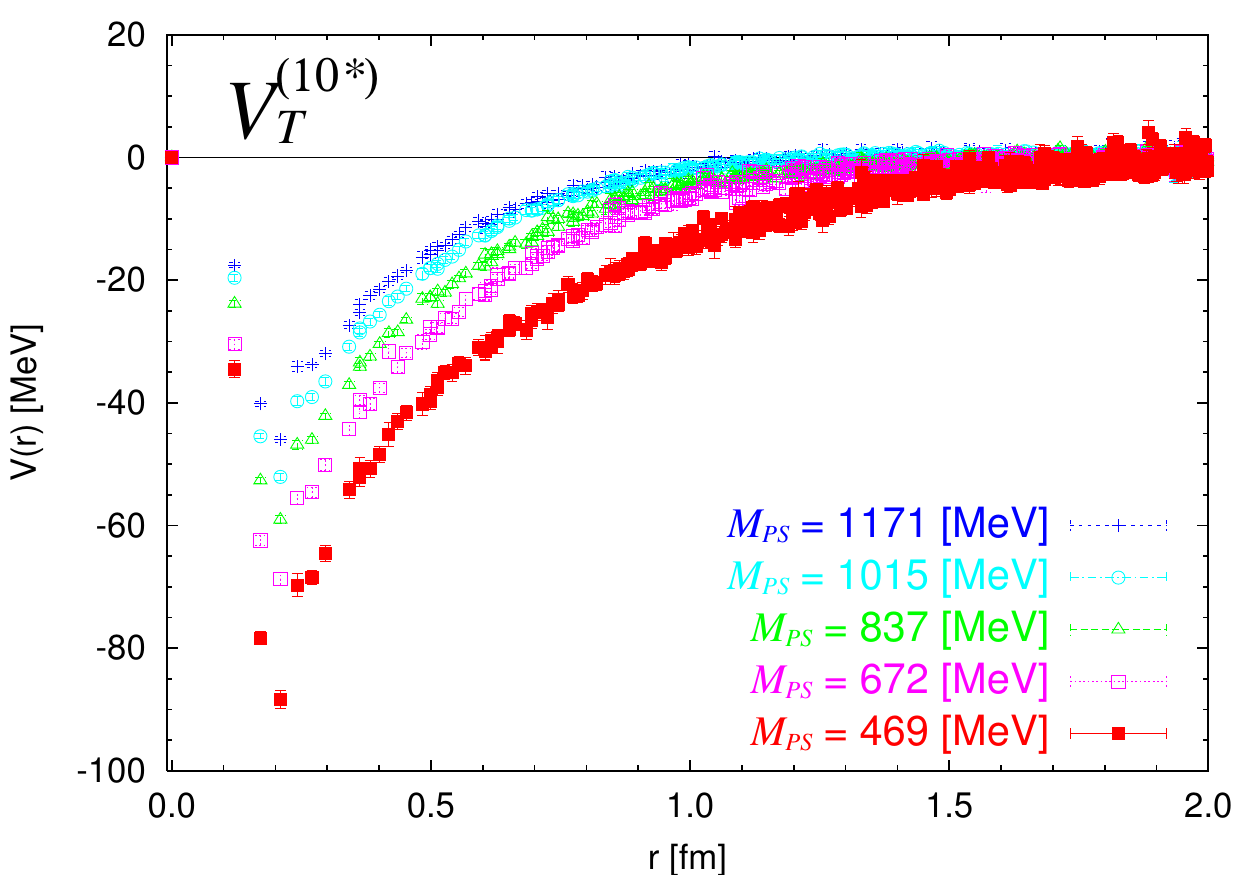}\\[5mm]
  \centering
  \includegraphics[width=0.42\textwidth]{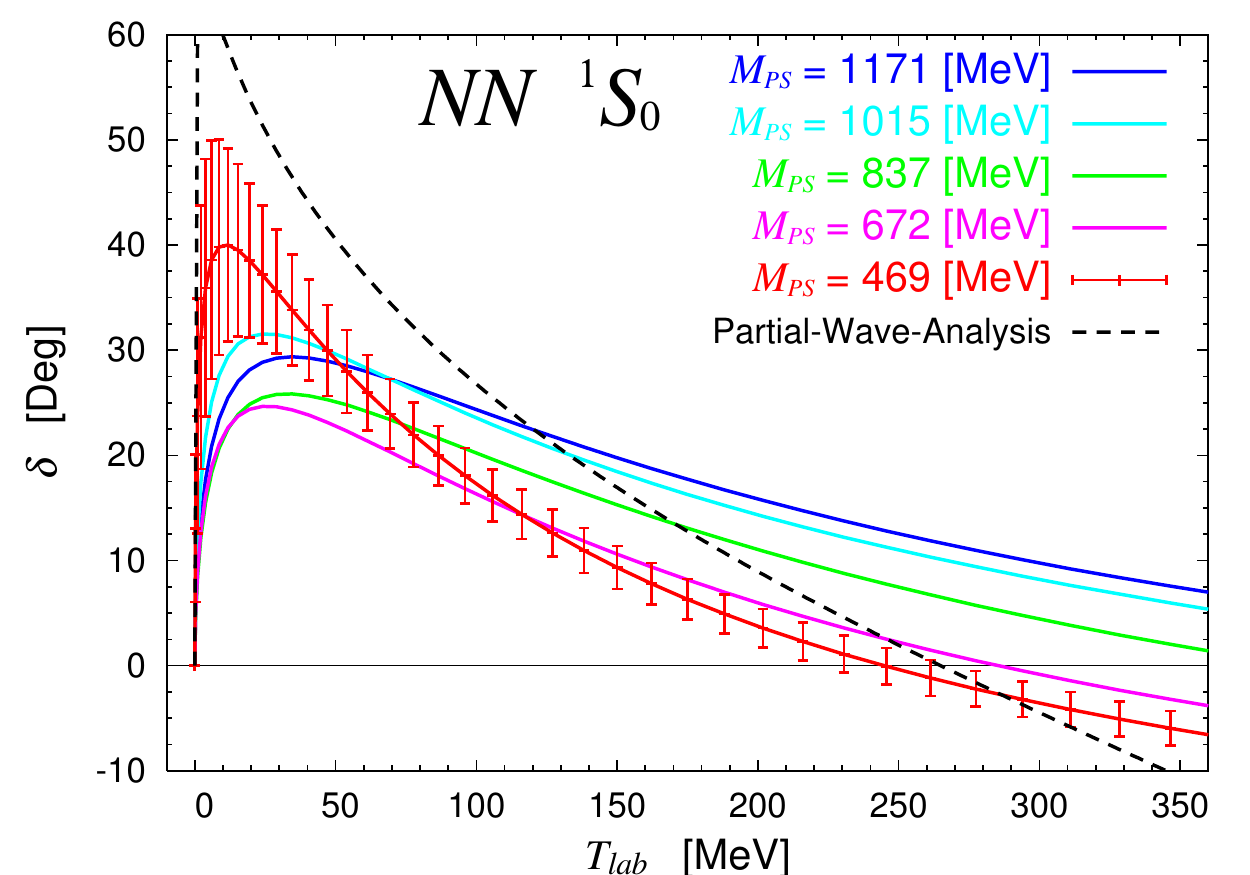}
  \includegraphics[width=0.42\textwidth]{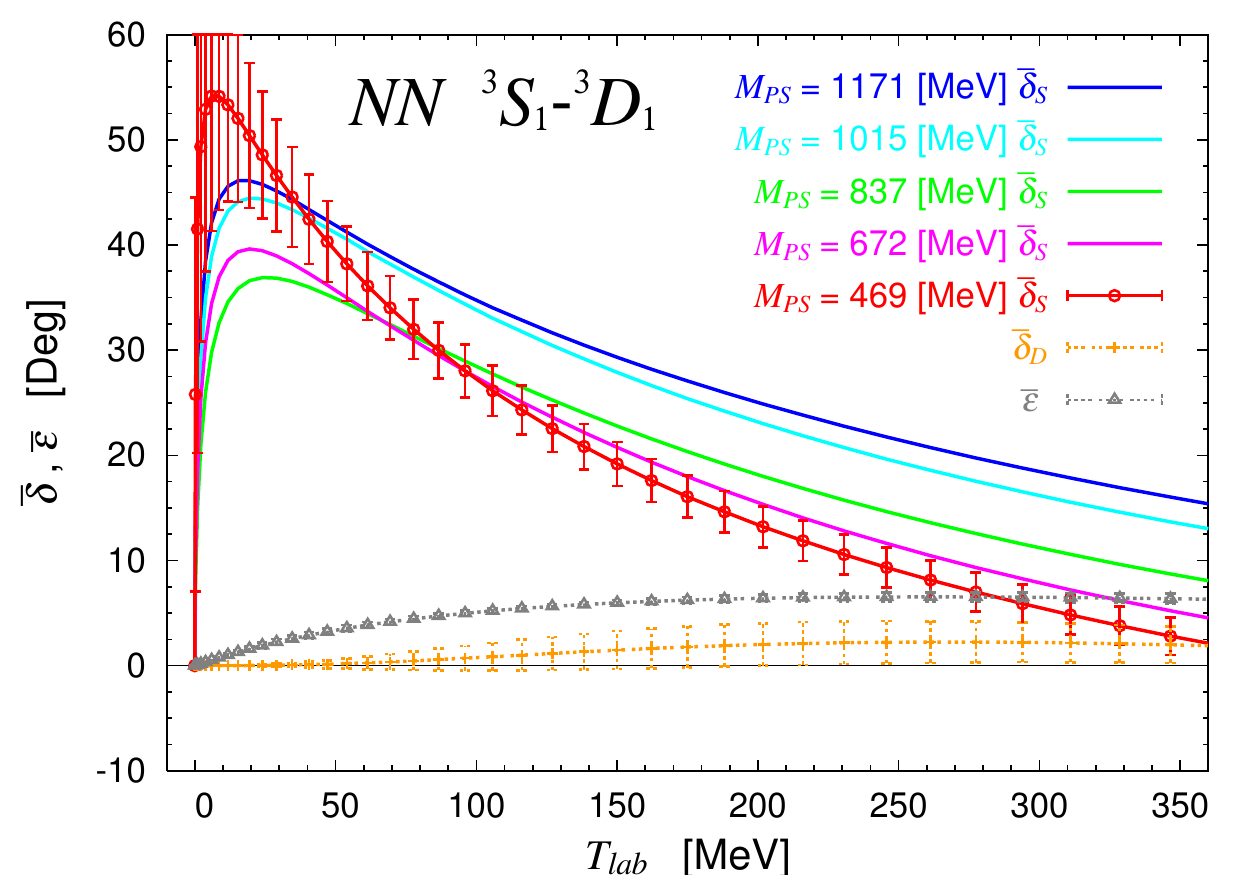}
  \caption{
    (Upper)
    Nuclear forces obtained from 3-flavor lattice QCD
    at $M_{\rm ps} =$ 469-1171 MeV.
    (Left) Central force in the $^1S_0$ channel (27-plet in SU(3)$_f$ representation).
    (Middle) Central force in the $^3S_1$-$^3D_1$ channel (10$^*$-plet in SU(3)$_f$ representation).
    (Right) Tensor force in the $^3S_1$-$^3D_1$ channel.
    (Lower)
  $NN$ scattering phase shifts as a function of energy in the laboratory frame (colored solid lines),
  obtained from 3-flavor lattice QCD at $M_{\rm ps} =$ 469-1171 MeV,
  together with those from experiments (black dashed lines).
  (Left) Results in the $^1S_0$ channel.
  (Right) Results in the $^3S_1$-$^3D_1$ channel (with Stapp's convention).
    Figures are taken from \citep{Inoue:2011ai}.
  }
\label{fig:su3_pot_phase}
\end{figure}

Shown in Fig.~\ref{fig:su3_pot_phase} (Upper) are the
lattice QCD results for the potentials.
We find that the results are insensitive to the Euclidean time $t$,
at which the NBS correlation function is evaluated,
indicating that the derivative expansion is well converged.
The obtained potentials are found to reproduce the qualitative features
of the phenomenological $NN$ potentials,
namely, attractive wells at long and medium
distances, central repulsive cores at short distance
and strong tensor force with a negative sign.
We also find intriguing features in the quark mass dependence of the potentials.
At long distances, it is observed that
the ranges of the tail structures in the central and tensor forces
become longer at lighter quark masses.
Such a behavior can be understood from the
viewpoint of one-boson-exchange potential.
{At} short distances, the repulsive cores in the central forces
are found to be enhanced at lighter quark masses.
This could be explained by the short-range repulsion due to the
one-gluon-exchange in the quark model, whose strength is proportional to
the inverse of the (constituent) quark mass.
In fact, our systematic studies including hyperon forces
with the same lattice setup revealed that the nature of repulsive core is
well described by the quark Pauli blocking effect
together with the one-gluon-exchange effect{~\citep{Inoue:2010hs, Inoue:2011ai, Oka:2000wj}}.

As noted before, the potentials themselves are not physical observables
and quantitative lattice QCD predictions shall be given in terms of scattering observables.
Shown in Fig.~\ref{fig:su3_pot_phase} (Lower) are the scattering phase shifts (and mixing angles)
obtained from lattice nuclear forces.
We find that $NN$ systems do not bound at these pseudoscalar masses
as discussed in Sec.~\ref{sec:comp}.
Behaviors of phase shifts are qualitatively similar to the experimental ones,
while the strength of the attraction is weaker due to the heavy quark masses in this calculation.
It is also observed that quark mass dependence of phase shifts is quite non-trivial.
In fact, if we decrease the quark masses,
there appear competing effects in the interaction:
the long-range attraction becomes stronger
and the short-range repulsive core also becomes stronger.
We also note that lighter quark masses correspond to
lighter nucleon mass, which leads to larger kinetic energies.


We also present the results
obtained in (2+1)-flavor lattice QCD
at quark masses corresponding to
$(m_\pi, m_N)\simeq $(701, 1584), (570, 1412) and (411,1215) MeV{~\citep{Ishii:2013ira}}.
Note that only up and down quark masses are varied
with a strange quark mass being fixed to the physical value in this study.
We employ the gauge configurations generated by the PACS-CS Collaboration
with the RG-improved Iwasaki gauge action and
non-perturbatively $\mathcal{O}(a)$-improved Wilson quark action
on a $L^3\times T = 32^3\times 64$ lattice.
The lattice spacing is $a \simeq 0.091$ fm ($a^{-1}=2.16(31)$GeV),
which leads to the spatial extension $L\simeq 2.9$ fm.

\begin{figure}[h]
  \centering
  \includegraphics[angle=270,width=0.43\textwidth]{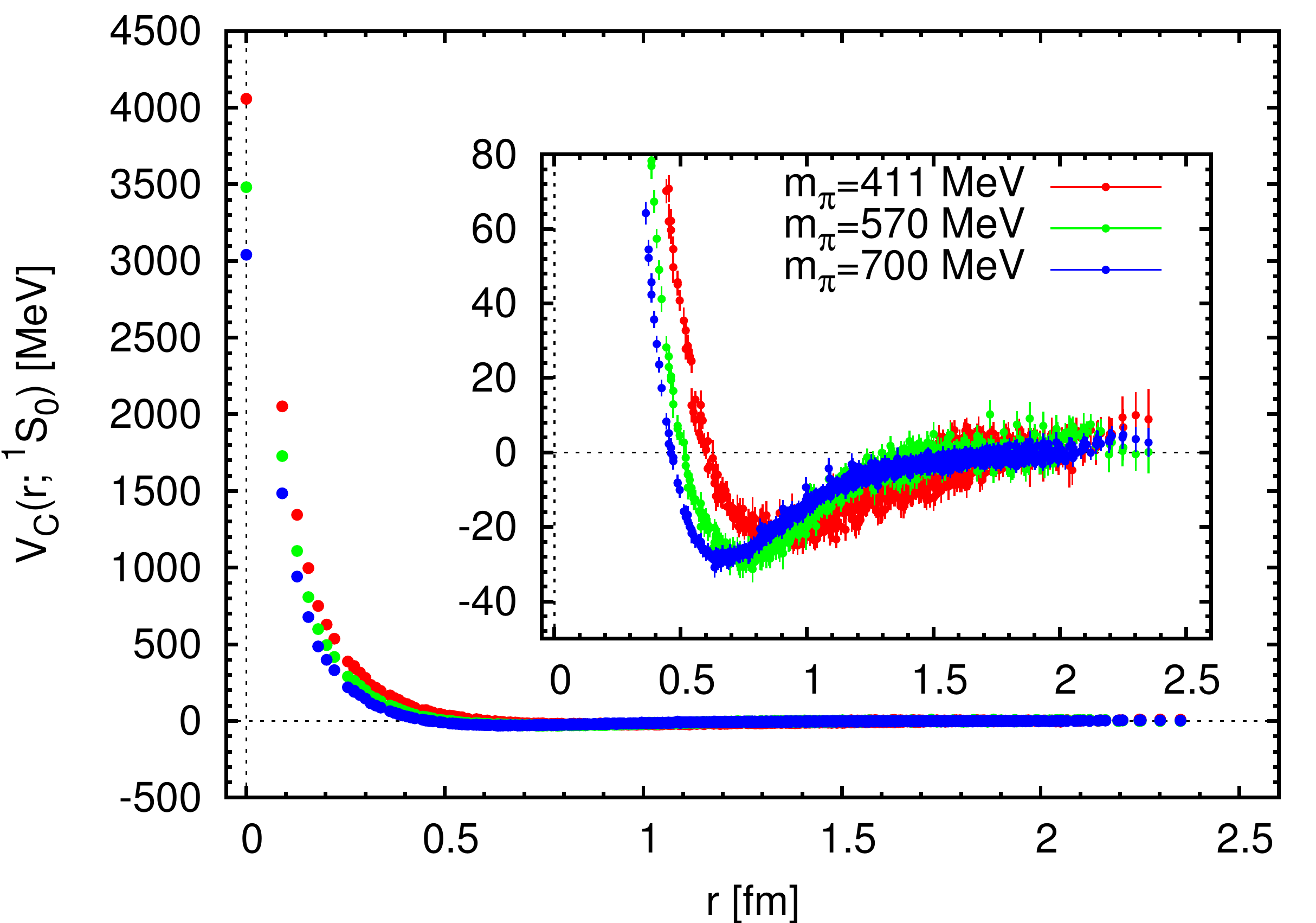}
  \includegraphics[angle=270,width=0.43\textwidth]{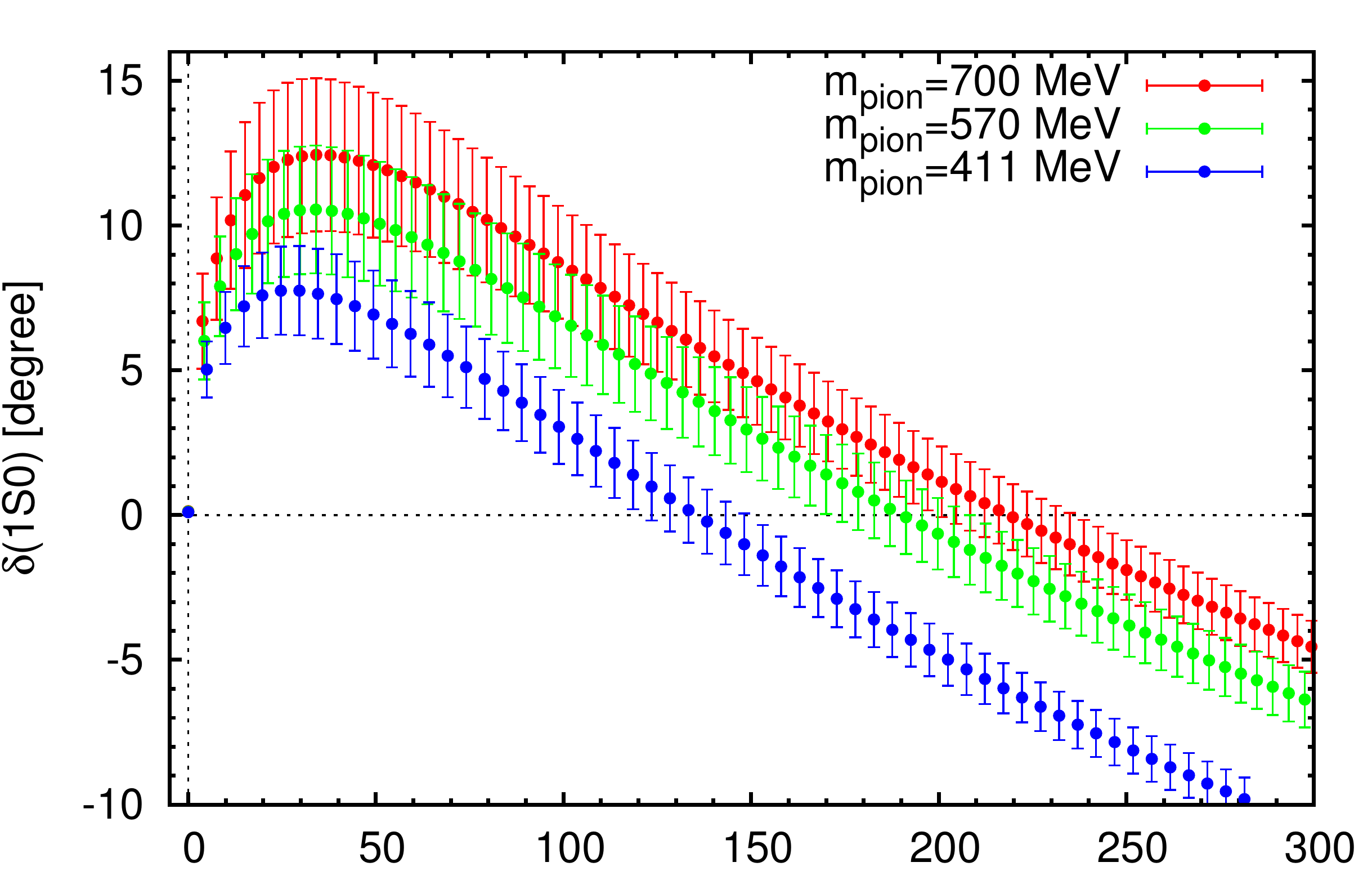}\\
  \includegraphics[angle=270,width=0.43\textwidth]{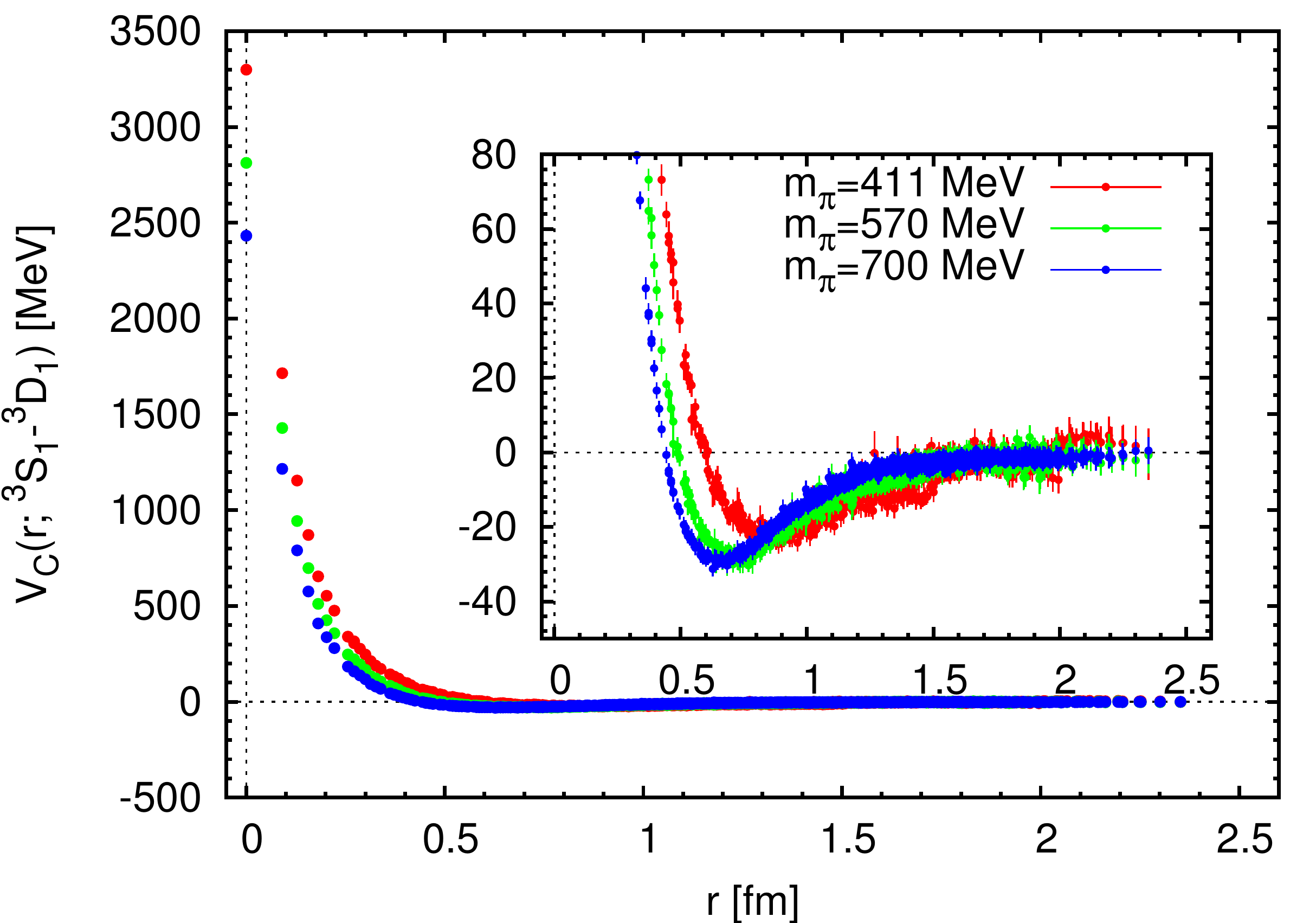}
  \includegraphics[angle=270,width=0.43\textwidth]{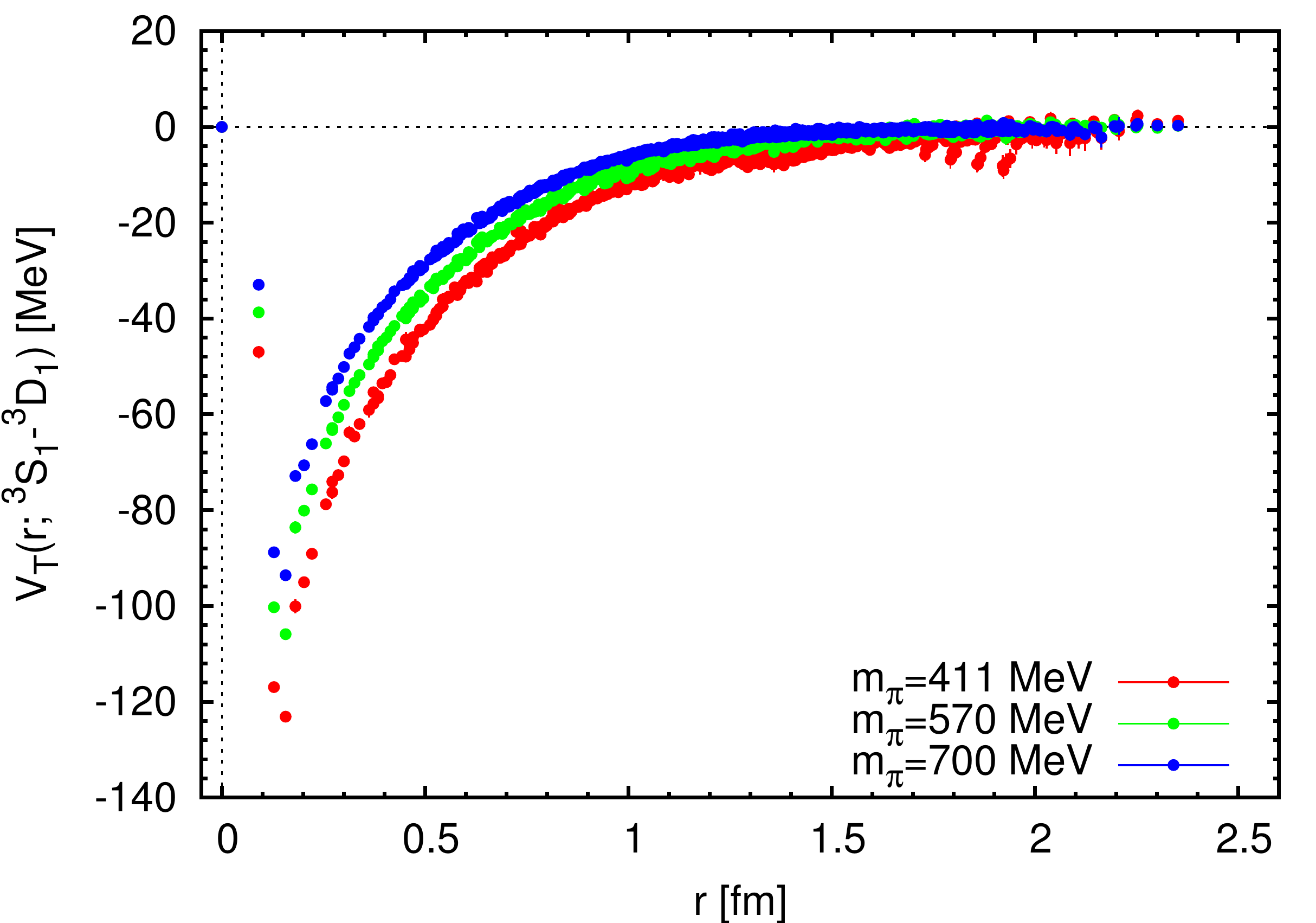}
  \caption{
    Nuclear {forces}
    obtained from (2+1)-flavor lattice QCD
    at $m_{\pi} \simeq$ 411 (red), 570 (green), 701 (blue) MeV:
    (Upper-Left) Central forces in the $^1S_0$ channel 
    (Lower) Central forces (left)  and tensor forces (right)  in the $^3S_1$-$^3D_1$ channel.
    (Upper-Right) The scattering  phase shifts in the $^1S_0$ channel 
    at $m_{\pi} \simeq$ 411 (blue), 570 (green), 701 (red) MeV.
     Figures are  taken from \citep{Ishii:2013ira}.
  }
\label{fig:pacs-cs_pot}
\end{figure}

In Fig.~\ref{fig:pacs-cs_pot},
we show the lattice QCD results for the potentials
in the $^1S_0$ and $^3S_1$-$^3D_1$ channels,
together with the corresponding phase shifts in the $^1S_0$ channel.
Qualitative features are similar to those in 3-flavor case:
(i) the central forces have repulsive cores at short distance
and attractive wells at long and medium distances,
both of which are enhanced at lighter quark masses
(ii) the tensor force is strong with a negative sign,
which increases at lighter quark masses.

\subsection{More structures: spin-orbit forces in {the} parity-odd channel and three nucleon forces}
\label{subsec:LS_3NF}

If we consider an interaction at higher order terms in the derivative expansion,
there appear more structures in the potentials.
In particular, the extension from LO analysis to NLO analysis enables
us to determine the spin-orbit (LS) force.
The LS force is known to play an important role
in the LS-splittings of nuclear spectra and the nuclear magic numbers.
In addition, the LS force in the $^3P_2$-$^3F_2$ channel
attracts great interest in nuclear astrophysics,
since it could lead to the $P$-wave superfluidity in the neutron stars
and affect the cooling process of neutron stars.

We here present the calculation
in parity-odd channels ($^1P_1$, $^3P_0$, $^3P_1$, $^3P_2$-$^3F_2$ channels)
at heavy quark masses and show the results of LS forces as well as central/tensor forces~\citep{Murano:2013xxa}.
In order to construct the source operator which couples to parity-odd states,
we employ the two nucleon operators as
\begin{eqnarray}
  \mathcal{J}_{\alpha\beta}(f_{i})
  \equiv
  N_{\alpha}(f^{(i)})
  N_{\beta }(f^{(i)\,*})
  \quad \mbox{for $i=\pm 1, \pm 2, \pm 3$}
\end{eqnarray}
where
$N$ denotes a nucleon operator with a momentum,
\begin{eqnarray}
  N_{\alpha} (f^{(i)})= \sum_{\vec x_1,\vec x_2,\vec x_3}
  \epsilon_{abc}
  \left(
    u_a^T(\vec x_1)
    C\gamma_5
    d_b(\vec x_2)
    \right)
    q_{c,\alpha}(\vec x_3)
    f^{(i)} (\vec x_3)
\end{eqnarray}
with
$f^{(\pm j)}(\vec x) \equiv \exp\left(\pm 2\pi i x_j / L\right)$.
A cubic group analysis shows that this source operator
contains the orbital contribution
$T_1^- \oplus  A_1^+ \oplus E^+$, whose dominant components have ${l} =1, 0, 2$, respectively,
and thus covers all the two-nucleon channels with $J\leq 2$.
Combined with the spin degrees of freedom,
we consider
the $T_1^-$ representation in the spin singlet channel
and
the $A_1^-$, $T_1^-$, $(E^- \oplus T_2^-)$ representations in the spin triplet channel.
At low energies, these representations correspond to
the $^1P_1$ channel and the $^3P_0$, $^3P_1$ and $^3P_2$-$^3F_2$ channels, respectively,
from which we extract the central force in the spin singlet channel ($V_{C,S=0}^{I=0}(r)$),
and the central, tensor and LS forces
($V_{C,S=1}^{I=1}(r), V_{T}^{I=1}(r), V_{LS}^{I=1}(r)$)
in the spin triplet channel.

Calculations are performed
in 2-flavor lattice QCD
at quark masses corresponding to
$(m_\pi, m_N)\simeq $(1133, 2158) MeV~\citep{Murano:2013xxa}.
We employ the gauge configurations generated by the CP-PACS Collaboration
with the RG-improved Iwasaki gauge action and
a mean field $\mathcal{O}(a)$-improved Wilson quark action
on a $16^3\times 32$ lattice.
The lattice spacing $a = 0.156(2)$ fm leads to the spatial extension $L\simeq 2.5$ fm.

\begin{figure}[t]
  \includegraphics[width=0.45\textwidth]{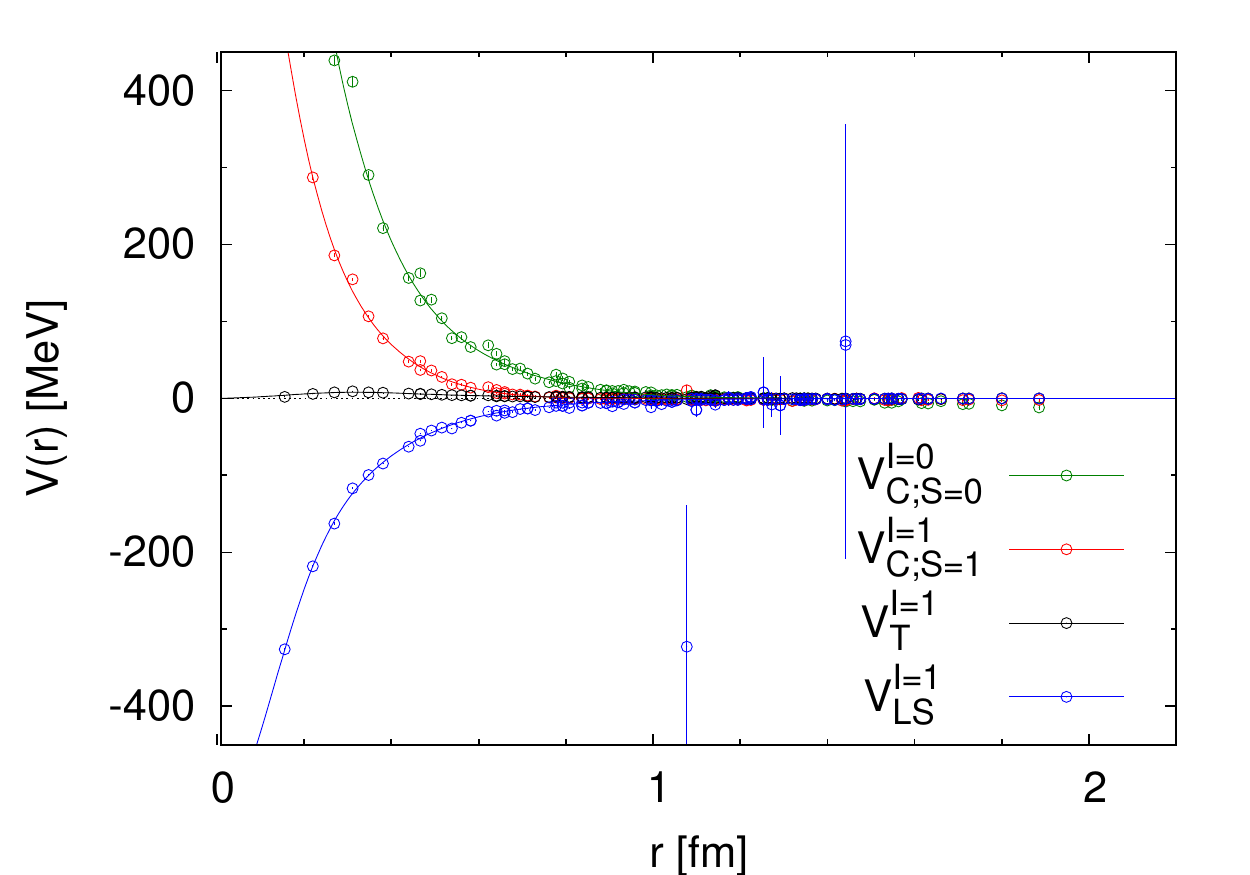}
  \includegraphics[width=0.45\textwidth]{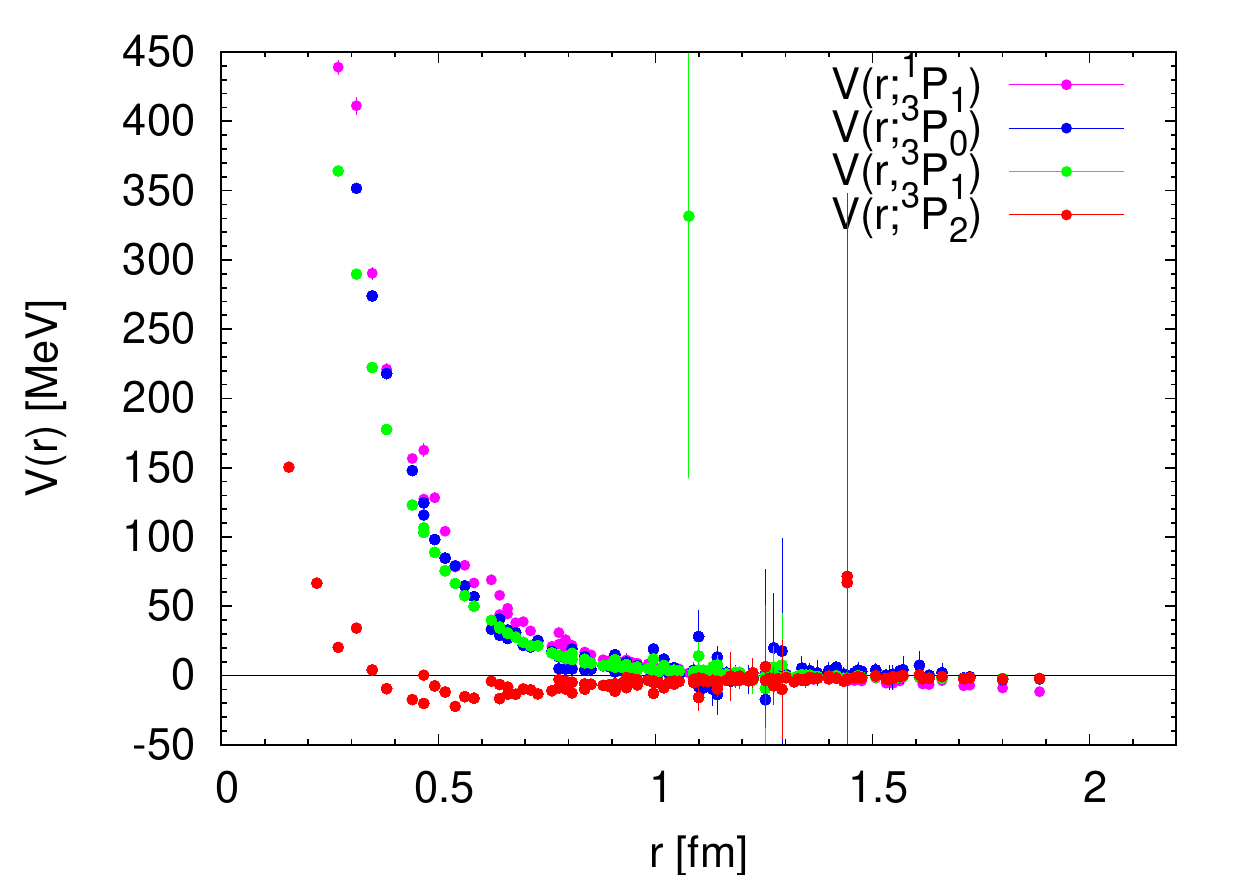}\\
  \includegraphics[width=0.45\textwidth]{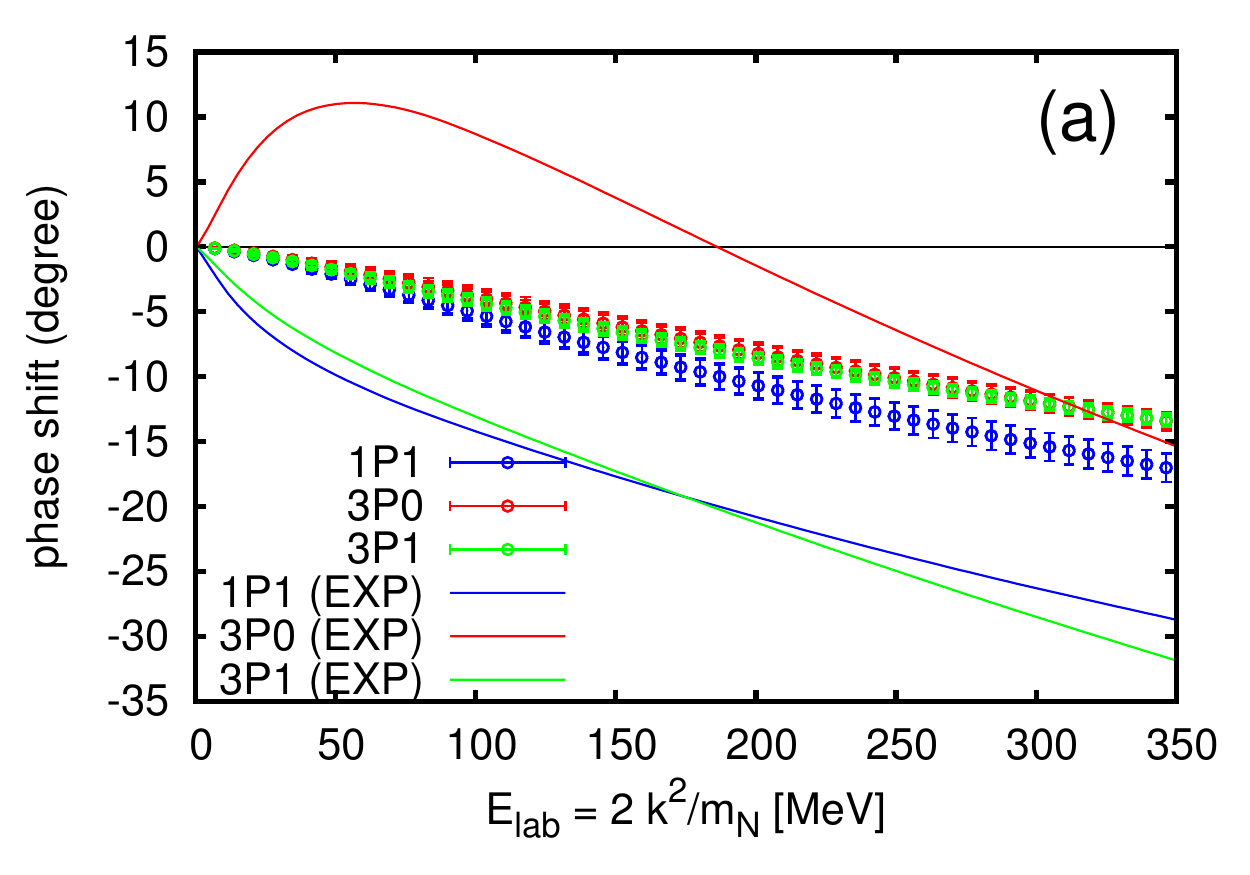}
  \includegraphics[width=0.45\textwidth]{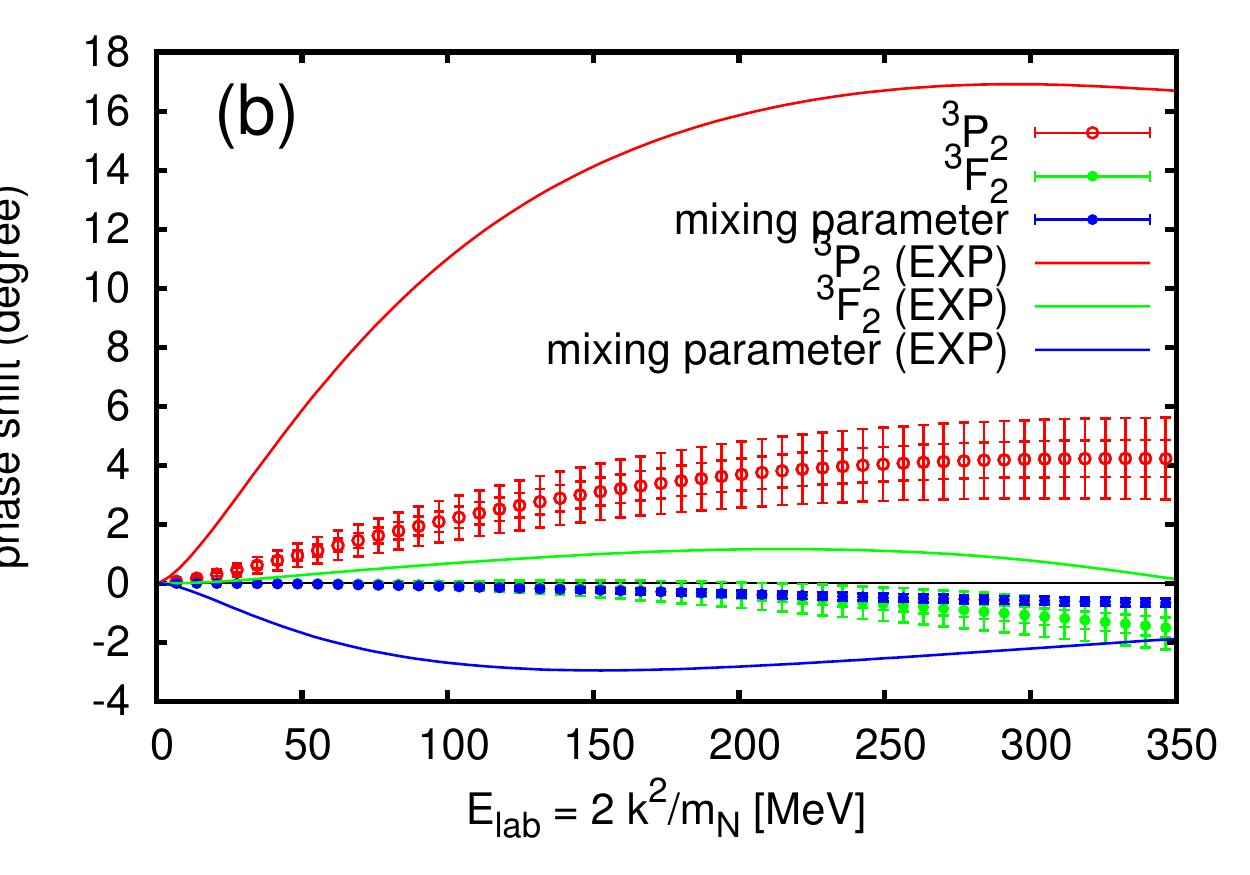}
  \caption{(Upper-Left)
    Central ($S=0$  and $1$), tensor  and spin-orbit potentials
    in parity-odd channels   obtained by 2-flavor lattice  QCD at $m_\pi \simeq 1133$ MeV.
    (Upper-Right)
    The potentials  for the $^1P_1$, $^3P_0$,  $^3P_1$ and  $^3P_2$
    channels.
    (Lower-Left) Phase shifts in the $^1P_1$, $^3P_0$ and $^3P_1$ channels, together with the
    experimental ones for comparisons.
    (Lower-Right) Phase shifts and  mixing parameter (with Stapp's convention)
    in the $^3P_2$--$^3F_2$
    channel, together  with the experimental ones.
     Figures are  taken from \citep{Murano:2013xxa}.  
}
 \label{fig:p-odd}
\end{figure}

Shown in Fig.~\ref{fig:p-odd} (Upper-Left) are
the lattice QCD results for the potential,
$V_{C,S=0}^{I=0}(r)$, $V_{C,S=1}^{I=1}(r), V_{T}^{I=1}(r), V_{LS}^{I=1}(r)$.
We find that
(i) the central forces $V_{C,S=0}^{I=0}(r)$ and  $V^{I=1}_{{\rm C};S=1}(r)$ are  repulsive,
(ii) the tensor force $V^{I=1}_{\rm T}(r)$  is positive  and 
weak  compared to $V^{I=1}_{{\rm C};S=1}(r)$ and $V^{I=1}_{\rm LS}(r)$,
and
(iii)  the LS force  $V^{I=1}_{\rm LS}(r)$
is  negative  and strong.
These  features are qualitatively in line well with
those of the phenomenological potential.
One can also see these properties in terms
of the potential in each channel.
In Fig.~\ref{fig:p-odd} (Upper-Right),
we plot the
potentials {in} the $^1P_1$, $^3P_0$, $^3P_1$ and
$^3P_2$ channels,
which are defined by
$  V(r;\ ^1P_1) = V^{I=0}_{{\rm C},S=0}(r) $, 
$ V(r;\ ^3P_0) = V^{I=1}_{{\rm C},S=1}(r) - 4V^{I=1}_{\rm T}(r) - 2 V^{I=1}_{\rm LS}(r)$,
$  V(r;\ ^3P_1) = V^{I=1}_{{\rm C},S=1}(r)  + 2V^{I=1}_{\rm T}(r) - V^{I=1}_{\rm LS}(r) $, 
  %
$  V(r;\ ^3P_2) = V^{I=1}_{{\rm C},S=1}(r)  - \frac{2}{5}V^{I=1}_{\rm T}(r) + V^{I=1}_{\rm LS}(r)$.

To obtain the scattering observables, 
we fit the potentials and solve the Schr\"odinger equation in the infinite volume.
In Fig.~\ref{fig:p-odd} (Lower), we show the results for the scattering phase shifts.
Compared with the experimental phase shifts,
we find that behaviors of phase shifts are generally well reproduced,
while the magnitudes are smaller due to the heavier pion mass
in lattice QCD calculations.
In the $^3P_0$  channel, we observe that the attraction is missing compared with
the experimental one, which however is also likely due to the
weak tensor force $V_{\rm T}$  caused by the heavier pion mass.
Among others,  the most interesting feature  
is the attraction  in the $^3P_2$ channel as shown in Fig.~\ref{fig:p-odd} (Lower-Right),
originated from the strong (and negative) LS forces.
As noted before, it is this interaction
which is relevant to the paring correlation of the neutrons
and possible $P$-wave superfluidity in the neutron stars.

We now turn to the study of three-nucleon forces.
Determination of three-nucleon forces is one of the most challenging
problems in nuclear physics:
Three-nucleon forces are known to play important role
in nuclear spectra/structures
such as the binding energies of (light) nuclei
and
properties of neutron-rich nuclei.
They are also essential ingredients to understand properties
of nuclear matters such as the equation of state (EoS) at high density,
which is relevant to the structures of neutron stars
and nucleosynthesis at the binary neutron star mergers.
While
there have been many studies to construct three-nucleon forces
by phenomenological approaches~\citep{Coon:2001pv,Pieper:2001ap}
or
by chiral EFT approaches~\citep{Weinberg:1992yk,Epelbaum:2008ga,Machleidt:2011zz,Hammer:2019poc},
it is most desirable to carry out the direct determination from QCD.

To study three-nucleon forces in lattice QCD,
we consider the NBS wave function for a $n (\geq 3)$-particle system, $\vert \alpha\rangle$,
\begin{eqnarray}
  \Psi^n_{\alpha}([{\bx}] ) e^{-W_\alpha t} &=&
  \langle 0 \vert N(\bx_1, t) N(\bx_2, t)\cdots N(\bx_n, t) \vert \alpha\rangle,
  \quad
  [\bx ] =\bx_1,\bx_2,\cdots,\bx_n
\end{eqnarray}
where $W_\alpha$ is the center of mass energy of the system
and we ignore the spins of nucleon for simplicity.
In~\citep{Aoki:2012bb,Aoki:2013cra,Gongyo:2018gou}, we show that the
asymptotic behavior of the NBS wave function
with the non-relativistic approximation
can be written as
\beqa
\Psi^n_{[L],[K]}(R, Q)
 & \propto &
 \sum_{[N]}
 U_{[L][N]}(Q) e^{i \delta_{[N]}(Q)} 
 \frac{\sin\left(Q R -\Delta_L +\delta_{[N]}(Q)\right)}{ (Q R)^{\frac{D-1}{2}}}
 U_{[N][K]}^\dagger(Q) 
 \label{eq:main_result}
\eeqa
where $D=3(n-1)$ is the dimension of a $n$-particle system,
$\Delta_L = (2L+D-3)\pi/4$,
$\Psi^n_{[L],[K]}(R, Q)$ is the radial component of
the NBS wave function in $D$-dimension
with $R$ and $Q$ being the hyper radius and momentum, respectively,
and
$[L], [K]$ denotes the quantum numbers of the angular momentum in $D$-dimension.
$\delta_{[N]}(Q)$ is the generalized ``phase shift'' for a $n$-particle system  and
$U_{[L][N]}(Q)$ is a unitary matrix, which parameterize the $T$-matrix as
\beqa
T_{[L][K]}(Q,Q) &=& \sum_{[N]} U_{[L][N]}(Q) T_{[N]}(Q) U_{[N][K]}^\dagger(Q), \\
T_{[N]}(Q) &=& - \frac{2 n^{3/2}}{m_N Q^{3n-5}} e^{i\delta_{[N]}(Q)} \sin \delta_{[N]}(Q) .
\label{eq:UTU}
\eeqa
Therefore, as in the case of $n=2$ system (See Sec.~\ref{subsubsec:strategy}),
the information of $T$-matrix is encoded in the
asymptotic behavior of the NBS wave function.
Based on this property,
we can define the energy-independent non-local potential for a $n$-particle system,
which can be extracted from the (time-dependent) HAL QCD method.

We calculate the six-point correlation function divided by two-point correlation function cubed,
\begin{eqnarray}
  R_{3N}(\vec{r},\vec{\rho},t-t_0) &\equiv& G_{3N} (\vec{r},\vec{\rho},t-t_0) / \{G_N(t-t_0)\}^3 \\
  G_{3N} (\vec{r},\vec{\rho},t-t_0)
  &\equiv& 
  \frac{1}{L^3}
  \sum_{\vec{R}}
  \langle 0 |
          (N(\vec{x}_1) N(\vec{x}_2) N (\vec{x}_3))(t) \
\overline{(N'       N'        N')}(t_0)
| 0 \rangle
\end{eqnarray}
where
$\vec{R} \equiv ( \vec{x}_1 + \vec{x}_2 + \vec{x}_3 )/3$,
$\vec{r} \equiv \vec{x}_1 - \vec{x}_2$, 
$\vec{\rho} \equiv \vec{x}_3 - (\vec{x}_1 + \vec{x}_2)/2$
are the Jacobi coordinates.
In the time-dependent HAL QCD method at the LO analysis for the derivative expansion
and with the non-relativistic approximation,
we can extract the three-nucleon forces
$V_{3NF}(\vec{r},\vec{\rho})$ through the following 
Schr\"odinger equation,
\begin{eqnarray}
%
\biggl[ 
- \frac{1}{2\mu_r} \nabla^2_{r} - \frac{1}{2\mu_\rho} \nabla^2_{\rho} 
+ \sum_{i<j} V_{2N} (\vec{r}_{ij})
+ V_{3NF} (\vec{r}, \vec{\rho})
\biggr] R_{3N}(\vec{r}, \vec{\rho},t)
= - \frac{\partial}{\partial t} R_{3N}(\vec{r}, \vec{\rho},t) , \ \ \ \ 
\label{eq:Sch_3N}
\end{eqnarray}
where
$V_{2N}(\vec{r}_{ij})$ with $\vec{r}_{ij} \equiv \vec{x}_i - \vec{x}_j$
denotes two-nucleon forces between $(i,j)$-pair,
$\mu_r = m_N/2$, $\mu_\rho = 2m_N/3$ the reduced masses.

In our first study of three-nucleon forces,
we consider the total 3N quantum numbers of $(I, J^P)=(1/2,1/2^+)$, the triton channel.
We also consider a particular spacial geometry of the 3N,
i.e., the ``linear setup'' ($\vec{\rho}=\vec{0}$),
where 3N are aligned linearly with equal spacing of 
$r_2 \equiv |\vec{r}|/2$.
This setup makes the analysis much simpler.
In addition, we consider the following channel,
$
\psi_S \equiv
\frac{1}{\sqrt{6}}
\Big[
-   \Pu \Nu \Nd + \Pu \Nd \Nu               
                - \Nu \Nd \Pu + \Nd \Nu \Pu 
+   \Nu \Pu \Nd               - \Nd \Pu \Nu
\Big]  ,
$
and calculate the corresponding matrix element of $V_{3NF}$,
so that we can suppress the statistical fluctuations
in subtracting the contribution from $V_{2N}$.

One of the biggest challenges in the lattice QCD study of three-nucleon forces
is the enormous computational cost required for the calculation of correlation functions.
In fact, in terms of a mass number $A$,
the cost grows with the multiplication of two factors,
one of which scales factorially in $A$ due to the Wick contraction (permutation of quarks),
and the other of which scales exponentially in $A$ due to the color/spinor contractions.
On this point, we have developed a novel computational algorithm, called the unified contraction algorithm (UCA),
in which two contractions are unified and redundant calculations are eliminated systematically~\citep{Doi:2012xd}. 
In particular, the computation becomes faster by a factor of 192
for a calculation of three-nucleon forces.

\begin{figure}[t]
  \includegraphics[width=0.45\textwidth]{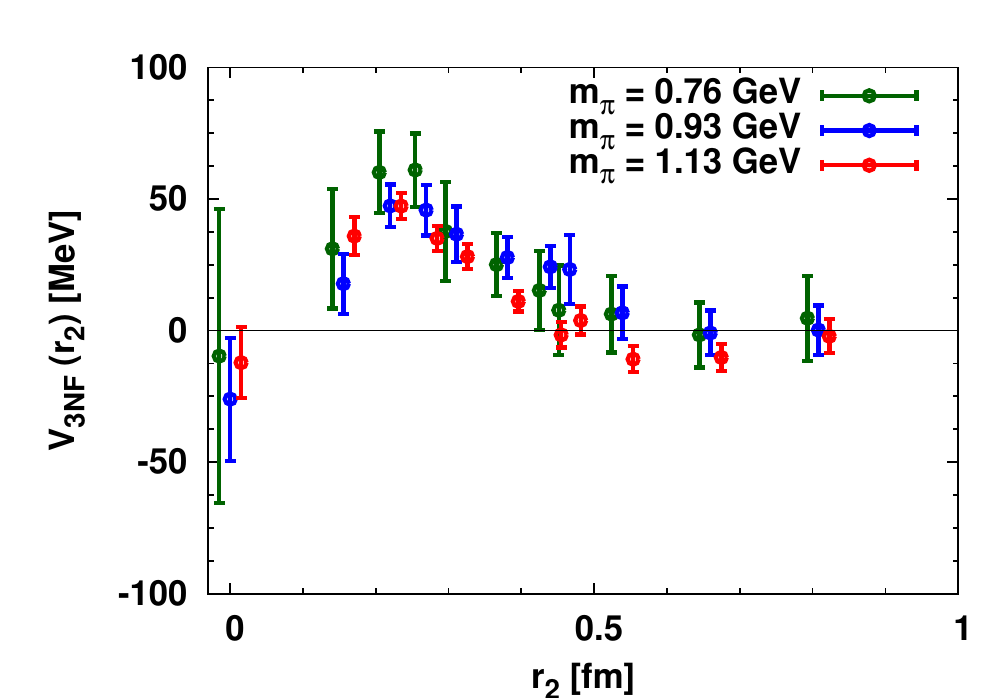}
  \includegraphics[width=0.45\textwidth]{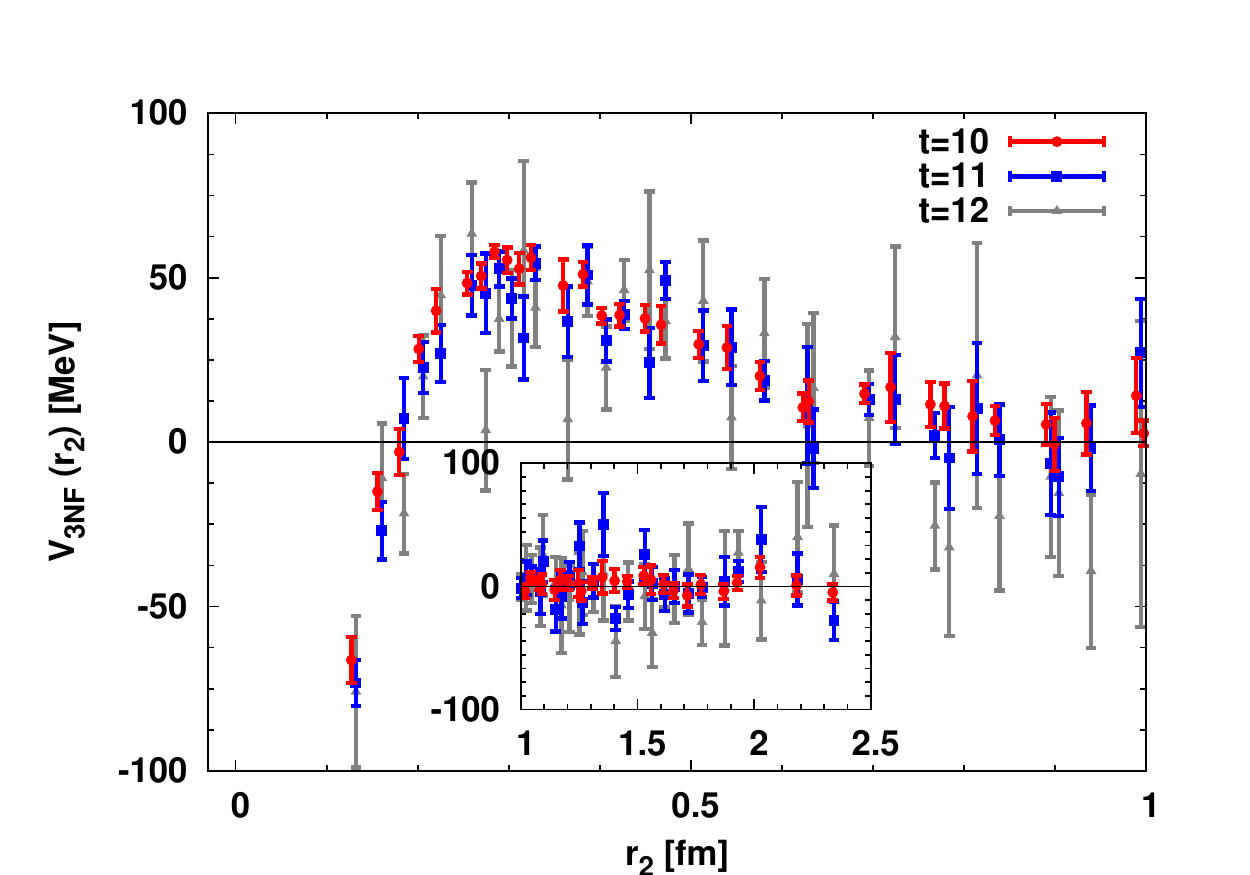}
  \caption{Three-nucleon forces in the triton channel with the linear setup.
    (Left)
    Results from 2-flavor lattice QCD at $m_\pi =$ 0.76-1.13 GeV.
    (Right)
    Results from (2+1)-flavor lattice QCD at $m_\pi = 0.51$ GeV.
}
 \label{fig:tnf}
\end{figure}

We perform the calculation 
in 2-flavor lattice QCD
at $(m_\pi, m_N) = (0.76, 1.81), (0.93, 1.85), (1.13, 2.15)$ GeV~\citep{Doi:2011gq}.
We employ the gauge configurations generated by
CP-PACS Collaboration
with mean field $\mathcal{O}(a)$-improved Wilson fermion 
and 
RG-improved Iwasaki gauge action
on a $L^3\times T = 16^3\times 32$ lattice.
The lattice spacing is $a=0.1555(17)$ fm 
and thus $L = 2.5$ fm.
Shown in Fig.~\ref{fig:tnf} (Left)
are the lattice QCD results for the three-nucleon forces.
We find a repulsive interaction at short-distances,
$r_2 \simeq $ 0.2-0.4 fm.
(Results at $r_2 \lesssim $ 0.2 fm would suffer
 from lattice discretization error.)
Note that a repulsive short-range three-nucleon force
is phenomenologically required 
to explain the properties of high density matter.
On the other hand, three nucleon forces are found to be suppressed at long distances.
This is in accordance with the suppression of two-pion-exchange
due to the heavier pion masses. 

Shown in Fig.~\ref{fig:tnf} (Right)
is the latest preliminary result obtained at $m_\pi = 510$ MeV.
In this calculation, we employ (2+1)-flavor lattice QCD gauge configurations
generated in~\citep{Yamazaki:2012hi}
with the RG-improved Iwasaki gauge action and
non-perturbatively $\mathcal{O}(a)$-improved Wilson quark action
on a $L^3\times T = 64^3\times 64$ lattice (work in progress).
The lattice spacing is $a = 0.090$ fm and $L = 5.8$ fm.
Avoiding the very short-distance region
where lattice discretization error could affect the results,
we again find the short-range repulsive three-nucleon forces
at $r_2 \simeq $ 0.2-0.7 fm.
We find that, while the pion mass dependence of three-nucleon forces
is not significant at $m_\pi =$ 0.76-1.13 GeV,
the range of repulsive three-nucleon forces tend to be enlarged 
at $m_\pi = 0.51$ GeV.
It is important to pursue the study at lighter pion masses
towards {the} physical pion mass.

\subsection{{Applications to nuclei, nuclear equation of state and structure of neutron stars}}
\label{subsec:application}

Once nuclear potentials are obtained by lattice QCD,
we can use them to study various phenomena in nuclear physics and astrophysics.
We here present the study of nuclear spectra/structures and
Equation of State (EoS) of dense matter relevant to neutron star physics.
Potentials used in this subsection are of the leading order only, and therefore are all hermitian. 
We can make non-hermitian higher order potentials in the HAL QCD method  hermitian in the derivative expansion\citep{Aoki:2019gqt}, which may be used for future applications in nuclear many body calculations.

In~\citep{McIlroy:2017ssf}, binding energies and structures of doubly magic nuclei,
$^4$He, $^{16}$O, $^{40}$Ca, 
are studied
by an ab initio nuclear many-body calculation based on lattice nuclear forces.
We employ the nuclear forces obtained in 3-flavor lattice QCD
at $M_{\rm ps}$ = 469 MeV (See Fig.~\ref{fig:su3_pot_phase}).
We consider two-body nuclear forces in $^1S_0$, $^3S_1$ and $^3D_1$ channels,
while nuclear forces in other channels and spin-orbit forces
as well as
three-nucleon forces are neglected.
For simplicity, the Coulomb force between protons is not taken into account, either.
As the ab initio many-body calculation,
we employ self-consistent Green's function (SCGF) method, 
  in which
  the single-particle propagator (Green's function) and
  the self-energy is solved self-consistently in a nonperturbative manner.
  In a practical calculation,
  the self-energy is calculated by
  Algebraic Diagrammatic Construction (ADC) formalism at third order
  for the so-called (low-momentum) $P$-space
  and
  Bethe-Goldstone equation (BGE) for the $Q=1-P$ space.
  (see~\citep{McIlroy:2017ssf} for details.)

In Tab.~\ref{tab:He4O16Ca40}, we summarize the results for the ground state energies,
together with the results from Brueckner Hartree-Fock (BHF) calculation~\citep{Inoue:2014ipa}
and exact stochastic variational calculation~\citep{Nemura:2014kea}
using the same lattice nuclear forces.
  For the results from SCGF,
  the first parentheses show the errors associated with the infrared (IR) extrapolation in the SCGF calculation.
  We also estimate the errors from many-body truncations using $^4$He as a benchmark.
  Since the SCGF result deviates from the exact solution by less than 10\% for $^4$He,
  and the SCGF approach is size extensive,
  we take a conservative estimate of 10\% error for $^{16}$O and $^{40}$Ca,
  which are quoted in the second parentheses.
  The BHF results are sensibly more bound than the SCGF results,
  and we interpret this as a limitation of BHF theory.
  For the results shown in Tab.~\ref{tab:He4O16Ca40},
  there exist additional errors associated with the statistical fluctuations
  in the input lattice nuclear forces, which are estimated to be $\sim$ 10\%~\citep{Inoue:2014ipa}.
  Note that statistical fluctuations are correlated among nuclei, so we expect our observations
  described below are rather robust against statistical errors.

We find that at $M_{ps}$=469~MeV in the SU(3) limit of QCD,
both $^4$He and $^{40}$Ca have bound ground states
while the deuteron is unbound. $^{16}$O  is likely to decay into four separate alpha particles,
though it is already close to become bound.
Moreover, we find that asymmetric isotopes, like $^{28}$O, are strongly unbound systems.
These results suggest that, when lowering the pion mass toward its physical value,
closed shell isotopes are created at first around the traditional magic numbers
and 
the region of $M_{ps} \sim$ 500 MeV marks a
transition between an unbound nuclear chart and the emergence of bound isotopes.

We calculate the root mean square radii,
which are given in Tab.~\ref{tab:radii},
where we show only the central values.
 Although the total binding energies are 15-20\% of the experimental value (Tab.~\ref{tab:He4O16Ca40}),
 the computed charge radii are about the same as the experiment. 
   We also find that the calculated one-nucleon spectral distributions are qualitatively close to
   those of real nuclei even for $M_{ps}$=469~MeV considered here.
   This is due to the fact that
 the heavy nucleon mass ($m_N$=1161.1 MeV) used here 
reduces the
motion of the nucleons inside the nuclei and counterbalances the effect of weak attraction of the lattice nuclear forces
at this pion mass.

\begin{table}[t]
\begin{center}
\begin{tabular}{lcccccc}
\hline
\hline
$E^A_0$ [MeV] &&  $^4$He  && $^{16}$O  && $^{40}$Ca  \\
\hline 
SCGF
&&   -4.80(0.03)  && -17.9 (0.3) (1.8)  &&  -75.4 (6.7) (7.5)  \\
BHF &&  -8.2  && -34.7   &&  -112.7  \\
Exact calc. && -5.09   && --  && --  \\
Experiment  &&  -28.3   &&  -127.7  &&   -342.0 \\
\hline
\multicolumn{3}{l}{Separation into $^4$He clusters: }   && -2.46 (0.3) (1.8)  && 24.5 (6.7) (7.5)  \\
\hline
\hline
\end{tabular}
\caption{
  Ground state energies of  $^4$He, $^{16}$O and $^{40}$Ca
  calculated by self-consistent Green's function (SCGF) method
  using nuclear forces at $M_{PS}$=469~MeV obtained from 3-flavor lattice QCD with the HAL QCD method.
  Comparison is given with those obtained with BHF~\citep{Inoue:2014ipa} and the exact calculation~\citep{Nemura:2014kea}.
The last line is the breakup energy for splitting the system in  $^4$He clusters (of total energy $A/4\times$5.09~MeV).
Taken from~\citep{McIlroy:2017ssf}.
}
\label{tab:He4O16Ca40}
\end{center} 
\end{table}

\begin{table}[t]
\begin{center}
\begin{tabular}{clccccccccc}
\hline
\hline
~~&&~& &~~~~& $^{4}$He &~~~~& $^{16}$O &~~~~& $^{40}$Ca & ~~\\
\hline
&$r_{pt-matter}$[fm]:&&
SCGF
&& 1.67 && 2.64  && 2.97  \\
&&& BHF &\qquad& 2.09  &\qquad& 2.35  &\qquad&  2.78  \\
&&& HF && 1.62 && 2.39  && 2.78  \\
\hline
&$r_{charge}$[fm]:
&&
SCGF
&& 1.89  && 2.79  && 3.10  \\
&&& Experiment && 1.67 && 2.73  && 3.48  \\
\hline
\hline
\end{tabular}
\end{center} 
\caption{ Matter and charge radii of ${}^4$He, $^{16}$O and $^{40}$Ca  at M$_{PS}$=469~MeV
  computed by the SCGF method,
  which are compared with
    those by BHF~\citep{Inoue:2014ipa}, by Hartree-Fock (HF) and by experiments~\citep{DeJager:1987qc,Angeli:2013epw}.
For charge radii, we assumed the physical charge distributions of the nucleons.
Taken from~\citep{McIlroy:2017ssf}.
}
\label{tab:radii}
\end{table}


\begin{figure}[t]
  \centering
  \includegraphics[width=0.32\textwidth]{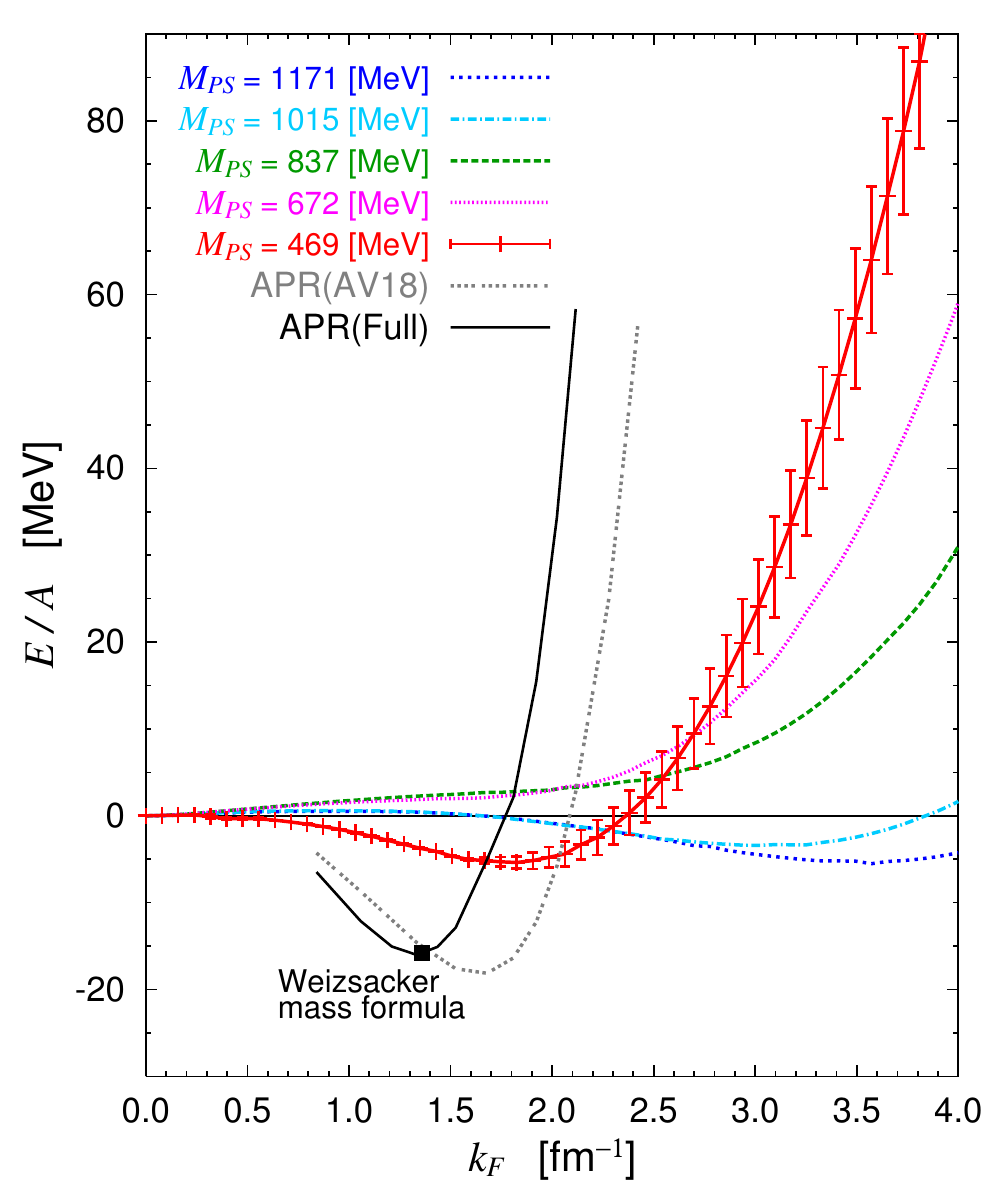}
  \includegraphics[width=0.32\textwidth]{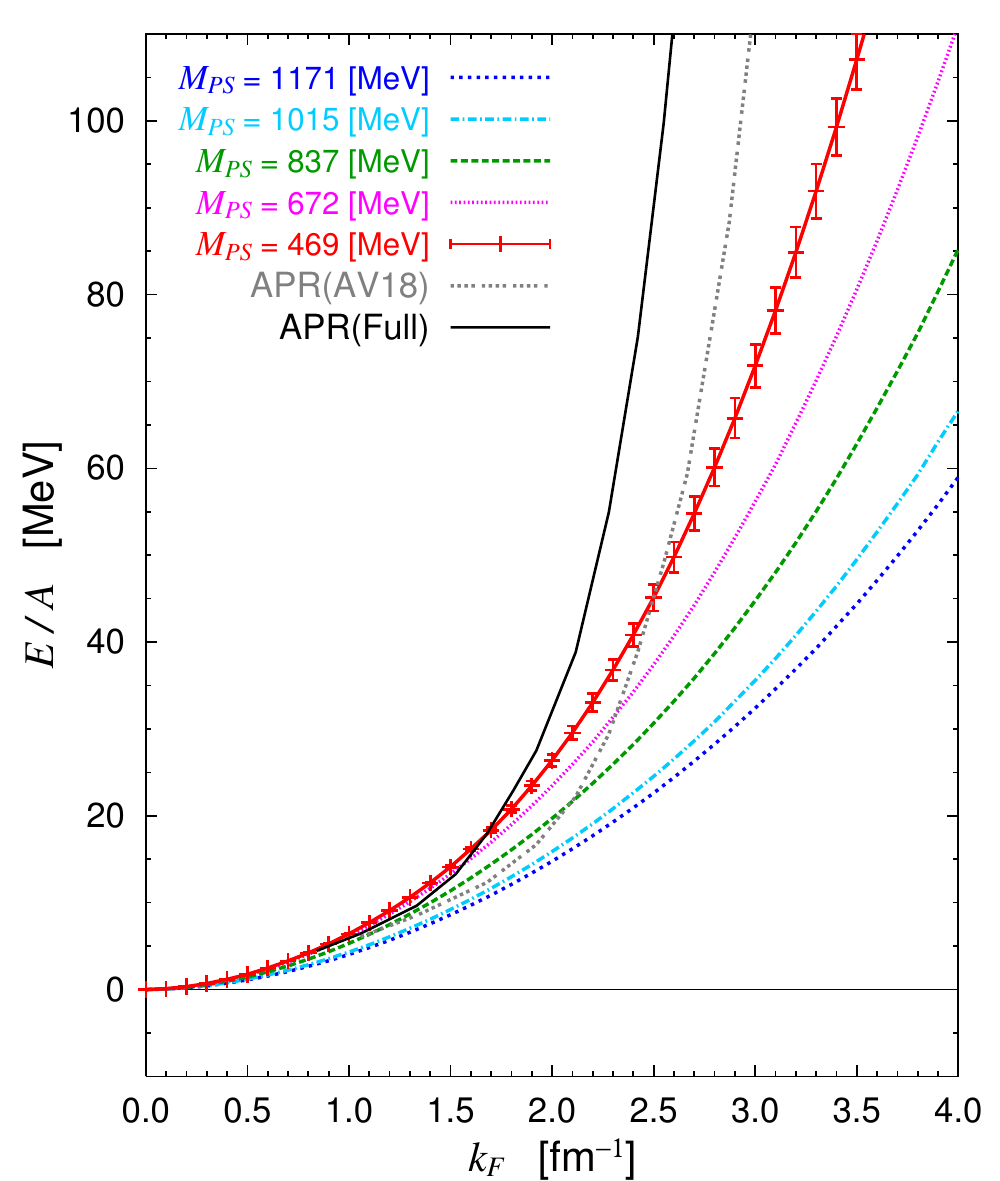}\\[5mm]
  \centering
  \includegraphics[width=0.45\textwidth]{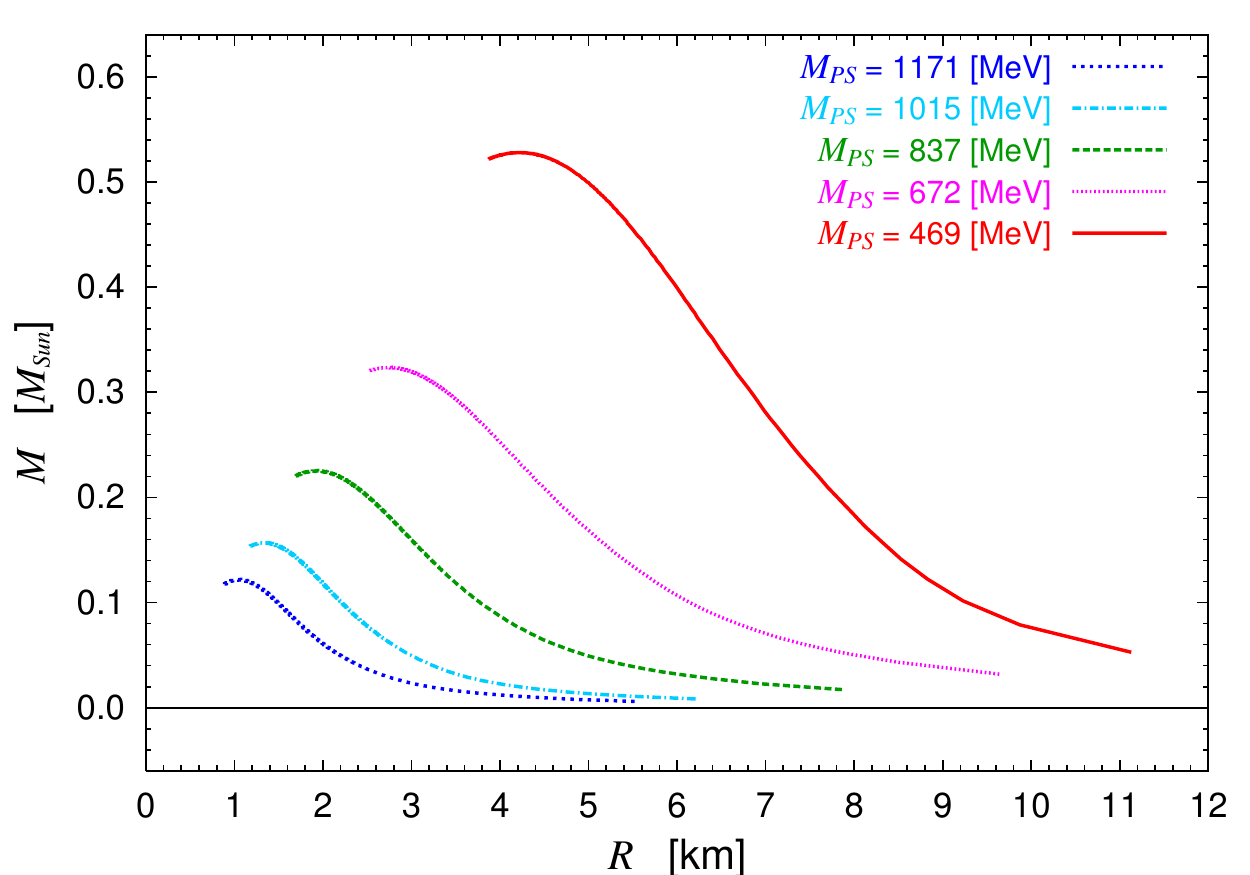}
  \caption{
    (Upper)
    Ground state energy per nucleon ($E/A$) as a function of the Fermi momentum $k_F$
    by the BHF theory with nuclear forces from 3-flavor lattice QCD at $M_{\rm ps}$ = 469-1171 MeV,
    together with that from APR~\citep{Akmal:1998cf} with and without phenomenological three-nucleon forces.
    (Left)
    Results for the symmetric nuclear matter. filled square indicates the empirical saturation point.
    (Right)
    Results for the pure neutron matter.
    (Lower)
    Mass-radius relation of the neutron star
    based on EoS obtained by the BHF theory with nuclear forces
    from 3-flavor lattice QCD at $M_{\rm ps}$ = 469-1171 MeV.
    Figures are  taken from \citep{Inoue:2013nfe}. 
}
 \label{fig:eos-ns}
\end{figure}

We next present the study of properties of dense matter,
namely, Equation of State (EoS) of nuclear matter.
We again employ the nuclear forces
in $^1S_0$, $^3S_1$ and $^3D_1$ channels
obtained in 3-flavor lattice QCD.
To study the quark mass dependence,
we use lattice results for all five quark masses,
at $M_{\rm ps}$ = 469, 672, 837, 1015, 1171 MeV, 
which are shown in Fig.~\ref{fig:su3_pot_phase}.
As a method for a many-body calculation,
we employ the Brueckner-Hartree-Fock (BHF) theory~\citep{Inoue:2013nfe},
which is known to be quantitative enough to give the essential underlying physics
for infinite matter.

In Fig.~\ref{fig:eos-ns} (Upper), we show
the results of
the ground state energy per nucleon ($E/A$)
as a function of the Fermi momentum $k_F$ 
for the symmetric nuclear matter and
the pure neutron matter.
  Shown together
  are the so-called APR EoS~\citep{Akmal:1998cf},
  which are obtained by the variational chain summation method
  from phenomenological nuclear forces
  with (APR(Full)) and without (APR(AV18)) three-nucleon forces.
In Fig.~\ref{fig:eos-ns} (Upper-Left),
we find that
the symmetric nuclear matter becomes a self-bound system with
a saturation point $(k_F, E/A) \simeq (1.83(16) \ {\rm fm}^{-1}, -5.4(5) \ {\rm MeV})$
at the lightest quark mass ($M_{\rm ps} = 469$ MeV).
This is the first time that the saturation in the symmetric nuclear matter
is obtained through first-principles lattice QCD simulations.
The saturation point, however, deviates from the empirical point primarily due to
heavy pion (pseudo-scalar meson) mass in lattice simulation and the lack of three-nucleon forces in BHF calculation.

We also find a nontrivial $M_{\rm ps}$ dependence of the EoS:
the saturation disappears at intermediate pion masses
($M_{\rm ps} = 672, 837$ MeV) and possibly appears again at the heavy pion mass region
($M_{\rm ps} = 1015, 1171$ MeV).  This implies that the saturation 
originates from a subtle balance between short-range repulsion and the intermediate 
attraction of the nuclear force, which have different $m_q$ dependence~\citep{Inoue:2011ai}.
A similar nontrivial $M_{\rm ps}$ dependence originated from the balance between repulsion and attraction is also observed for $NN$ scattering phase shifts, as was discussed in Sec.~\ref{subsec:parity_even}.

In Fig.~\ref{fig:eos-ns} (Upper-Right),
we find that neutron matter is not self-bound due to large Fermi energy.
If we decrease the pion mass, EoS is found to become stiffer.
To further study the impact on phenomena in nuclear astrophysics,
we calculate the mass ($M$) versus the radius ($R$) relation of neutron stars
at each pion mass.
Here, we solve the Tolman-Oppenheimer-Volkoff (TOV) equation 
by using the EoS of the neutron-star matter with
neutron, proton, electron and muon  under the charge neutrality and beta equilibrium,
where we use the standard parabolic approximation for asymmetric nuclear matters.

Shown in Fig.~\ref{fig:eos-ns} (Lower) is the $M$-$R$ relation of the neutron star for different pion masses.
As $M_{\rm ps}$ decreases, the $M$-$R$ curve shifts to the upper right direction,
due to the stiffening of the EoS.
While the maximum mass of the neutron star ($M_{\rm max}$)
in this calculation
is much smaller than the recent observations, $M_{\rm max} \simeq 2 M_{\odot}$,
the deviation is most likely due to the heavy pion masses and
lack of interactions as three-nucleon forces.
A naive extrapolation of $M_{\rm max}$ and the corresponding radius to $M_{\rm ps}=137$ MeV
would give  $M_{\rm max} \sim 2.2 M_{\odot}$ and $R \sim 12 $ km,
which are encouraging for more quantitative studies {in future}.
Another hottest topic in the context of neutron star physics
is the effect of hyperon on the EoS at high density (so-called ``hyperon puzzle'').
Lattice QCD can play an unique role to study this effect by determining 
the hyperon forces which suffer from large uncertainties in experiments to date.
For the on-going study in this direction, see~\citep{Inoue:2018axd}.

\subsection{Challenge: nuclear forces near the physical pion {mass}}
\label{subsec:NN_pot_phys}

While the results of nuclear forces at heavy pion masses
are very intriguing and useful to extract the physical picture of nuclear forces,
the quantitative results require the study at the {physical} pion  mass.
Note that the pion mass dependence of nuclear forces is quite non-trivial
as discussed in Secs.~\ref{subsec:parity_even} and \ref{subsec:application}, so the direct calculation
near the physical point is desirable.

\begin{figure}[t]
\begin{center}
\vspace*{-4mm}
\includegraphics[width=0.95\textwidth]{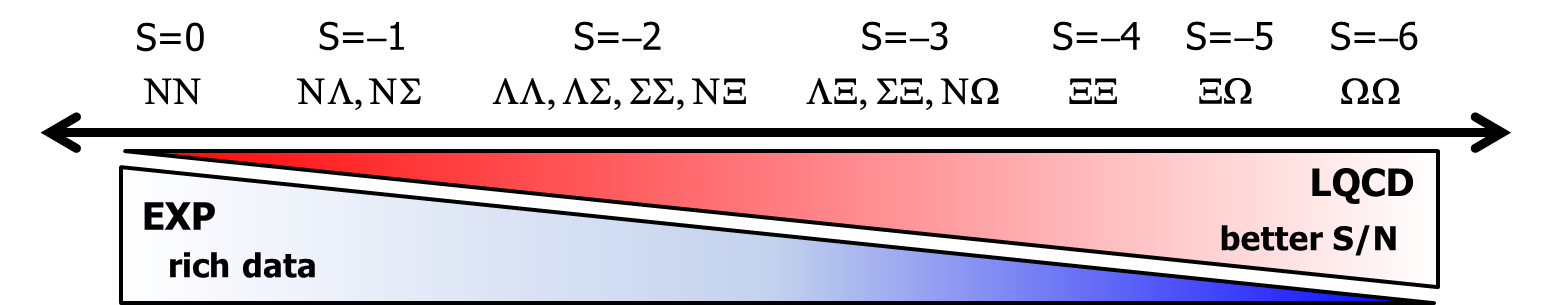}
\vspace*{-2mm}
\caption{
\label{fig:lat_exp}
An illustration of
the complementary role of lattice QCD and experiments
in the determination of baryon forces.
}
\end{center}
\vspace*{-3mm}
\end{figure}

To this end, we have recently performed the first calculation of nuclear forces
near the physical {up, down} and strange quark masses.
Actually,
our aim is to calculate not only nucleon forces but also hyperon forces,
hereby achieve the comprehensive determination of
two-baryon interactions from the strangeness $S = 0$ to $-6$
in parity-even channels ($S$- and $D$-waves).
As mentioned before, the statistical fluctuations in lattice QCD 
are smaller (larger) for larger (smaller) quark masses,
and thus the results have better precision in sectors involving more number of strange quarks
(larger strangeness $|S|$).
On the other hand, experiments in such larger $|S|$ sectors are more difficult
due to the short life time of hyperons.
Therefore, lattice QCD studies and experiments are complementary with each other
in the determination of baryon forces (See Fig.~\ref{fig:lat_exp}).

(2+1)-flavor gauge configurations are generated on
a $L^3\times T = 96^3\times 96$ lattice
with the RG-improved Iwasaki gauge action and
non-perturbatively $\mathcal{O}(a)$-improved Wilson quark action 
and APE stout smearing.
The lattice spacing is $a \simeq 0.0846$ fm ($a^{-1} \simeq 2.333$ GeV),
so that spatial extent, $L = 8.1$ fm, is sufficiently large to accommodate two baryons in a box.
Quark masses are tuned so as to be near the physical point,
and the hadron masses are found to be $(m_\pi, m_K, m_N) \simeq (146, 525, 955)$ MeV.
NBS correlation functions for two-baryon systems are calculated
for 55 channels in total and we extract the central and tensor forces
in parity-even channel at the LO analysis for the derivative expansion (work in progress, and see also~\citep{Doi:2017zov}).
In order to make this first calculation a reality,
``trinity'' of state-of-the-art developments was crucial:
(a) time-dependent HAL QCD method (theory),
(b) unified contraction algorithm (software)
and
(c) K-computer, HOKUSAI and HA-PACS supercomputers (hardware).

\begin{figure}[t]
\centering
\includegraphics[angle=0,width=0.32\textwidth]{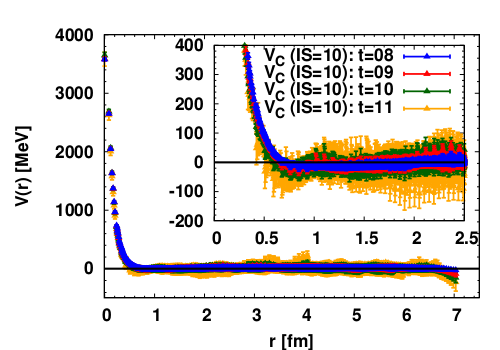}
\includegraphics[angle=0,width=0.32\textwidth]{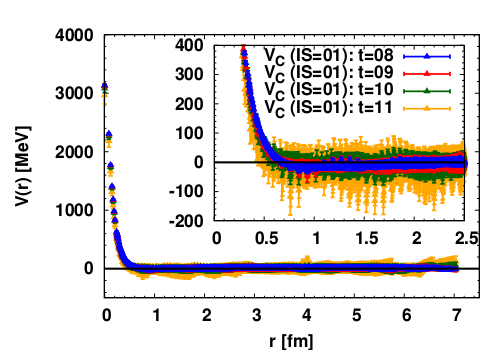}
\includegraphics[angle=0,width=0.32\textwidth]{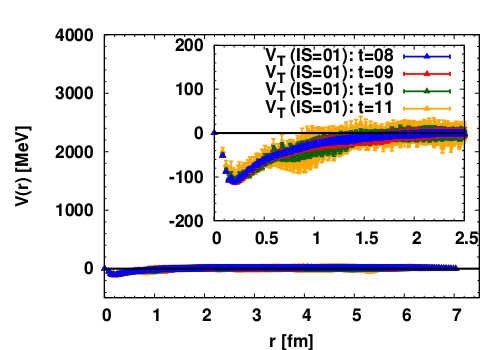}
\caption{
\label{fig:pot:NN:phys}
Nuclear forces from (2+1)-flavor lattice QCD near the physical point, $m_\pi = 146$ MeV.
 The central force in the $^1S_0$ channel (Left). 
 The central force (Middle) and the tensor  force (Right)
 in the $^3S_1$-$^3D_1$ channel.
}
\end{figure}

Shown in Fig.~\ref{fig:pot:NN:phys}
are the results for
the central force in the $^1S_0$ channel (Left),
and the central force (Middle) and tensor force (Right) in the $^3S_1$-$^3D_1$ channel.
As noted above, nuclear forces are the most challenging interactions
in lattice QCD calculation,
and one can see that the results suffer from large statistical fluctuations.
Nevertheless, the obtained results exhibit several interesting properties.

First of all, the repulsive core at short-range is clearly observed
in central forces.
In order to clarify the physical picture for the repulsive core,
it is useful to compare them with hyperon forces obtained in the same lattice setup.
We find that the strength of repulsive core (or attractive core) highly depends on
the flavor SU(3) (SU(3)$_f$) classification, in a consistent way with
the quark Pauli blocking effect.
In addition, if we compare interactions which belong to the same
SU(3)$_f$ classification, such as $NN (^1S_0)$ and $\Xi\Xi (^1S_0)$
both of which belong to 27-plet,
we find that the strength differs in a way
which can be understood from the viewpoint of one-gluon-exchange
(e.g., repulsive core in $NN (^1S_0)$ is stronger than that in $\Xi\Xi (^1S_0)$).
These observations confirm the physical picture for the repulsive core
obtained in the 3-flavor lattice QCD (Sec.~\ref{subsec:parity_even}),
the quark Pauli blocking effect and the one-gluon-exchange,
is relevant even at physical quark masses.
See also~\citep{Park:2019bsz} for a more detailed study on this point.

At middle and long distances, while statistical errors are quite large,
we observe that the central force is attractive,
resembling the phenomenological potential as one-pion-exchange potential (OPEP).
The tensor force has relatively smaller  statistical errors than the central forces,
showing that 
the tensor force becomes stronger  (with a negative sign)
and has a longer tail,  
as compared with the tensor forces at heavier pion masses (Sec.~\ref{subsec:parity_even}).
This property can be understood by the {picture of OPEP.
These results are encouraging and
serve as the first step to establish a direct connection between QCD and nuclear physics.
At the same time, statistical errors remain to be large and there also exist systematic errors
associated with inelastic state contaminations.
The studies to resolve these issues are in progress,
and the second generation calculation is planned on the forthcoming Exascale computer, ``Fugaku'
(See https://postk-web.r-ccs.riken.jp/).

\section{Dibaryons}
\label{sec:dibaryon}

Before closing this review, we present our latest results on dibaryon  searches in lattice QCD near the physical pion mass~\citep{Doi:2017zov}. 
A dibaryon, a bound-state (or a resonance) with a baryon number $B=2$ in QCD, can be classified 
in the SU(3)$_f$  as
\begin{equation}
{\bf 8} \otimes {\bf 8} = {\bf 27} \oplus {\bf 8_s}  \oplus {\bf 1}   \oplus {\bf \overline{10}}  \oplus {\bf 10}  \oplus {\bf 8_a}  
\end{equation}
for the octet-octet system, where the deuteron, {the only} stable dibaryon observed  in nature so far,
appears in the ${\bf \overline{10}}$ representation while $H$ dibaryon has been predicted in the {\bf 1} representation~\citep{Jaffe:1976yi} and actively investigated in lattice QCD~\citep{Inoue:2010hs,Inoue:2010es,Inoue:2011ai,Beane:2010hg,Francis:2018qch}.
{For the decuplet-octet system, the classification leads to}
\begin{equation}
{\bf 10} \otimes {\bf 8} = {\bf 35} \oplus {\bf 8}  \oplus {\bf 10}    \oplus {\bf 27} 
\end{equation}
{and} $N\Omega$ ( $N\Delta$) dibaryon 
has been predicted in the {\bf 8} ({\bf 27}) representation~\citep{Goldman:1987ma,Oka:1988yq,Dyson:1964xwa},
and 
\begin{equation}
{\bf 10} \otimes {\bf 10} = {\bf 28} \oplus {\bf 27}  \oplus {\bf 35}   \oplus {\bf \overline{10}} 
\end{equation}
for the decuplet-decuplet system, where $\Omega\Omega$ dibaryon has been predicted in the {\bf 28} representation~\citep{Zhang:1997ny} while 
$\Delta\Delta$ has been predicted in the ${\bf \overline{10}}$~\citep{Dyson:1964xwa,Kamae:1976at}
and the corresponding $d^*(2380)$ was indeed observed~\citep{Adlarson:2011bh}.
Note that among decuplet baryons, only $\Omega$ is stable against strong decays.

\subsection{The most strange dibaryon}
\label{subsec:OmgOmg}

\begin{figure}[t]
\centering
 \includegraphics[width=0.48\textwidth]{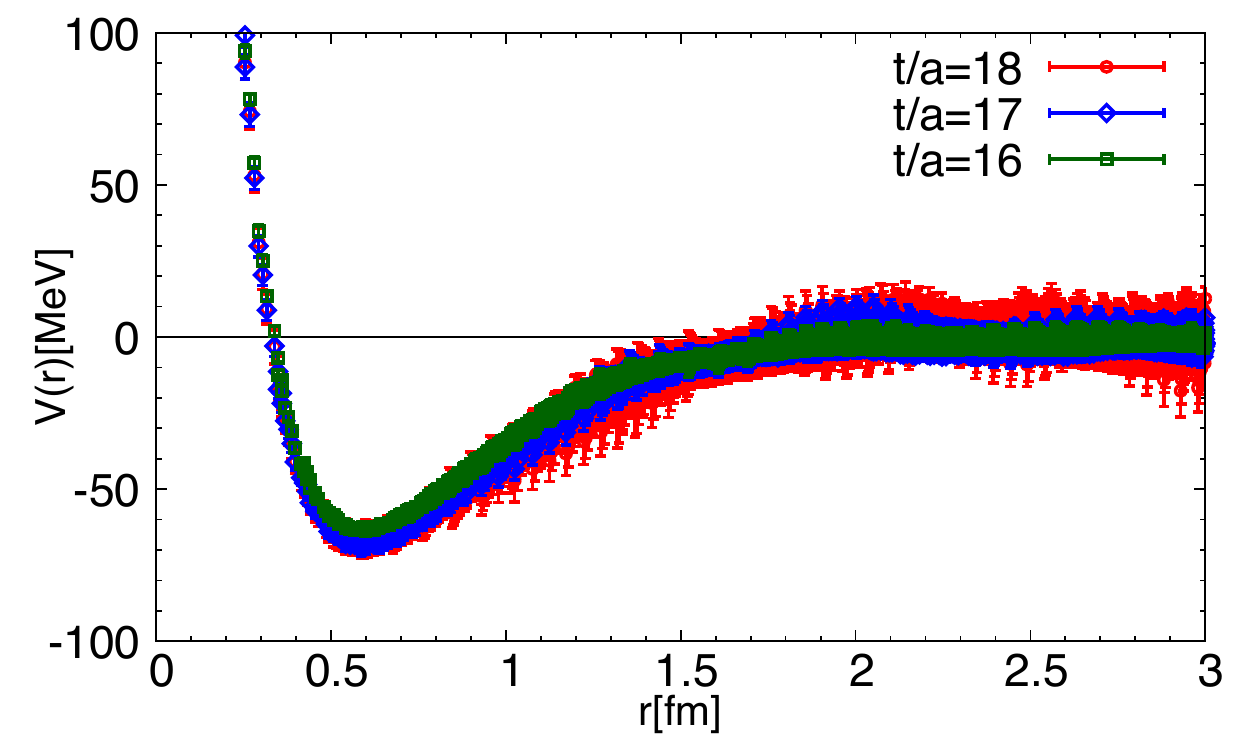}
 \includegraphics[width=0.48\textwidth]{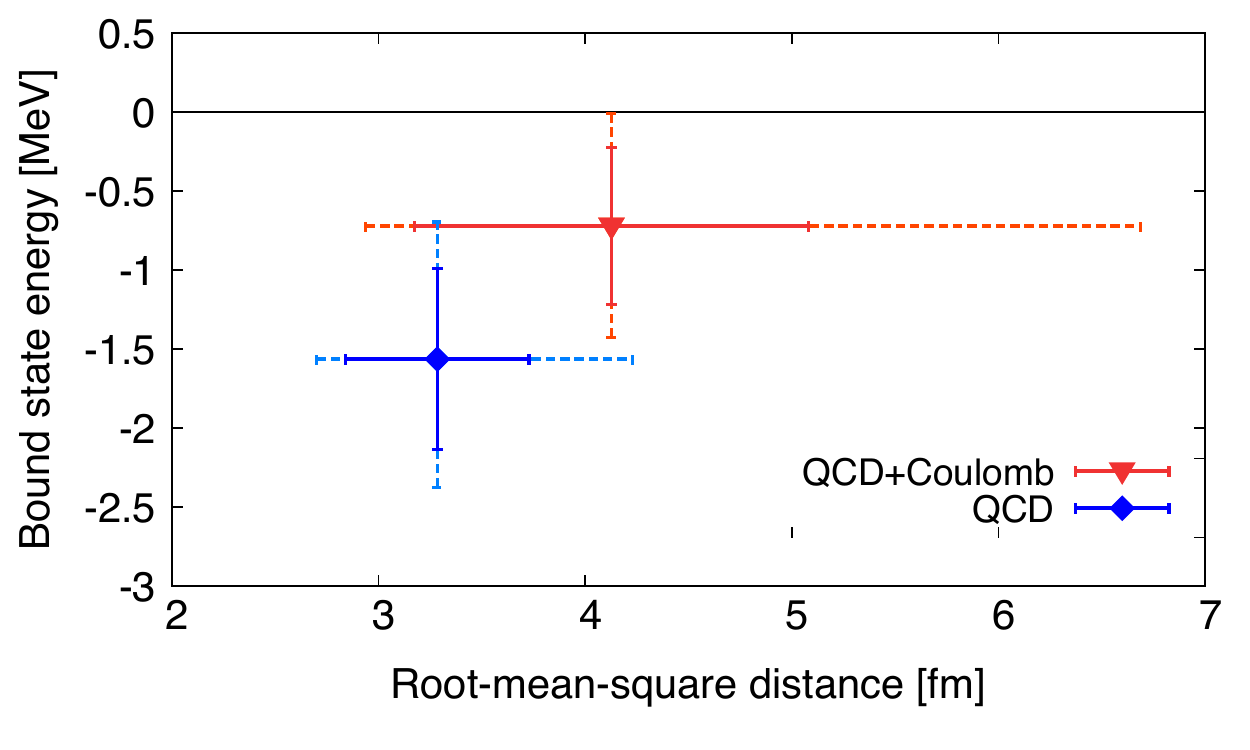} \\[5mm]
\centering
 \includegraphics[width=0.48\textwidth]{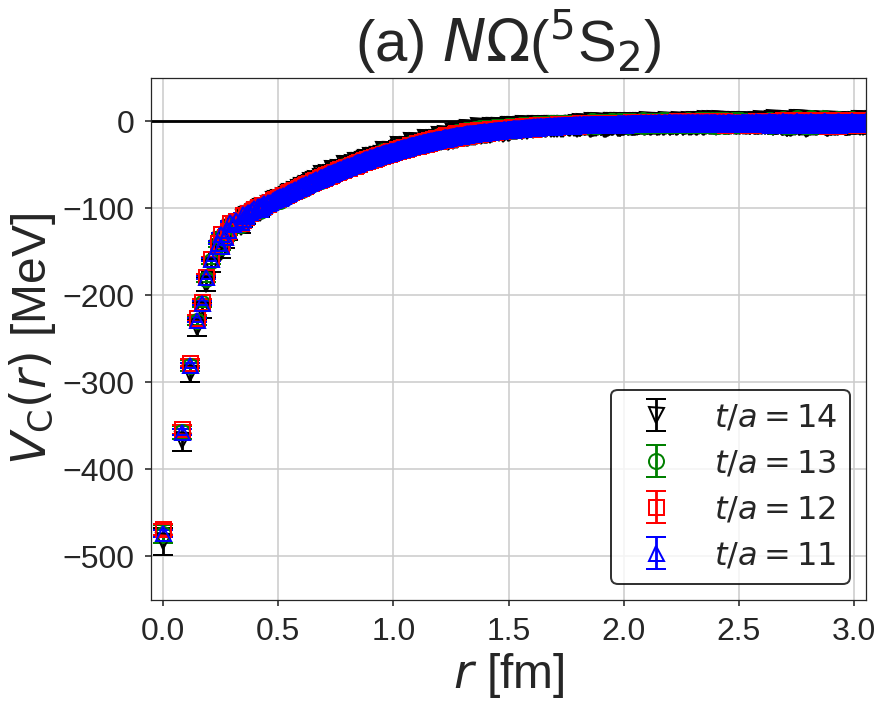}
 \includegraphics[width=0.48\textwidth]{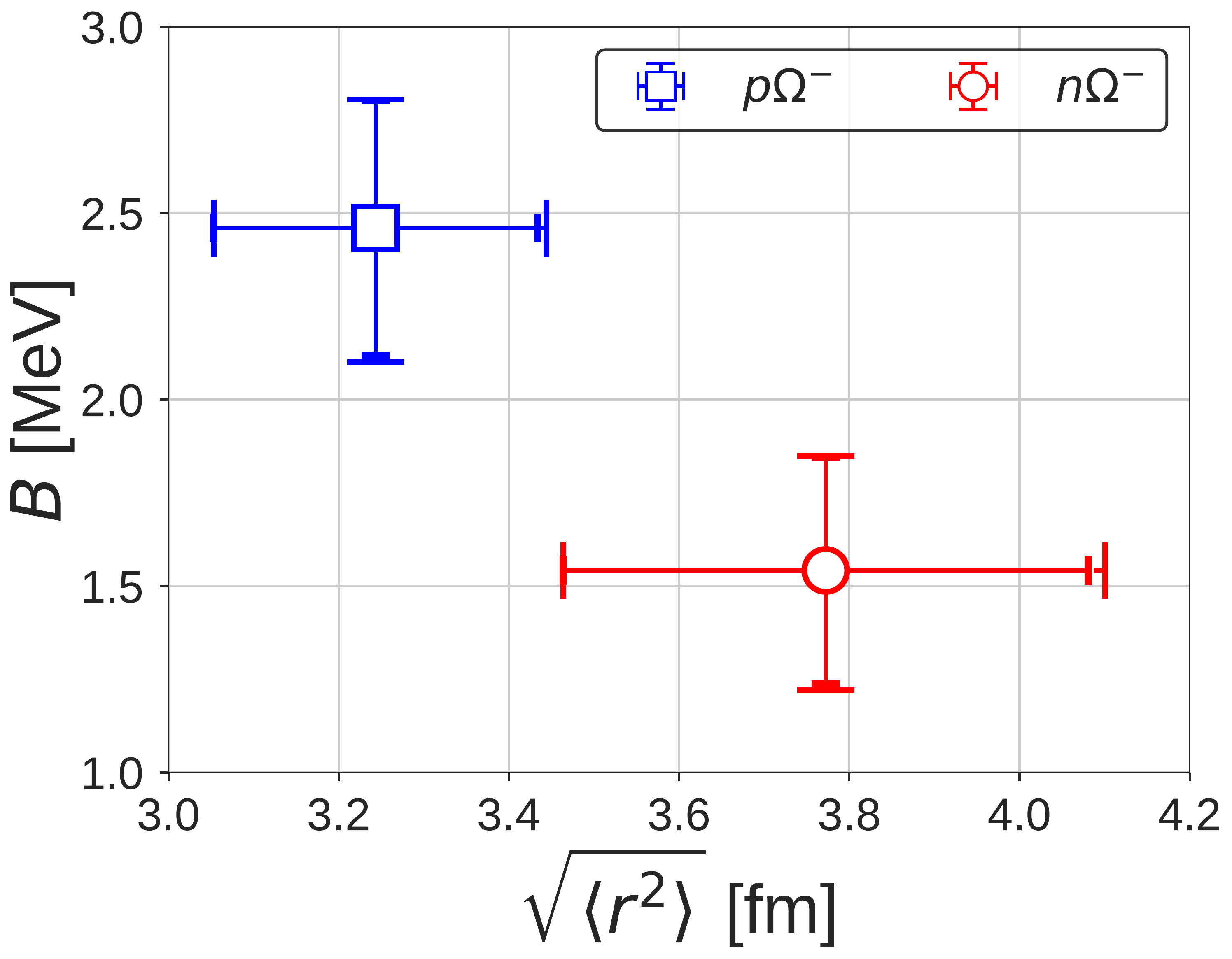}
   \caption{
     (Upper)
       The $\Omega\Omega$ system in the $^1S_0$ channel
       in $2+1$ flavor QCD at $m_\pi \simeq 146$ MeV and $a\simeq 0.0846$ fm on a (8.1 fm)$^3$ box. 
 (Left) The $\Omega\Omega$ potential $V(r)$ 
 at $t/a =16,17,18$. 
 (Right) The binding energy of the $\Omega\Omega$ system and the root-mean-square distance between two $\Omega$'s
{are shown by blue solid diamond (red solid triangle),}
 calculated from the $\Omega\Omega$ potential $V(r)$ at $t/a=17$
 without {(with)} the Coulomb repulsion. 
 Taken from \citep{Gongyo:2017fjb}.
 (Lower)
   The $N\Omega$ system in the $^5S_2$ channel
   with the same lattice setup for $\Omega\Omega$.
    (Left) The  $N\Omega$ potential $V_C(r)$ 
   at $t/a=11,12,13,14$.
 (Right) The binding energy and the root-mean-square distance 
 for the $n\Omega^-$ (red open circle) and 
 $p\Omega^-$ (blue open square).
 Taken from \citep{Iritani:2018sra}.  
 }
 \label{fig:OmegaOmega-NOmega}
\end{figure} 
We first consider the $\Omega\Omega$ system in the {\bf 28} representation of SU(3)$_f$
{in the $^1S_0$ channel}~\citep{Gongyo:2017fjb}.

Fig.~\ref{fig:OmegaOmega-NOmega} (Upper-Left) shows $\Omega\Omega$ potentials at $t/a=16,17,18$,
which has qualitative features similar to the central potentials for $NN$
but whose repulsion is weaker and attraction is shorter-ranged.
This potential predicts an existence of one shallow bound state,
whose binding energy
is plotted in Fig.~\ref{fig:OmegaOmega-NOmega} (Upper-Right) as a function of the root-mean-square distance,  with (red) and  without (blue) Coulomb repulsion between $\Omega\Omega$. 
  We may call  this $\Omega\Omega$ bound state {\it ``the most strange dibaryon''}.
Such a system can be best
searched experimentally by two-particle correlations in relativistic heavy-ion collisions~\citep{Morita:2019rph}.

\subsection{$N\Omega$ {dibaryon}}
\label{subsec:NOmg}

We next consider the $N\Omega$ system with $S=-3$ in the {\bf 8} representation
{in the $^5S_2$ channel}~\citep{Iritani:2018sra}.
Near the physical point, $N\Omega$($^5$S$_2$) may couple to
{$D$}-wave octet-octet channels below the $N\Omega$ threshold ($\Lambda\Xi$ and $\Sigma\Xi$),
but such couplings are assumed to be small in this calculation.

Fig.~\ref{fig:OmegaOmega-NOmega} (Lower-Left) shows the $N\Omega$ potential at $t/a=11$--$14$, 
which {is attractive} at all distances without repulsive core, so that
one bound state appears in this channel.
In Fig.~\ref{fig:OmegaOmega-NOmega} (Lower-Right),  the binding energy (vertical) and the the root-mean-square distance (horizontal) are plotted for $n\Omega^-$ with no Coulomb interaction (red) and $p\Omega^-$ with  Coulomb attraction (blue). 
These binding energies are much smaller than $B=18.9(5.0)(^{+12.1}_{-1.8})$ MeV at heavy pion mass $m_\pi = 875$ MeV~\citep{Etminan:2014tya}.
  Such a $N\Omega$ state
  can be searched through two-particle correlations in relativistic nucleus-nucleus collisions~\citep{Morita:2019rph}
  and an experimental indication was also reported~\citep{STAR:2018uho}.

\subsection{Comparison {among dibaryons}}
\label{subsec:dibaryon}

\begin{figure}[t]
\centering
 \includegraphics[width=0.5\textwidth]{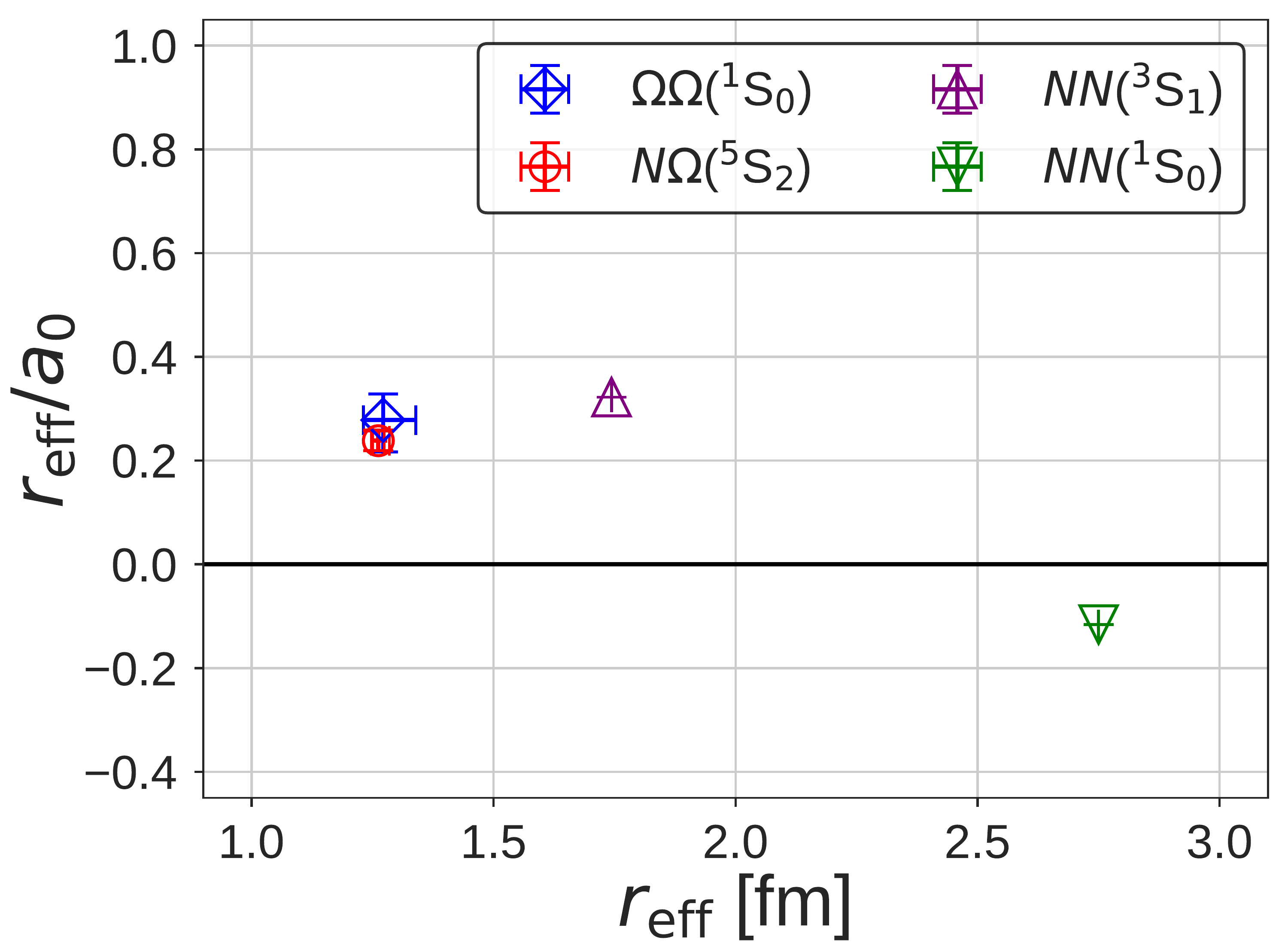}
 \caption{The ratio of the effective range and the scattering length $r_{\rm eff}/a_0$ as a function of $r_{\rm eff}$
     for $\Omega\Omega (^1S_0)$ (blue open diamond) and $N\Omega(^5S_2)$ (red open circle)
   obtained in lattice QCD, as well as for $NN(^3S_1)$ (purple open up-triangle) and 
   $NN(^1S_0)$ (green open down-triangle) in experiments.
   Taken from  \citep{Iritani:2018sra}.
   The sign convention for the scattering length is opposite to Eq.~(\ref{eq:ERE_nlo}) in this figure.
  }
 \label{fig:Unitary}
\end{figure} 
Let us consider the scattering length $a_0$ and the effective range $r_{\rm eff}$ for 
$\Omega\Omega$($^1$S$_0$) and $N\Omega$($^{{5}}$S$_2$).
In Fig.~\ref{fig:Unitary}, the ratio $r_{\rm eff}/a_0$ as a function of $r_{\rm eff}$ are plotted for 
$\Omega\Omega$($^1$S$_0$) and $N\Omega$($^{{5}}$S$_2$) obtained in lattice QCD near {the} physical pion mass, together with the experimental values for $NN$($^3$S$_1$) (deuteron) and
$NN$($^1$S$_0$) (di-neutron). Small values of $\vert r_{\rm eff}/a_0\vert$ in all cases indicate
that these systems are located close to the unitary limit.

\section{Conclusions}
\label{sec:conclusion}

In this paper, we have reviewed the recent progress in lattice QCD
study of baryon-baryon interactions by the HAL QCD method.
We first presented the detailed account on how to define the
potentials in quantum field theories such as QCD.
The key observation is that
the Nambu-Bethe-Salpeter (NBS) wave functions
contain the information of scattering phase shifts below inelastic threshold
in their asymptotic behaviors outside the range of the interactions.
The potentials at the interaction region can then be defined
through the NBS wave functions
so as to be 
faithful to the phase shifts by construction,
where the non-locality of the potential
is defined by the derivative expansion.
In addition, by constructing the potentials in energy-independent way,
the potentials can be extracted from
two-baryon correlation functions without the requirement of
the ground state saturation.

We then made a detailed comparison between the HAL QCD method
and the conventional method,
in which phase shifts are obtained from the finite volume energies
through  the L\"uscher's formula.
We pointed out that, while the validity of the latter method
relies on the ground state 
saturation
of the correlation function,
its practical procedure for multi-baryon systems (``direct method'') 
so far has utilized
only the plateau-like structures of the effective energies
at Euclidean times much earlier than
the inverse of the lowest excitation energy.
We showed theoretical and numerical evidences that
such a procedure generally leads to unreliable results
due to the contaminations from the  elastic excited states:
For instance,
the results were found to be dependent on the operators
and
unphysical behaviors were exposed by the normality check.
This invalidates the claim of the literature in the direct method
that $NN$ bound states exist for pion masses heavier than 300 MeV.

On the other hand,
HAL QCD method is free from such a serious problem
since the signal of potentials can be extracted
from not only the ground state but also elastic excited states.
While there instead exists the truncation error of the derivative expansion of the potential,
the calculation of the higher order term in the derivative expansion
showed that the convergence of the expansion is sufficiently good at low energies.
Furthermore, utilizing the finite volume eigenmodes of the HAL QCD Hamiltonian,
the excited state contaminations in the direct method were explicitly quantified.
It turns out that
the plateau-like structures of effective energies
at early time slices are indeed pseudo-plateaux
contaminated by elastic excited states and that
the plateau for the ground state is realized only at a much larger time
corresponding to the inverse of the lowest excitation energy gap.
We also showed that,
 by employing an optimized operator utilizing
the finite volume eigenmodes,
the effective energies from the correlation functions 
give consistent results with those from the HAL QCD potential.
Thus the long-standing issue on the consistency between
the conventional method based on the L\"uscher's formula
and the HAL QCD method was positively resolved.


After establishing the reliability of the HAL QCD method,
we presented the numerical results of nuclear forces from the HAL QCD method
at various lattice QCD setups.
At heavy pion masses, where good signal-to-noise ratio can be achieved in lattice QCD,
we observed that the obtained $NN$ potentials
in the parity-even channel ($^1S_0$, $^3S_1$-$^3D_1$) reproduce
the qualitative features
of the phenomenological potentials,
namely, attractive wells at long and medium
distances, accompanied with  repulsive cores at short distance in the central potentials
and the strong tensor force. 
The net interactions were found to be attractive,
which however are not strong enough to form a bound $NN$ state, probably
due to the heavy pion masses.
We observed that
the tail structures are enhanced at lighter pion masses,
which can be understood from the viewpoint of one-pion exchange contributions.
We also found 
the repulsive cores are enhanced at lighter pion masses.
Combined with our systematic studies including hyperon forces,
the nature of repulsive cores was found to be well described
by the quark Pauli blocking effect together with the one-gluon-exchange contribution.

The HAL QCD method can be extended to determine more complicated nuclear forces,
such as spin-orbit forces and three-nucleon forces.
In this paper, we considered
two-nucleon systems in the parity-odd channels ($^1P_1$, $^3P_0$, $^3P_1$, $^3P_2$-$^3F_2$ channels)
and calculated spin-orbit forces as well as central and tensor forces.
We found that qualitative features of experimental results are generally well reproduced,
while the magnitudes differ due to the heavy pion mass.
In particular, we observed the strong (and negative) spin-orbit forces,
which lead to the attraction in the $^3P_2$ channel.
Three-nucleon forces were studied in the triton channel,
$(I, J^P)=(1/2,1/2^+)$,
thank to  the unified contraction algorithm (UCA),
which can enormously speed up calculations of multi-baryon correlation functions.
It was found that there exists a repulsive three-nucleon forces at short distances.
These observations are of interest
in the context of not only the structures of nuclei
but also those of neutron stars,
e.g., $P$-wave superfluidity and the maximum mass of neutron stars.


We carried out the
applications to nuclei, nuclear equation of state (EoS) and structure of neutron stars
based on lattice nuclear forces at heavy quark masses.
We performed ab initio self-consistent Green's function (SCGF) calculations
for closed shell nuclei 
with nuclear forces at $M_{ps}$=469~MeV in the SU(3) limit of QCD.
We found that $^4$He, $^{40}$Ca nuclei are bound
and $^{16}$O is close to become bound,
while asymmetric isotopes are strongly unbound.
The results suggest that, when lowering the pion mass toward its physical value,
islands of stability appear at first around the traditional doubly magic numbers.
The nuclear EoS was also studied
by the BHF theory
with nuclear forces in the flavor SU(3) limit.
We found that the saturation property appears in the symmetric nuclear matter
at $M_{ps}$=469~MeV.
A mass-radius
relation of the neutron star was also studied
based on the EoS obtained from lattice nuclear forces
and
we observed a tendency that
the maximum mass of a neutron star increases as the pion mass decreases.


Finally, we presented the first lattice QCD study of baryon forces
near the physical pion mass in the parity-even channel.
The computation is quite challenging particularly for nuclear forces
due to bad signal-to-noise ratio near the physical point.
Nevertheless, we observed prominent characteristics of nuclear forces,
such as
repulsive cores at short distances as well as 
attractive interactions at mid and long distances
in central forces,
and 
a strong (and negative) tensor force.
We also presented the results for the hyperon forces obtained near the physical point.
We found that both $\Omega\Omega (^1S_0)$ and $N\Omega (^5S_2)$ systems
have   strong attractions,
and 
(quasi) bound dibaryons are formed near the unitary limit.
These systems could be searched experimentally
through two-particle correlations in relativistic nucleus-nucleus collisions.


Present results shown in this paper
already indicate a clear pathway which connects nuclear physics with
its underlying theory of the strong interaction, QCD.
While there remain many challenges to accomplish researches in this direction,
there is no doubt that successive theoretical developments together
with next-generation supercomputers will further deepen the connection
between the two.
The outcome is also expected to play a crucial role to understand the
nuclear astrophysical phenomena such as supernova explosions 
and
mergers of binary neutron stars,
as well as the nucleosynthesis associated with these explosive phenomena.

\section*{Acknowledgements}
\label{sec:ack}

This work is supported in part by the Grant-in-Aid of the Japanese Ministry of Education, Sciences and Technology, Sports and Culture (MEXT) for Scientific Research (Nos. JP16H03978, JP18H05236, JP18H05407, JP19K03879),
by HPCI programs (hp120281, hp130023, hp140209, hp150223, hp150262, hp150085, hp160211, hp160093, hp170230, hp170170, hp180179, hp180117, hp190160, hp190103),
by a priority issue (Elucidation of the fundamental laws and evolution of the universe) to be tackled by using Post ``K" Computer, and by Joint Institute for Computational Fundamental Science (JICFuS).
The authors thank members of the HAL QCD Collaboration
  for providing lattice QCD results
  and for fruitful collaborations based on which this paper is prepared.
  Figure~\ref{fig:su3_pot_phase} is
    reprinted from
    \citep{Inoue:2011ai}
    with permission from Elsevier.
    Figure~\ref{fig:p-odd} is  taken from \citep{Murano:2013xxa},
    and
    Figures~\ref{fig:OmegaOmega-NOmega} and \ref{fig:Unitary} are taken from \citep{Iritani:2018sra},
    under the term of the
    https://creativecommons.org/licenses/by/3.0/ .



\bibliographystyle{frontiersinHLTH_FPHY} 
\bibliography{HAL}

\end{document}